\documentclass[longauth]{aa}

\usepackage{graphicx}
\usepackage{natbib}
\usepackage{scalerel}
\usepackage[compatibility=false, labelfont=bf, font=small]{caption}

\usepackage[table]{xcolor}

\newcommand{\Muv}{M_{\rm{UV}}}
\newcommand{\spitzer}{\textit{Spitzer}/IRAC\xspace}
\newcommand{\zphot}{z_{\rm{phot}}}

\usepackage{txfonts}
\usepackage[pdfencoding=auto,psdextra]{hyperref}
\hypersetup{
    colorlinks=true,
    linkcolor=blue,
    filecolor=red,      
    urlcolor=blue,
    citecolor=blue
}
\makeatletter
\renewcommand*\aa@pageof{, page \thepage{} of \pageref*{LastPage}}
\makeatother

\usepackage[utf8]{inputenc}

\usepackage[switch, modulo]{lineno}

\usepackage{euclid}

\defcitealias{Finkelstein15}{F15}
\defcitealias{Varadaraj23}{V23}
\defcitealias{Bowler17}{B17}
\defcitealias{Bouwens21}{B21}

\newif\ifrevision
\revisionfalse

\usepackage{xcolor}

\ifrevision
  
\else
  
\fi

\begin{document}

\title{\Euclid: Discovery of bright $\boldsymbol{z\simeq7}$ Lyman-break galaxies in UltraVISTA and \Euclid COSMOS\thanks{This paper is published on
       behalf of the Euclid Consortium}}

\newcommand{\orcid}[1]{}                           
\author{R.~G.~Varadaraj\orcid{0009-0006-9953-6471}\thanks{\email{rohan.varadaraj@physics.ox.ac.uk}}\inst{\ref{aff1}}
\and R.~A.~A.~Bowler\orcid{0000-0003-3917-1678}\inst{\ref{aff2}}
\and M.~J.~Jarvis\orcid{0000-0001-7039-9078}\inst{\ref{aff1},\ref{aff3}}
\and J.~R.~Weaver\orcid{0000-0003-1614-196X}\inst{\ref{aff4}}
\and E.~Ba\~nados\orcid{0000-0002-2931-7824}\inst{\ref{aff5}}
\and P.~Holloway\orcid{0009-0002-8896-6100}\inst{\ref{aff1},\ref{aff6}}
\and K.~I.~Caputi\orcid{0000-0001-8183-1460}\inst{\ref{aff7},\ref{aff8}}
\and S.~M.~Wilkins\orcid{0000-0003-3903-6935}\inst{\ref{aff9}}
\and D.~Yang\orcid{0000-0002-6769-0910}\inst{\ref{aff10}}
\and B.~Milvang-Jensen\orcid{0000-0002-2281-2785}\inst{\ref{aff8},\ref{aff11}}
\and L.~Gabarra\orcid{0000-0002-8486-8856}\inst{\ref{aff1}}
\and P.~A.~Oesch\orcid{0000-0001-5851-6649}\inst{\ref{aff12},\ref{aff8},\ref{aff11}}
\and A.~Amara\inst{\ref{aff13}}
\and S.~Andreon\orcid{0000-0002-2041-8784}\inst{\ref{aff14}}
\and N.~Auricchio\orcid{0000-0003-4444-8651}\inst{\ref{aff15}}
\and C.~Baccigalupi\orcid{0000-0002-8211-1630}\inst{\ref{aff16},\ref{aff17},\ref{aff18},\ref{aff19}}
\and M.~Baldi\orcid{0000-0003-4145-1943}\inst{\ref{aff20},\ref{aff15},\ref{aff21}}
\and S.~Bardelli\orcid{0000-0002-8900-0298}\inst{\ref{aff15}}
\and A.~Biviano\orcid{0000-0002-0857-0732}\inst{\ref{aff17},\ref{aff16}}
\and E.~Branchini\orcid{0000-0002-0808-6908}\inst{\ref{aff22},\ref{aff23},\ref{aff14}}
\and M.~Brescia\orcid{0000-0001-9506-5680}\inst{\ref{aff24},\ref{aff25}}
\and S.~Camera\orcid{0000-0003-3399-3574}\inst{\ref{aff26},\ref{aff27},\ref{aff28}}
\and G.~Ca\~nas-Herrera\orcid{0000-0003-2796-2149}\inst{\ref{aff29},\ref{aff10}}
\and V.~Capobianco\orcid{0000-0002-3309-7692}\inst{\ref{aff28}}
\and C.~Carbone\orcid{0000-0003-0125-3563}\inst{\ref{aff30}}
\and J.~Carretero\orcid{0000-0002-3130-0204}\inst{\ref{aff31},\ref{aff32}}
\and M.~Castellano\orcid{0000-0001-9875-8263}\inst{\ref{aff33}}
\and G.~Castignani\orcid{0000-0001-6831-0687}\inst{\ref{aff15}}
\and S.~Cavuoti\orcid{0000-0002-3787-4196}\inst{\ref{aff25},\ref{aff34}}
\and K.~C.~Chambers\orcid{0000-0001-6965-7789}\inst{\ref{aff35}}
\and A.~Cimatti\inst{\ref{aff36}}
\and C.~Colodro-Conde\inst{\ref{aff37}}
\and G.~Congedo\orcid{0000-0003-2508-0046}\inst{\ref{aff38}}
\and C.~J.~Conselice\orcid{0000-0003-1949-7638}\inst{\ref{aff2}}
\and L.~Conversi\orcid{0000-0002-6710-8476}\inst{\ref{aff39},\ref{aff40}}
\and Y.~Copin\orcid{0000-0002-5317-7518}\inst{\ref{aff41}}
\and F.~Courbin\orcid{0000-0003-0758-6510}\inst{\ref{aff42},\ref{aff43},\ref{aff44}}
\and H.~M.~Courtois\orcid{0000-0003-0509-1776}\inst{\ref{aff45}}
\and M.~Cropper\orcid{0000-0003-4571-9468}\inst{\ref{aff46}}
\and A.~Da~Silva\orcid{0000-0002-6385-1609}\inst{\ref{aff47},\ref{aff48}}
\and H.~Degaudenzi\orcid{0000-0002-5887-6799}\inst{\ref{aff12}}
\and G.~De~Lucia\orcid{0000-0002-6220-9104}\inst{\ref{aff17}}
\and H.~Dole\orcid{0000-0002-9767-3839}\inst{\ref{aff49}}
\and F.~Dubath\orcid{0000-0002-6533-2810}\inst{\ref{aff12}}
\and C.~A.~J.~Duncan\orcid{0009-0003-3573-0791}\inst{\ref{aff38}}
\and X.~Dupac\inst{\ref{aff40}}
\and S.~Dusini\orcid{0000-0002-1128-0664}\inst{\ref{aff50}}
\and S.~Escoffier\orcid{0000-0002-2847-7498}\inst{\ref{aff51}}
\and M.~Farina\orcid{0000-0002-3089-7846}\inst{\ref{aff52}}
\and R.~Farinelli\inst{\ref{aff15}}
\and F.~Faustini\orcid{0000-0001-6274-5145}\inst{\ref{aff33},\ref{aff53}}
\and S.~Ferriol\inst{\ref{aff41}}
\and F.~Finelli\orcid{0000-0002-6694-3269}\inst{\ref{aff15},\ref{aff54}}
\and P.~Fosalba\orcid{0000-0002-1510-5214}\inst{\ref{aff55},\ref{aff56}}
\and N.~Fourmanoit\orcid{0009-0005-6816-6925}\inst{\ref{aff51}}
\and M.~Frailis\orcid{0000-0002-7400-2135}\inst{\ref{aff17}}
\and E.~Franceschi\orcid{0000-0002-0585-6591}\inst{\ref{aff15}}
\and M.~Fumana\orcid{0000-0001-6787-5950}\inst{\ref{aff30}}
\and S.~Galeotta\orcid{0000-0002-3748-5115}\inst{\ref{aff17}}
\and K.~George\orcid{0000-0002-1734-8455}\inst{\ref{aff57}}
\and B.~Gillis\orcid{0000-0002-4478-1270}\inst{\ref{aff38}}
\and C.~Giocoli\orcid{0000-0002-9590-7961}\inst{\ref{aff15},\ref{aff21}}
\and J.~Gracia-Carpio\inst{\ref{aff58}}
\and A.~Grazian\orcid{0000-0002-5688-0663}\inst{\ref{aff59}}
\and F.~Grupp\inst{\ref{aff58},\ref{aff60}}
\and L.~Guzzo\orcid{0000-0001-8264-5192}\inst{\ref{aff61},\ref{aff14},\ref{aff62}}
\and S.~V.~H.~Haugan\orcid{0000-0001-9648-7260}\inst{\ref{aff63}}
\and J.~Hoar\inst{\ref{aff40}}
\and H.~Hoekstra\orcid{0000-0002-0641-3231}\inst{\ref{aff10}}
\and W.~Holmes\inst{\ref{aff64}}
\and I.~M.~Hook\orcid{0000-0002-2960-978X}\inst{\ref{aff65}}
\and F.~Hormuth\inst{\ref{aff66}}
\and A.~Hornstrup\orcid{0000-0002-3363-0936}\inst{\ref{aff67},\ref{aff68}}
\and K.~Jahnke\orcid{0000-0003-3804-2137}\inst{\ref{aff5}}
\and M.~Jhabvala\inst{\ref{aff69}}
\and B.~Joachimi\orcid{0000-0001-7494-1303}\inst{\ref{aff70}}
\and E.~Keih\"anen\orcid{0000-0003-1804-7715}\inst{\ref{aff71}}
\and S.~Kermiche\orcid{0000-0002-0302-5735}\inst{\ref{aff51}}
\and A.~Kiessling\orcid{0000-0002-2590-1273}\inst{\ref{aff64}}
\and M.~Kilbinger\orcid{0000-0001-9513-7138}\inst{\ref{aff72}}
\and B.~Kubik\orcid{0009-0006-5823-4880}\inst{\ref{aff41}}
\and M.~K\"ummel\orcid{0000-0003-2791-2117}\inst{\ref{aff60}}
\and M.~Kunz\orcid{0000-0002-3052-7394}\inst{\ref{aff73}}
\and H.~Kurki-Suonio\orcid{0000-0002-4618-3063}\inst{\ref{aff74},\ref{aff75}}
\and A.~M.~C.~Le~Brun\orcid{0000-0002-0936-4594}\inst{\ref{aff76}}
\and S.~Ligori\orcid{0000-0003-4172-4606}\inst{\ref{aff28}}
\and P.~B.~Lilje\orcid{0000-0003-4324-7794}\inst{\ref{aff63}}
\and V.~Lindholm\orcid{0000-0003-2317-5471}\inst{\ref{aff74},\ref{aff75}}
\and I.~Lloro\orcid{0000-0001-5966-1434}\inst{\ref{aff77}}
\and G.~Mainetti\orcid{0000-0003-2384-2377}\inst{\ref{aff78}}
\and D.~Maino\inst{\ref{aff61},\ref{aff30},\ref{aff62}}
\and E.~Maiorano\orcid{0000-0003-2593-4355}\inst{\ref{aff15}}
\and O.~Mansutti\orcid{0000-0001-5758-4658}\inst{\ref{aff17}}
\and O.~Marggraf\orcid{0000-0001-7242-3852}\inst{\ref{aff79}}
\and M.~Martinelli\orcid{0000-0002-6943-7732}\inst{\ref{aff33},\ref{aff80}}
\and N.~Martinet\orcid{0000-0003-2786-7790}\inst{\ref{aff81}}
\and F.~Marulli\orcid{0000-0002-8850-0303}\inst{\ref{aff82},\ref{aff15},\ref{aff21}}
\and R.~J.~Massey\orcid{0000-0002-6085-3780}\inst{\ref{aff83}}
\and E.~Medinaceli\orcid{0000-0002-4040-7783}\inst{\ref{aff15}}
\and S.~Mei\orcid{0000-0002-2849-559X}\inst{\ref{aff84},\ref{aff85}}
\and M.~Melchior\inst{\ref{aff86}}
\and Y.~Mellier\inst{\ref{aff87},\ref{aff88}}
\and M.~Meneghetti\orcid{0000-0003-1225-7084}\inst{\ref{aff15},\ref{aff21}}
\and E.~Merlin\orcid{0000-0001-6870-8900}\inst{\ref{aff33}}
\and G.~Meylan\inst{\ref{aff89}}
\and A.~Mora\orcid{0000-0002-1922-8529}\inst{\ref{aff90}}
\and M.~Moresco\orcid{0000-0002-7616-7136}\inst{\ref{aff82},\ref{aff15}}
\and L.~Moscardini\orcid{0000-0002-3473-6716}\inst{\ref{aff82},\ref{aff15},\ref{aff21}}
\and R.~Nakajima\orcid{0009-0009-1213-7040}\inst{\ref{aff79}}
\and C.~Neissner\orcid{0000-0001-8524-4968}\inst{\ref{aff91},\ref{aff32}}
\and S.-M.~Niemi\orcid{0009-0005-0247-0086}\inst{\ref{aff29}}
\and C.~Padilla\orcid{0000-0001-7951-0166}\inst{\ref{aff91}}
\and S.~Paltani\orcid{0000-0002-8108-9179}\inst{\ref{aff12}}
\and F.~Pasian\orcid{0000-0002-4869-3227}\inst{\ref{aff17}}
\and K.~Pedersen\inst{\ref{aff92}}
\and W.~J.~Percival\orcid{0000-0002-0644-5727}\inst{\ref{aff93},\ref{aff94},\ref{aff95}}
\and V.~Pettorino\orcid{0000-0002-4203-9320}\inst{\ref{aff29}}
\and S.~Pires\orcid{0000-0002-0249-2104}\inst{\ref{aff72}}
\and G.~Polenta\orcid{0000-0003-4067-9196}\inst{\ref{aff53}}
\and M.~Poncet\inst{\ref{aff96}}
\and L.~A.~Popa\inst{\ref{aff97}}
\and L.~Pozzetti\orcid{0000-0001-7085-0412}\inst{\ref{aff15}}
\and F.~Raison\orcid{0000-0002-7819-6918}\inst{\ref{aff58}}
\and A.~Renzi\orcid{0000-0001-9856-1970}\inst{\ref{aff98},\ref{aff50}}
\and J.~Rhodes\orcid{0000-0002-4485-8549}\inst{\ref{aff64}}
\and G.~Riccio\inst{\ref{aff25}}
\and E.~Romelli\orcid{0000-0003-3069-9222}\inst{\ref{aff17}}
\and M.~Roncarelli\orcid{0000-0001-9587-7822}\inst{\ref{aff15}}
\and E.~Rossetti\orcid{0000-0003-0238-4047}\inst{\ref{aff20}}
\and R.~Saglia\orcid{0000-0003-0378-7032}\inst{\ref{aff60},\ref{aff58}}
\and Z.~Sakr\orcid{0000-0002-4823-3757}\inst{\ref{aff99},\ref{aff100},\ref{aff101}}
\and D.~Sapone\orcid{0000-0001-7089-4503}\inst{\ref{aff102}}
\and B.~Sartoris\orcid{0000-0003-1337-5269}\inst{\ref{aff60},\ref{aff17}}
\and M.~Schirmer\orcid{0000-0003-2568-9994}\inst{\ref{aff5}}
\and P.~Schneider\orcid{0000-0001-8561-2679}\inst{\ref{aff79}}
\and T.~Schrabback\orcid{0000-0002-6987-7834}\inst{\ref{aff103}}
\and A.~Secroun\orcid{0000-0003-0505-3710}\inst{\ref{aff51}}
\and G.~Seidel\orcid{0000-0003-2907-353X}\inst{\ref{aff5}}
\and S.~Serrano\orcid{0000-0002-0211-2861}\inst{\ref{aff55},\ref{aff104},\ref{aff56}}
\and P.~Simon\inst{\ref{aff79}}
\and C.~Sirignano\orcid{0000-0002-0995-7146}\inst{\ref{aff98},\ref{aff50}}
\and G.~Sirri\orcid{0000-0003-2626-2853}\inst{\ref{aff21}}
\and L.~Stanco\orcid{0000-0002-9706-5104}\inst{\ref{aff50}}
\and J.-L.~Starck\orcid{0000-0003-2177-7794}\inst{\ref{aff72}}
\and J.~Steinwagner\orcid{0000-0001-7443-1047}\inst{\ref{aff58}}
\and P.~Tallada-Cresp\'{i}\orcid{0000-0002-1336-8328}\inst{\ref{aff31},\ref{aff32}}
\and A.~N.~Taylor\inst{\ref{aff38}}
\and H.~I.~Teplitz\orcid{0000-0002-7064-5424}\inst{\ref{aff105}}
\and I.~Tereno\orcid{0000-0002-4537-6218}\inst{\ref{aff47},\ref{aff106}}
\and N.~Tessore\orcid{0000-0002-9696-7931}\inst{\ref{aff70},\ref{aff46}}
\and S.~Toft\orcid{0000-0003-3631-7176}\inst{\ref{aff8},\ref{aff11}}
\and R.~Toledo-Moreo\orcid{0000-0002-2997-4859}\inst{\ref{aff107}}
\and F.~Torradeflot\orcid{0000-0003-1160-1517}\inst{\ref{aff32},\ref{aff31}}
\and I.~Tutusaus\orcid{0000-0002-3199-0399}\inst{\ref{aff56},\ref{aff55},\ref{aff100}}
\and L.~Valenziano\orcid{0000-0002-1170-0104}\inst{\ref{aff15},\ref{aff54}}
\and J.~Valiviita\orcid{0000-0001-6225-3693}\inst{\ref{aff74},\ref{aff75}}
\and T.~Vassallo\orcid{0000-0001-6512-6358}\inst{\ref{aff17},\ref{aff57}}
\and A.~Veropalumbo\orcid{0000-0003-2387-1194}\inst{\ref{aff14},\ref{aff23},\ref{aff22}}
\and Y.~Wang\orcid{0000-0002-4749-2984}\inst{\ref{aff105}}
\and J.~Weller\orcid{0000-0002-8282-2010}\inst{\ref{aff60},\ref{aff58}}
\and G.~Zamorani\orcid{0000-0002-2318-301X}\inst{\ref{aff15}}
\and F.~M.~Zerbi\inst{\ref{aff14}}
\and E.~Zucca\orcid{0000-0002-5845-8132}\inst{\ref{aff15}}
\and J.~Mart\'{i}n-Fleitas\orcid{0000-0002-8594-569X}\inst{\ref{aff108}}
\and V.~Scottez\orcid{0009-0008-3864-940X}\inst{\ref{aff87},\ref{aff109}}
\and M.~Viel\orcid{0000-0002-2642-5707}\inst{\ref{aff16},\ref{aff17},\ref{aff19},\ref{aff18},\ref{aff110}}}

\institute{Department of Physics, Oxford University, Keble Road, Oxford OX1 3RH, UK\label{aff1}
\and
Jodrell Bank Centre for Astrophysics, Department of Physics and Astronomy, University of Manchester, Oxford Road, Manchester M13 9PL, UK\label{aff2}
\and
Department of Physics and Astronomy, University of the Western Cape, Bellville, Cape Town, 7535, South Africa\label{aff3}
\and
MIT Kavli Institute for Astrophysics and Space Research, Massachusetts Institute of Technology, Cambridge, MA 02139, USA\label{aff4}
\and
Max-Planck-Institut f\"ur Astronomie, K\"onigstuhl 17, 69117 Heidelberg, Germany\label{aff5}
\and
Institute of Cosmology and Gravitation, University of Portsmouth, Portsmouth PO1 3FX, UK\label{aff6}
\and
Kapteyn Astronomical Institute, University of Groningen, PO Box 800, 9700 AV Groningen, The Netherlands\label{aff7}
\and
Cosmic Dawn Center (DAWN)\label{aff8}
\and
Department of Physics \& Astronomy, University of Sussex, Brighton BN1 9QH, UK\label{aff9}
\and
Leiden Observatory, Leiden University, Einsteinweg 55, 2333 CC Leiden, The Netherlands\label{aff10}
\and
Niels Bohr Institute, University of Copenhagen, Jagtvej 128, 2200 Copenhagen, Denmark\label{aff11}
\and
Department of Astronomy, University of Geneva, ch. d'Ecogia 16, 1290 Versoix, Switzerland\label{aff12}
\and
School of Mathematics and Physics, University of Surrey, Guildford, Surrey, GU2 7XH, UK\label{aff13}
\and
INAF-Osservatorio Astronomico di Brera, Via Brera 28, 20122 Milano, Italy\label{aff14}
\and
INAF-Osservatorio di Astrofisica e Scienza dello Spazio di Bologna, Via Piero Gobetti 93/3, 40129 Bologna, Italy\label{aff15}
\and
IFPU, Institute for Fundamental Physics of the Universe, via Beirut 2, 34151 Trieste, Italy\label{aff16}
\and
INAF-Osservatorio Astronomico di Trieste, Via G. B. Tiepolo 11, 34143 Trieste, Italy\label{aff17}
\and
INFN, Sezione di Trieste, Via Valerio 2, 34127 Trieste TS, Italy\label{aff18}
\and
SISSA, International School for Advanced Studies, Via Bonomea 265, 34136 Trieste TS, Italy\label{aff19}
\and
Dipartimento di Fisica e Astronomia, Universit\`a di Bologna, Via Gobetti 93/2, 40129 Bologna, Italy\label{aff20}
\and
INFN-Sezione di Bologna, Viale Berti Pichat 6/2, 40127 Bologna, Italy\label{aff21}
\and
Dipartimento di Fisica, Universit\`a di Genova, Via Dodecaneso 33, 16146, Genova, Italy\label{aff22}
\and
INFN-Sezione di Genova, Via Dodecaneso 33, 16146, Genova, Italy\label{aff23}
\and
Department of Physics "E. Pancini", University Federico II, Via Cinthia 6, 80126, Napoli, Italy\label{aff24}
\and
INAF-Osservatorio Astronomico di Capodimonte, Via Moiariello 16, 80131 Napoli, Italy\label{aff25}
\and
Dipartimento di Fisica, Universit\`a degli Studi di Torino, Via P. Giuria 1, 10125 Torino, Italy\label{aff26}
\and
INFN-Sezione di Torino, Via P. Giuria 1, 10125 Torino, Italy\label{aff27}
\and
INAF-Osservatorio Astrofisico di Torino, Via Osservatorio 20, 10025 Pino Torinese (TO), Italy\label{aff28}
\and
European Space Agency/ESTEC, Keplerlaan 1, 2201 AZ Noordwijk, The Netherlands\label{aff29}
\and
INAF-IASF Milano, Via Alfonso Corti 12, 20133 Milano, Italy\label{aff30}
\and
Centro de Investigaciones Energ\'eticas, Medioambientales y Tecnol\'ogicas (CIEMAT), Avenida Complutense 40, 28040 Madrid, Spain\label{aff31}
\and
Port d'Informaci\'{o} Cient\'{i}fica, Campus UAB, C. Albareda s/n, 08193 Bellaterra (Barcelona), Spain\label{aff32}
\and
INAF-Osservatorio Astronomico di Roma, Via Frascati 33, 00078 Monteporzio Catone, Italy\label{aff33}
\and
INFN section of Naples, Via Cinthia 6, 80126, Napoli, Italy\label{aff34}
\and
Institute for Astronomy, University of Hawaii, 2680 Woodlawn Drive, Honolulu, HI 96822, USA\label{aff35}
\and
Dipartimento di Fisica e Astronomia "Augusto Righi" - Alma Mater Studiorum Universit\`a di Bologna, Viale Berti Pichat 6/2, 40127 Bologna, Italy\label{aff36}
\and
Instituto de Astrof\'{\i}sica de Canarias, V\'{\i}a L\'actea, 38205 La Laguna, Tenerife, Spain\label{aff37}
\and
Institute for Astronomy, University of Edinburgh, Royal Observatory, Blackford Hill, Edinburgh EH9 3HJ, UK\label{aff38}
\and
European Space Agency/ESRIN, Largo Galileo Galilei 1, 00044 Frascati, Roma, Italy\label{aff39}
\and
ESAC/ESA, Camino Bajo del Castillo, s/n., Urb. Villafranca del Castillo, 28692 Villanueva de la Ca\~nada, Madrid, Spain\label{aff40}
\and
Universit\'e Claude Bernard Lyon 1, CNRS/IN2P3, IP2I Lyon, UMR 5822, Villeurbanne, F-69100, France\label{aff41}
\and
Institut de Ci\`{e}ncies del Cosmos (ICCUB), Universitat de Barcelona (IEEC-UB), Mart\'{i} i Franqu\`{e}s 1, 08028 Barcelona, Spain\label{aff42}
\and
Instituci\'o Catalana de Recerca i Estudis Avan\c{c}ats (ICREA), Passeig de Llu\'{\i}s Companys 23, 08010 Barcelona, Spain\label{aff43}
\and
Institut de Ciencies de l'Espai (IEEC-CSIC), Campus UAB, Carrer de Can Magrans, s/n Cerdanyola del Vall\'es, 08193 Barcelona, Spain\label{aff44}
\and
UCB Lyon 1, CNRS/IN2P3, IUF, IP2I Lyon, 4 rue Enrico Fermi, 69622 Villeurbanne, France\label{aff45}
\and
Mullard Space Science Laboratory, University College London, Holmbury St Mary, Dorking, Surrey RH5 6NT, UK\label{aff46}
\and
Departamento de F\'isica, Faculdade de Ci\^encias, Universidade de Lisboa, Edif\'icio C8, Campo Grande, PT1749-016 Lisboa, Portugal\label{aff47}
\and
Instituto de Astrof\'isica e Ci\^encias do Espa\c{c}o, Faculdade de Ci\^encias, Universidade de Lisboa, Campo Grande, 1749-016 Lisboa, Portugal\label{aff48}
\and
Universit\'e Paris-Saclay, CNRS, Institut d'astrophysique spatiale, 91405, Orsay, France\label{aff49}
\and
INFN-Padova, Via Marzolo 8, 35131 Padova, Italy\label{aff50}
\and
Aix-Marseille Universit\'e, CNRS/IN2P3, CPPM, Marseille, France\label{aff51}
\and
INAF-Istituto di Astrofisica e Planetologia Spaziali, via del Fosso del Cavaliere, 100, 00100 Roma, Italy\label{aff52}
\and
Space Science Data Center, Italian Space Agency, via del Politecnico snc, 00133 Roma, Italy\label{aff53}
\and
INFN-Bologna, Via Irnerio 46, 40126 Bologna, Italy\label{aff54}
\and
Institut d'Estudis Espacials de Catalunya (IEEC),  Edifici RDIT, Campus UPC, 08860 Castelldefels, Barcelona, Spain\label{aff55}
\and
Institute of Space Sciences (ICE, CSIC), Campus UAB, Carrer de Can Magrans, s/n, 08193 Barcelona, Spain\label{aff56}
\and
University Observatory, LMU Faculty of Physics, Scheinerstrasse 1, 81679 Munich, Germany\label{aff57}
\and
Max Planck Institute for Extraterrestrial Physics, Giessenbachstr. 1, 85748 Garching, Germany\label{aff58}
\and
INAF-Osservatorio Astronomico di Padova, Via dell'Osservatorio 5, 35122 Padova, Italy\label{aff59}
\and
Universit\"ats-Sternwarte M\"unchen, Fakult\"at f\"ur Physik, Ludwig-Maximilians-Universit\"at M\"unchen, Scheinerstrasse 1, 81679 M\"unchen, Germany\label{aff60}
\and
Dipartimento di Fisica "Aldo Pontremoli", Universit\`a degli Studi di Milano, Via Celoria 16, 20133 Milano, Italy\label{aff61}
\and
INFN-Sezione di Milano, Via Celoria 16, 20133 Milano, Italy\label{aff62}
\and
Institute of Theoretical Astrophysics, University of Oslo, P.O. Box 1029 Blindern, 0315 Oslo, Norway\label{aff63}
\and
Jet Propulsion Laboratory, California Institute of Technology, 4800 Oak Grove Drive, Pasadena, CA, 91109, USA\label{aff64}
\and
Department of Physics, Lancaster University, Lancaster, LA1 4YB, UK\label{aff65}
\and
Felix Hormuth Engineering, Goethestr. 17, 69181 Leimen, Germany\label{aff66}
\and
Technical University of Denmark, Elektrovej 327, 2800 Kgs. Lyngby, Denmark\label{aff67}
\and
Cosmic Dawn Center (DAWN), Denmark\label{aff68}
\and
NASA Goddard Space Flight Center, Greenbelt, MD 20771, USA\label{aff69}
\and
Department of Physics and Astronomy, University College London, Gower Street, London WC1E 6BT, UK\label{aff70}
\and
Department of Physics and Helsinki Institute of Physics, Gustaf H\"allstr\"omin katu 2, University of Helsinki, 00014 Helsinki, Finland\label{aff71}
\and
Universit\'e Paris-Saclay, Universit\'e Paris Cit\'e, CEA, CNRS, AIM, 91191, Gif-sur-Yvette, France\label{aff72}
\and
Universit\'e de Gen\`eve, D\'epartement de Physique Th\'eorique and Centre for Astroparticle Physics, 24 quai Ernest-Ansermet, CH-1211 Gen\`eve 4, Switzerland\label{aff73}
\and
Department of Physics, P.O. Box 64, University of Helsinki, 00014 Helsinki, Finland\label{aff74}
\and
Helsinki Institute of Physics, Gustaf H{\"a}llstr{\"o}min katu 2, University of Helsinki, 00014 Helsinki, Finland\label{aff75}
\and
Laboratoire d'etude de l'Univers et des phenomenes eXtremes, Observatoire de Paris, Universit\'e PSL, Sorbonne Universit\'e, CNRS, 92190 Meudon, France\label{aff76}
\and
SKAO, Jodrell Bank, Lower Withington, Macclesfield SK11 9FT, United Kingdom\label{aff77}
\and
Centre de Calcul de l'IN2P3/CNRS, 21 avenue Pierre de Coubertin 69627 Villeurbanne Cedex, France\label{aff78}
\and
Universit\"at Bonn, Argelander-Institut f\"ur Astronomie, Auf dem H\"ugel 71, 53121 Bonn, Germany\label{aff79}
\and
INFN-Sezione di Roma, Piazzale Aldo Moro, 2 - c/o Dipartimento di Fisica, Edificio G. Marconi, 00185 Roma, Italy\label{aff80}
\and
Aix-Marseille Universit\'e, CNRS, CNES, LAM, Marseille, France\label{aff81}
\and
Dipartimento di Fisica e Astronomia "Augusto Righi" - Alma Mater Studiorum Universit\`a di Bologna, via Piero Gobetti 93/2, 40129 Bologna, Italy\label{aff82}
\and
Department of Physics, Institute for Computational Cosmology, Durham University, South Road, Durham, DH1 3LE, UK\label{aff83}
\and
Universit\'e Paris Cit\'e, CNRS, Astroparticule et Cosmologie, 75013 Paris, France\label{aff84}
\and
CNRS-UCB International Research Laboratory, Centre Pierre Bin\'etruy, IRL2007, CPB-IN2P3, Berkeley, USA\label{aff85}
\and
University of Applied Sciences and Arts of Northwestern Switzerland, School of Engineering, 5210 Windisch, Switzerland\label{aff86}
\and
Institut d'Astrophysique de Paris, 98bis Boulevard Arago, 75014, Paris, France\label{aff87}
\and
Institut d'Astrophysique de Paris, UMR 7095, CNRS, and Sorbonne Universit\'e, 98 bis boulevard Arago, 75014 Paris, France\label{aff88}
\and
Institute of Physics, Laboratory of Astrophysics, Ecole Polytechnique F\'ed\'erale de Lausanne (EPFL), Observatoire de Sauverny, 1290 Versoix, Switzerland\label{aff89}
\and
Telespazio UK S.L. for European Space Agency (ESA), Camino bajo del Castillo, s/n, Urbanizacion Villafranca del Castillo, Villanueva de la Ca\~nada, 28692 Madrid, Spain\label{aff90}
\and
Institut de F\'{i}sica d'Altes Energies (IFAE), The Barcelona Institute of Science and Technology, Campus UAB, 08193 Bellaterra (Barcelona), Spain\label{aff91}
\and
DARK, Niels Bohr Institute, University of Copenhagen, Jagtvej 155, 2200 Copenhagen, Denmark\label{aff92}
\and
Waterloo Centre for Astrophysics, University of Waterloo, Waterloo, Ontario N2L 3G1, Canada\label{aff93}
\and
Department of Physics and Astronomy, University of Waterloo, Waterloo, Ontario N2L 3G1, Canada\label{aff94}
\and
Perimeter Institute for Theoretical Physics, Waterloo, Ontario N2L 2Y5, Canada\label{aff95}
\and
Centre National d'Etudes Spatiales -- Centre spatial de Toulouse, 18 avenue Edouard Belin, 31401 Toulouse Cedex 9, France\label{aff96}
\and
Institute of Space Science, Str. Atomistilor, nr. 409 M\u{a}gurele, Ilfov, 077125, Romania\label{aff97}
\and
Dipartimento di Fisica e Astronomia "G. Galilei", Universit\`a di Padova, Via Marzolo 8, 35131 Padova, Italy\label{aff98}
\and
Institut f\"ur Theoretische Physik, University of Heidelberg, Philosophenweg 16, 69120 Heidelberg, Germany\label{aff99}
\and
Institut de Recherche en Astrophysique et Plan\'etologie (IRAP), Universit\'e de Toulouse, CNRS, UPS, CNES, 14 Av. Edouard Belin, 31400 Toulouse, France\label{aff100}
\and
Universit\'e St Joseph; Faculty of Sciences, Beirut, Lebanon\label{aff101}
\and
Departamento de F\'isica, FCFM, Universidad de Chile, Blanco Encalada 2008, Santiago, Chile\label{aff102}
\and
Universit\"at Innsbruck, Institut f\"ur Astro- und Teilchenphysik, Technikerstr. 25/8, 6020 Innsbruck, Austria\label{aff103}
\and
Satlantis, University Science Park, Sede Bld 48940, Leioa-Bilbao, Spain\label{aff104}
\and
Infrared Processing and Analysis Center, California Institute of Technology, Pasadena, CA 91125, USA\label{aff105}
\and
Instituto de Astrof\'isica e Ci\^encias do Espa\c{c}o, Faculdade de Ci\^encias, Universidade de Lisboa, Tapada da Ajuda, 1349-018 Lisboa, Portugal\label{aff106}
\and
Universidad Polit\'ecnica de Cartagena, Departamento de Electr\'onica y Tecnolog\'ia de Computadoras,  Plaza del Hospital 1, 30202 Cartagena, Spain\label{aff107}
\and
Aurora Technology for European Space Agency (ESA), Camino bajo del Castillo, s/n, Urbanizacion Villafranca del Castillo, Villanueva de la Ca\~nada, 28692 Madrid, Spain\label{aff108}
\and
ICL, Junia, Universit\'e Catholique de Lille, LITL, 59000 Lille, France\label{aff109}
\and
ICSC - Centro Nazionale di Ricerca in High Performance Computing, Big Data e Quantum Computing, Via Magnanelli 2, Bologna, Italy\label{aff110}}    

%
%
\abstract{
We present a search for $z\simeq7$ Lyman-break galaxies using the $1.72 \, \rm{deg}^2$ near-infrared (NIR) UltraVISTA survey in the COSMOS field, reaching $5\,\sigma$ depths in $Y$ of 26.2. 
We incorporated deep {\Euclid optical and \Euclid + \textit{Spitzer} NIR imaging} for a full spectral energy distribution (SED) fitting analysis.
We found 289 candidate galaxies at $6.5\leq z \leq 7.5$ covering $-22.6 \leq \Muv \leq -20.2$, faint enough to overlap with \textit{Hubble} Space Telescope studies.
We conducted a separate selection by including complementary \Euclid performance verification imaging (reaching $5\,\sigma$ depths of $26.3$), yielding 140 galaxies in $0.65 \, \rm{deg}^2$, with 38 sources unique to this sample.
We computed the rest-frame UV luminosity function (UV LF) from our samples, extending below the knee ($M^*=-21.14^{+0.28}_{-0.25}$). 
We find that the shape of the UV LF is consistent with both a Schechter function and a double power law (DPL) at the magnitudes probed by this sample, with a DPL preferred at $\Muv<-22.5$ when bright-end results are included.
The UltraVISTA+\Euclid sample provides a clean measurement of the LF due to the overlapping NIR filters identifying molecular absorption features in the SEDs of ultra-cool dwarf interlopers, and additional faint galaxies were recovered.
A comparison with JWST LFs at $z>7$ suggests a gentle evolution in the bright-end slope, although this is limited by a lack of robust bright-end measurements at $z>9$.
We forecast that in the Euclid Deep Fields, the removal of contaminant ultra-cool dwarfs as point sources will be possible at $\JE < 24.5$.
Finally, we present a high-equivalent-width Lyman-$\alpha$ emitter candidate identified by combining HSC, VISTA, and \Euclid broadband photometry, highlighting the synergistic power these instruments will have in the Euclid Auxiliary Fields for identifying extreme sources in the epoch of reionisation.
    }


   \titlerunning{\Euclid: Discovery of bright $z\simeq7$ LBGs in UltraVISTA and \Euclid COSMOS}
   \authorrunning{Varadaraj et al.}
   
   \maketitle

\section{\label{sc:Intro}Introduction}

A central goal of astrophysics is to unveil the formation and evolution of the first galaxies in the Universe \citep{Stark16, Adamo24}.
The luminosity function (LF), or number density of galaxies as a function of luminosity/magnitude, is a key statistic for this endeavour.
In particular, at high redshift ($z>5$), observations of rare, luminous galaxies at the bright end ($L > L^{\ast}$) of the UV LF provide key insight into astrophysical effects such as feedback and dust build-up \citep[e.g.][]{Bowler15, Finkelstein22_agn, Nikopoulos24, Algera25}.
Degree-scale ground-based imaging has been central to discovering and characterising these rare luminous Lyman-break galaxies (LBGs), so-called because of their strong redshifted Lyman-$\alpha$ spectral break at $\lambda_{\rm rest}=1216\,\AA$ \citep{Guhathakurta90, Steidel96}.
Near-infrared (NIR) {ground-based} surveys such the Ultra Deep Survey (UDS) field of the UKIRT Infrared Deep Sky Survey \citep[UKIDSS,][]{Lawrence07} and UltraVISTA \citep{ultravista} have provided the means to discover rare $L>L^{\ast}$ LBG candidates at $z>5$ \citep{McLure09, Bowler12}.
Subsequent ground-based studies have confirmed a double power law (DPL) LF with little evolution in the bright end at $z=6$--$10$ \citep[][hereafter \citetalias{Varadaraj23}]{Bowler14, Ono18, Stefanon19, Bowler20, harikane22, Kauffmann22, Donnan23, Varadaraj23}.
The \textit{Hubble} Space Telescope (HST) has allowed strong complementary constraints to be placed on the faint end ($L < L^{\ast}$, \citealt{McLure13}; \citealt[]{Finkelstein15}, hereafter \citetalias{Finkelstein15}; \citealt{Bouwens21}, hereafter \citetalias{Bouwens21}).
However, ground-based telescopes and HST can only probe out to $\lambda = 2\,\micron$, placing a barrier at $z=10$.
The unparalleled NIR capabilities of JWST have substantially advanced the redshift frontier, revealing an abundance of luminous blue galaxies and a markedly slow evolution of the UV LF over the redshift range $z=10$--$14$ \citep[e.g.][]{Donnan23, McLeod24, Adams24, Chemerynska24, Whitler25, harikane24}. 
This has invoked scenarios such as increased star-formation efficiency or Pop. III stars \citep{Harikane23jwst}, ejection of dust by radiation-driven outflows \citep{Ferrara23}, and even tension with Lambda cold dark matter \citep[$\Lambda$CDM][]{Labbe23}.
The star-formation histories (SFHs) of some luminous $z\simeq7$ sources can require substantial star formation at $z>9$ \citep[e.g.][]{Whitler23}.
This, combined with the slow evolution of the LF, means it is natural to return to the most massive, luminous sources at $z\simeq7$ in order to understand their connection to bright JWST galaxies at $z>10$.

The study of $z\simeq7$ sources hinges on the ability to robustly detect them and remove interloper sources.
A major issue facing ground-based studies at $z\simeq7$ has been contamination by Galactic M-, L-, and T-type ultra-cool dwarfs \citep[UCDs;* e.g.][]{Stanway08, Bowler12, Wilkins14}.
These sources have high number densities at the typical apparent magnitudes of $L > L^*$ LBGs, often matching or even exceeding the number of LBG candidates in most degree-scale extragalactic surveys \citep{Ryan11}.
The same molecular species responsible for making the Earth's atmosphere opaque at certain wavelengths in the NIR are also present in UCD atmospheres, causing deep molecular absorption complexes at wavelengths impossible to probe from the ground.
This means that in certain cases the NIR photometry of a UCD can be confused with a flat rest-UV LBG continuum.
Furthermore, while luminous $z\simeq7$ LBGs are generally marginally resolved \citep[][hereafter \citetalias{Bowler17}]{Bowler17}, atmospheric seeing often prevents distinguishing between UCDs and LBGs based on their morphology.

\begin{figure*}
    \sidecaption
    \includegraphics[width=12cm]{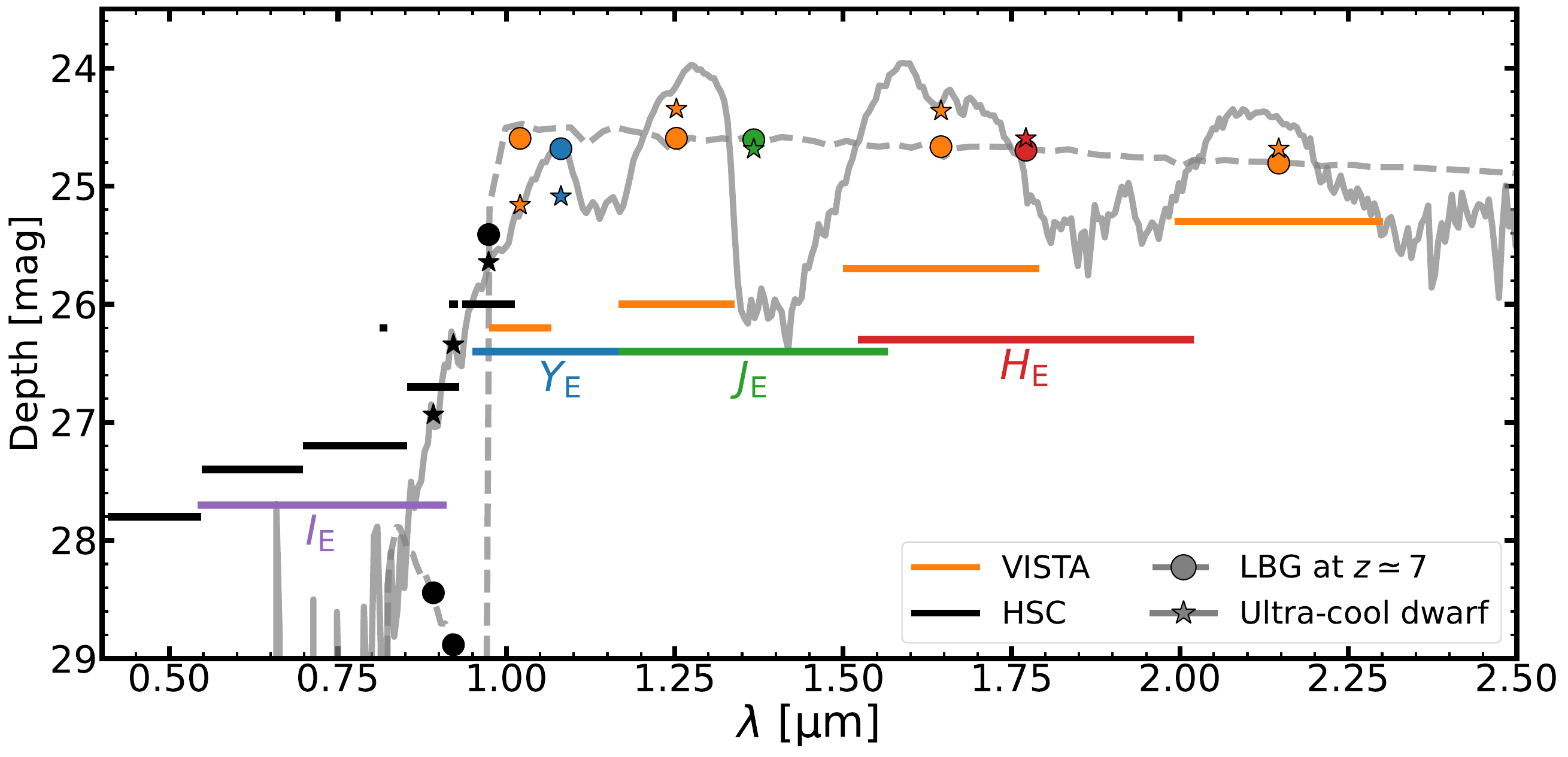}
    \caption{Limiting magnitudes or modal depths ($5\,\sigma$) of the photometric filters used in this work within the COSMOS field. 
    The line widths represent the FWHM of the filter transmission curves, and the depths are reported in Table \ref{tab:Depths}.
    The HSC and VISTA filters are shown in black and orange, respectively, and the four \Euclid filters are labelled.
    We also show example SEDs of a UCD and a LBG with the solid and dashed lines, respectively. 
    The model photometry for the UCD and the LBG are shown by the stars and circles respectively and are colour-coded by their filters.
    Note that the \Euclid NIR filters cover the gaps between the VISTA filters. These wavelengths are inaccessible from the ground due to the atmospheric absorption.}
    \label{fig:filters}
\end{figure*}

The launch of \Euclid signals the first time astronomers have had access to degree-scale, space-based NIR imaging.
\Euclid is a European Space Agency medium-class mission launched in July 2023. The \Euclid spacecraft is equipped with a $1.2\,\text{m}$ primary mirror \citep{EuclidSkyOverview}, and the 
main goal of the mission is to probe dark matter and dark energy through weak lensing and galaxy clustering, with a significant focus also on non-cosmological science.
The Visible Camera instrument on \Euclid \citep[VIS;][]{EuclidSkyVIS} features a high-resolution -- $\ang{;;0.16}$ full width half maximum (FWHM)  -- optical filter, $\IE$, equivalent to the ground-based $riz$ filters.
The Near-Infrared Spectrometer and Photometer \citep[NISP;][]{EuclidSkyNISP} features three NIR filters, \YE, \JE, and \HE, that can probe NIR wavelengths inaccessible from the ground (see Fig. \ref{fig:filters}).
The field of view of \Euclid is $0.55 \, \rm{deg}^2$, and the main survey will eventually map out $14\,000 \, \rm{deg}^2$ of the extragalactic sky \citep{EuclidSkyOverview}.
The Early Release Observations \citep[EROs;][]{EROData} have led to the identification of the first \Euclid-selected $z>6$ LBG candidates \citep{EROLensData, EROLensVISDropouts}, with the very deep \IE filter allowing for the removal of M dwarfs and low-redshift galaxy interlopers by requiring a strong break in $\IE - \YE$. 
However, the availability of only four relatively wide photometric filters and the fact that the NISP images in the ERO reduction have a pixel scale of $\ang{;;0.3} \,\text{pix}^{-1}$ means there were probably still high levels of contamination by L- and T-type dwarfs.

\Euclid will dedicate approximately $10\%$ of its observing time to imaging the Euclid Deep Fields \citep[][]{Scaramella-EP1, EuclidSkyOverview}, covering $53 \, \deg^2$ to depths of around $26$.
The resulting imaging will represent an approximately $30$-fold increase in area compared to previous NIR surveys reaching this depth \citep[UltraVISTA,][]{ultravista}.
The considerably wider area will lead to the discovery of thousands of $z \simeq 7$ galaxies brighter than $m_{\rm{AB}}=26$ \citepalias{Bowler17}, thus allowing for definitive measurements of the bright end of the UV LF. 
Until the Euclid Deep Survey is complete, early imaging from the Euclid Auxiliary Fields (EAFs), used for calibration and reaching depths comparable to the EDFs, serves as an ideal test bed for selecting high-redshift galaxies with \Euclid.
In the COSMOS field, by combining NISP with complementary NIR photometry from VISTA, it may be possible to break the degeneracy between the colours of L- and T-type dwarfs and genuine high-redshift LBGs.
The construction of pure $z\simeq7$ samples is critical because there is still some tension between ground-based studies, which suggest a shallower decline in the bright end akin to a DPL (\citealt{Bowler14}; \citetalias{Bowler17}; \citealt{harikane22}; \citetalias{Varadaraj23}), and the widest-area HST-based study of \citetalias{Bouwens21}, who find a marked drop in the LF between $\Muv = -21.5$ and $\Muv = -22$, suggesting an exponential decline.
The final data release of the {NIR UltraVISTA survey, covering 1.72 deg$^2$ in the COSMOS field at $\lambda =1$--$2.5\,\micron$,} provides the necessary depth to connect ground-based observations with space-based observations at the knee of the LF, and its combination with \Euclid imaging provides the means to construct clean samples of UltraVISTA-selected galaxies.
Furthermore, with the higher resolution of \Euclid, we can resolve luminous ground-selected sources for the first time without the need for dedicated follow-up imaging from HST (\citetalias{Bowler17}; \citealt{Stefanon17}).
This could allow for the removal of UCDs as point sources (and perhaps even through proper motion) and will unveil the morphologies of thousands of galaxies at $z>6$.

This paper is structured as follows. In Sect. \ref{sec:data and image processing}, we describe the ground-based and space-based imaging used in this work as well as the image preparation and photometric catalogue construction. 
Section \ref{sec: candidate selection} outlines the selection of our LBG candidates using spectral energy distribution (SED) fitting. For the SED fitting, we conducted two separate selections: one with and one without the additional \Euclid photometry.
We present each of our UltraVISTA-only and UltraVISTA+\Euclid samples in Sect. \ref{sec: candidate galaxies}, and we compute the UV LF at $z\simeq7$ in Sect. \ref{sec: rest-UV LF}, comparing our results with JWST measurements of the UV LF at $z>7$. 
Then, in Sect. \ref{sec: Outlook with euclid} we discuss the improved SED constraints with \Euclid, investigate the ability of \Euclid to remove UCDs as point sources, and highlight the unique capabilities of several overlapping filters to identify extreme Lyman-$\alpha$ emitters. 
Finally, we conclude and summarise in Sect. \ref{sec: conclusions}.
The uncertainties presented in this paper denote $\pm\,1\,\sigma$ uncertainties where Gaussian or enclose 68.3\% of the data when the underlying distribution is asymmetric.
All magnitudes are reported in the AB system \citep{Oke83}. 
We assumed a standard $\Lambda$CDM cosmology, with $H_0 = 70 \, \kmsMpc$,  $\Omega_{\rm{m}}=0.3$, and $\Omega_{\Lambda}=0.7$.

\section{Data and image processing}
\label{sec:data and image processing}

\begin{figure}
    \centering
    \includegraphics[trim=100 0 190 0, clip, width=\columnwidth]{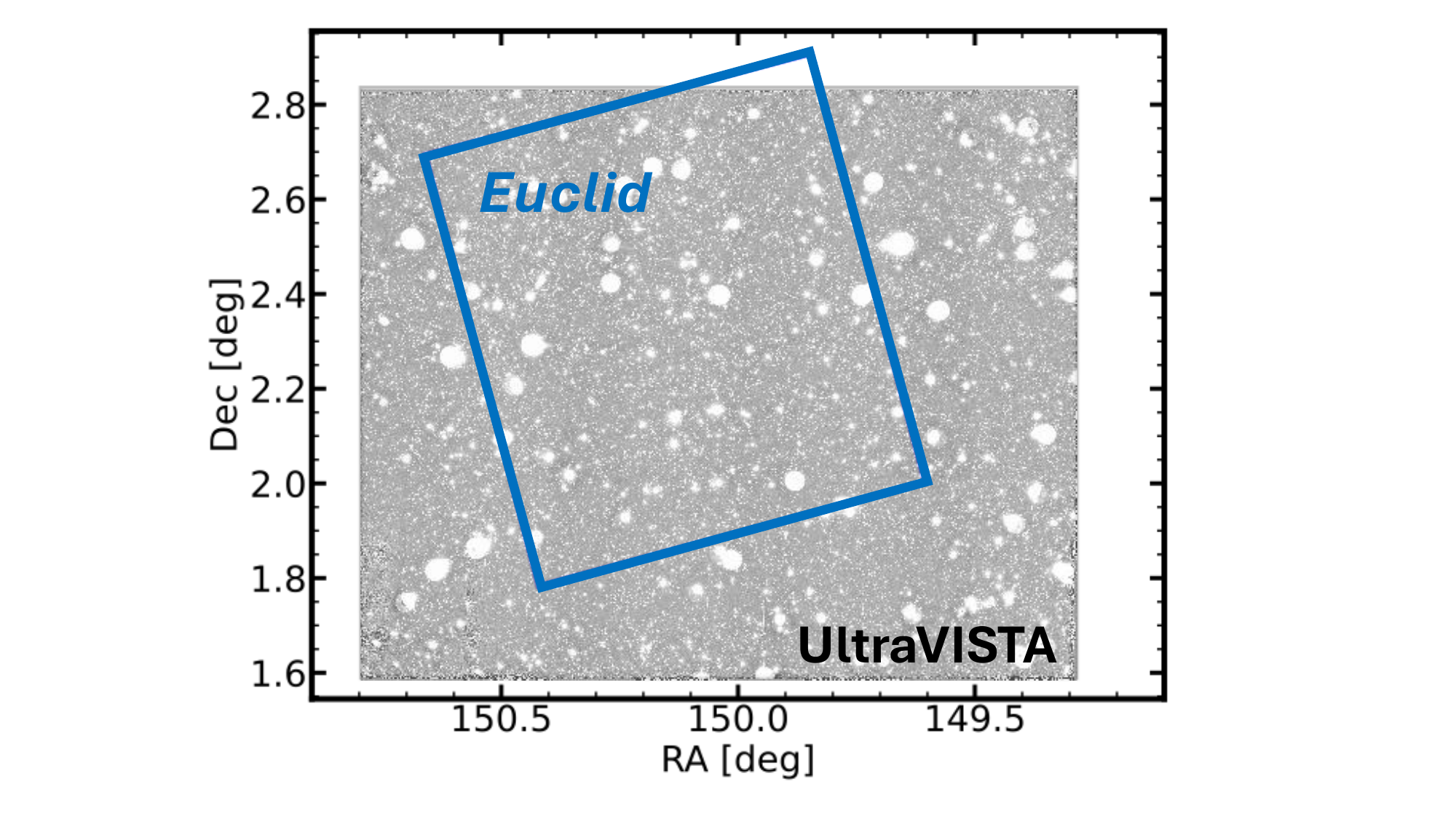}
    \caption{COSMOS UltraVISTA \citep{ultravista} footprint with the \Euclid COSMOS performance verification footprint overlaid. Optical imaging from HSC-SSP DR3 \citep{Aihara22} covers the full UltraVISTA area.
    The overlapping area between \Euclid and UltraVISTA covers $0.65 \deg^2$.}
    \label{fig:footprint}
\end{figure}

In this section we present the ground- and space-based data used in this work, and the steps taken to prepare the images for creating catalogues and conducting a high-redshift galaxy selection.

\subsection{Ground-based imaging and \textit{Spitzer}}

We made use of extensive multi-wavelength imaging in the COSMOS field \citep{scoville2007}. 
The UltraVISTA survey \citep{ultravista} is an ultra-deep NIR survey covering $1.72\,\mathrm{deg}^{2}$ at $\lambda=1$--$2.5\,\micron$, vital for characterising the UV continuum of high-redshift LBGs. 
Data Release 6 (DR6; see \citealt{UVISTA_overview} for an overview), the final data release,\footnote{https://eso.org/rm/api/v1/public/releaseDescriptions/221} provides uniform depths across {the full area}, bringing the `deep' stripes up to the same depth as the `ultradeep' stripes {($5\,\sigma$ depths in $Y$ of 26.2; see Sect. \ref{sec:depths}).
These stripes differed in depth in previous data releases, effectively halving the available ultradeep area.} 
The footprint of the UltraVISTA survey is shown in Fig. \ref{fig:footprint}.
Deep optical data are available from the Hyper Suprime-Cam Subaru Strategic Program DR3 \citep[HSC-SSP,][]{Aihara22} in the $grizy$ bands at $\lambda=0.4$--$1 \, \micron$.
{The HSC-SSP data cover the full UltraVISTA footprint.}
This deep optical imaging is critical for ensuring non-detections bluewards of the Lyman break, where flux bluewards of $1216\,\AA$ (rest-frame) is absorbed by neutral hydrogen in the intergalactic medium.
{HSC-SSP} also provides two narrowband filters, namely NB0816 and NB0921, powerful for reducing uncertainties in photometric redshifts {(in combination with $i$, $z$, and $y$ for constraining the break position, and for detecting Lyman-$\alpha$ with NB0921 at $z=6.57$).}
The HSC and VISTA filter widths and wavelength coverage are shown in Fig. \ref{fig:filters}.
Additionally, infrared data from \spitzer are available as part of the Cosmic Dawn Survey (DAWN, \citealt{Moneti-EP17}) which takes data from S-COSMOS \citep{S-COSMOS} and SMUVS \citep{SMUVS} in the $3.6 \, \micron$ and $4.5 \, \micron$ filters. These lie beyond the Balmer break of $z\simeq7$ galaxies, providing a measurement of their rest-frame optical flux.
The depths of the imaging (measured in Sect. \ref{sec:depths}) in the above filters are presented in Table \ref{tab:Depths}.

\subsection{\Euclid}

We made use of performance verification imaging, taken for calibration of the instrument, in the COSMOS field over three months.
The imaging used in this work is the result of 184 individual exposures in a single observation.
Each observation is composed of six exposures for VIS (four long exposures separated on the sky by the dither pattern and two short exposures on two of the dithers) and four exposures for NISP.
The calibration involves computing the astrometric solution using \textit{Gaia} DR3 \citep{Gaia_DR3}, flat fielding, and photometric calibration.
These calibrated frames are then background-subtracted, coadded and split into mosaics with common astrometry and pixel scale ($17\arcmin\times17\arcmin$ for the EDFs and EAFs with $\ang{;;0.1}\,\rm{pix}^{-1}$).
After just a few months of operation, the \Euclid imaging in COSMOS reaches deeper limiting magnitudes than the UltraVISTA survey, which was conducted over nearly 15 years.
The \Euclid imaging covers $0.65 \, \mathrm{deg}^{2}$ of the UltraVISTA footprint, as shown in Fig. \ref{fig:footprint}.
The NISP \YE, \JE, and \HE bands are $0.2$--$0.6$ mag deeper than their VISTA counterparts.
Additionally, the high-resolution \IE filter from the VIS instrument, which covers the same wavelengths as HSC $r$, $i$, and $z$, is $0.3$--$1.0$ mag deeper than these filters.
We show the \Euclid filter widths and wavelength coverage in Fig. \ref{fig:filters}.
The depths of the \Euclid imaging are presented in both Table \ref{tab:Depths} and Fig. \ref{fig:filters}.
We do not use data from the NISP spectroscopic channel (NISP-S, grism spectroscopy) since these sources are too faint for spectral extraction by the SIR pipeline ($\HE < 22.5$ is currently required).

\subsection{Image preparation}

We followed a similar procedure as in \citetalias{Varadaraj23}.
The UltraVISTA data were matched to the \textit{Gaia} EDR3 reference catalogue \citep[][]{Gaia}.
We used \texttt{Scamp} \citep{scamp} and \texttt{Swarp} \citep{swarp} to shift all other auxiliary data into the same frame as the UltraVISTA imaging.
The \Euclid `MER' (merged) mosaics imaging over the UltraVISTA field consisted of 23 of the $17\arcmin\times17\arcmin$ tiles.
\texttt{Swarp} was also used to resample the \Euclid images to match the pixel scale of UltraVISTA, $\ang{;;0.15}\,\mathrm{pix}^{-1}$, and to produce one large mosaic for each \Euclid image matching the plate scale of UltraVISTA.

\subsection{PSF homogenisation}
\label{sec:psf homogenisation}

In the ground-based seeing-dominated imaging, sources tend to be close to unresolved (PSF FWHM $=\ang{;;0.8}$).
The differences in PSF between VISTA and \Euclid and the possibility of sources being both resolved and unresolved depending on the instrument means that the fraction of flux falling within a fixed-size aperture depends strongly on the filter. 
We therefore homogenised all of the space-based imaging to match the VISTA $Y$ band PSF, which was chosen since it is the main detection band for $z\simeq7$ LBGs and because it has the largest PSF of VISTA.
We used \texttt{PSFEx} \citep{psfex} to construct an empirical PSF model. 
Stars were selected from the magnitude-FWHM diagram, with FWHMs determined by \texttt{SExtractor} \citep{sextractor}.
We used \texttt{PyPher} \citep{pypher} to find the convolution kernel between the \Euclid PSFs and the VISTA $Y$ band PSF.
The space-based images are convolved with this kernel, leading to PSF-homogenised images that are pixel-matched to UltraVISTA.
We rerun \texttt{PSFEx} \citep{psfex} on the homogenised images using the same stars selected in the original \Euclid images. 
The new PSF model is used to determine the enclosed flux within a $\ang{;;1.8}$ diameter circular aperture, assuming a point source.

Prior to PSF homogenisation, the FWHMs of the \Euclid PSFs are $\ang{;;0.20}$ in \IE, $\ang{;;0.49}$ in \YE, $\ang{;;0.51}$ in \JE, and $\ang{;;0.53}$ in \HE, as measured with \texttt{PSFEx}.
These values are in agreement with PSF FWHMs measured by the \Euclid pipeline.
After homogenisation of the \Euclid images, all PSF FWHMs match that of VISTA $Y$, $\ang{;;0.85}$.
We note that all \Euclid postage stamp cutouts presented in this work are from the original images, prior to PSF homogenisation.
We then check the PSF-homogenised \Euclid photometry to ensure it is consistent with UltraVISTA.
This is done by taking the stars selected in Sect. \ref{sec:psf homogenisation} and imposing the cut $| Y - J| < 0.05 \wedge |J - H| < 0.05$ to obtain stars with flat colours across the VISTA $YJH$ bands.
We then check the colour of each \Euclid filter and its nearest VISTA counterpart.
This check is shown in Appendix \ref{sec: euclid photometry check}.
Just as with the VISTA photometry, a minimum uncertainty of $5\%$ is placed on the \Euclid photometry.
There are no large colour differences between \Euclid and VISTA filters for stars with flat NIR colours, indicating that there are no major zeropoint or flux calibration issues in the \Euclid COSMOS imaging.

\subsection{Depths}
\label{sec:depths}

\begin{table}
        \centering
        \caption{Measured $5\,\sigma$ limiting magnitudes for each band used in this work in the COSMOS field.}
        \label{tab:Depths}
        \begin{tabular}{llc} 
                \hline
        \hline
        \noalign{\vskip 1pt}Filter & Depth ($5\,\sigma$) & Instrument  \\[1ex]
        \hline
        \noalign{\vskip 1pt}\textit{g} & 27.8 & HSC \\
        \textit{r} & 27.4 & HSC \\
        \textit{i} & 27.2 & HSC \\
        NB0816 & 26.2 & HSC \\
        \textit{z} & 26.7 & HSC \\
        NB0921 & 26.0 & HSC \\
        \textit{y} &  26.0 & HSC \\
        \noalign{\vskip 1pt}\hline
        \noalign{\vskip 1pt}\textit{Y} & 26.2 & VISTA \\
        \textit{J} & 26.0 & VISTA \\
        \textit{H} & 25.7 & VISTA \\
        \textit{K$_s$} & 25.3 & VISTA \\
        \hline
        \IE & 27.7 & \Euclid \\
        \YE & 26.4 & \Euclid \\
        \JE & 26.4 & \Euclid \\
        \HE & 26.3 & \Euclid \\
        \hline
        \noalign{\vskip 1pt}3.6 & 25.2 & \spitzer \\
        4.5 & 25.2 & \spitzer \\
         \hline
        \end{tabular}
    \tablefoot{
    The local depths were measured by placing $\ang{;;1.8}$ diameter circular apertures on empty regions of the images. The depth quoted here is the mode of these local depths. Depths for \Euclid were measured on images pixel-matched and PSF-homogenised to the VISTA \textit{Y} band. IRAC depths were measured in $\ang{;;2.8}$ diameter circular apertures on the original resolution image to account for the poorer resolution.
    }
\end{table}

We computed $5\,\sigma$ depths across the images by placing {roughly $2\times10^6$} apertures with diameters of $\ang{;;1.8}$ on empty regions of the image (determined using the \texttt{SExtractor} segmentation map).
For the \spitzer images we use a $\ang{;;2.8}$ diameter circular aperture to account for the broader PSF.
We determine local depth maps by taking the closest 300 apertures to each point and measuring the median absolute deviation of the aperture fluxes.
The depths reported in Table \ref{tab:Depths} are the mode of the local depths.
The depths of the PSF homogenised \Euclid images match the depths of the original image.

\subsection{Catalogues}
\label{sec:catalogues}

We ran \texttt{SExtractor} in dual image mode on a VISTA $Y+J$ stacked image, using the same parameters as in \citetalias{Varadaraj23}. 
Photometry was performed in a $\ang{;;1.8}$ diameter circular aperture, enclosing 70--80\% of the total flux assuming a point source.
This provides a balance between a high signal-to-noise ratio and preventing the need for a large aperture correction.
We used $\ang{;;2.8}$ diameter circular apertures for \spitzer to account for its broader PSF, enclosing a similar fraction of the total flux.
The raw aperture flux measurements must be corrected to account for light falling outside of the aperture in order to obtain a measure of the total flux.
As discussed in Sect. \ref{sec:psf homogenisation}, we use \texttt{PSFEx} to determine a PSF model with  stars selected from the FWHM versus MAG\_AUTO diagram. 
This empirical model is used to measure the flux enclosed in a $\ang{;;1.8}$ diameter aperture (or $\ang{;;2.8}$ for \spitzer), providing the PSF correction.
A minimum uncertainty of 5\% is placed on photometry in all bands.
We note that for extended sources, aperture photometry may underestimate the total flux.
To check this, we compare our aperture fluxes to MAG\_AUTO measurements from \texttt{SExtractor}, which uses flexible elliptical apertures to measure the total flux \citep{Kron1980}.
We find that the aperture photometry can underestimate the total flux by around 0.1 mag, consistent with a similar analysis conducted in (\citealt{Bowler14}; \citetalias{Bowler17}).
We scale the SEDs by the ratio between MAG\_AUTO and the aperture photometry before measuring quantities which rely on the SED, such as $\Muv$ and $V_{\rm max}$ (see Sects. \ref{sec: candidate galaxies} and \ref{sec: the rest-frame uv lf}).
When creating the catalogues, we also mask regions of low signal-to-noise and bright stars in the UltraVISTA images.
This is accounted for in the field area of $1.72 \, \rm{deg}^2$.

\section{Candidate selection}
\label{sec: candidate selection}

In this section we outline the methods used to select the $z\simeq7$ candidates.
We conducted two different selections: The UltraVISTA-only selection (U-only) and the UltraVISTA+\Euclid selection (U+E). For the first (U-only) selection, we ran the SED fitting steps with the HSC+VISTA+\spitzer photometry but without \Euclid. For the second (U+E), we added in the \Euclid photometry for the SED fitting and visual selection steps. We stress that the sample is still UltraVISTA-selected since sources were first selected based on their VISTA photometry (see Sect. \ref{sec:initial selection}).
By keeping the selection based on ground-based imaging, we avoided the introduction of artefacts from \Euclid, such as persistence \citep[see][]{EROLensVISDropouts}.
Persistence is an issue for LBG searches at $z>6$ since it only appears in NISP, masquerading as a VIS-dropout source.
Additionally, selecting from the shallower VISTA imaging usually ensures a detection in the deeper \Euclid imaging.\footnote{{Sources detected in VISTA but not in \Euclid are usually located in regions where the \Euclid imaging is shallower than VISTA, for example near the edges of the \Euclid footprint. However, the possibility of extremely high proper motion cannot be excluded.}}
The \Euclid photometry therefore provides additional information for the SED fitting of an UltraVISTA-selected sample.
In both selections, ancillary HSC and \textit{Spitzer} data are also used.
The HSC data are used to assert non-detections bluewards of the Lyman break, and if unconfused, the \spitzer data are used to remove low-redshift contaminants, {which generally show redder slopes at these wavelengths than LBGs (see Sect. \ref{sec: removing interlopers}).}

In the following sections, we outline the steps for the U-only sample. 
Then, in Sect. \ref{sec: incorporating euclid photometry}, we present the U+E SED fitting, highlighting the differences in the selection steps from the U-only sample.
The selection steps and number of sources removed at each step for both selections are shown in Table \ref{tab:selection steps}.

\subsection{Initial selection}
\label{sec:initial selection}

\begin{figure*}
    \centering
    \includegraphics[width=0.49\linewidth]{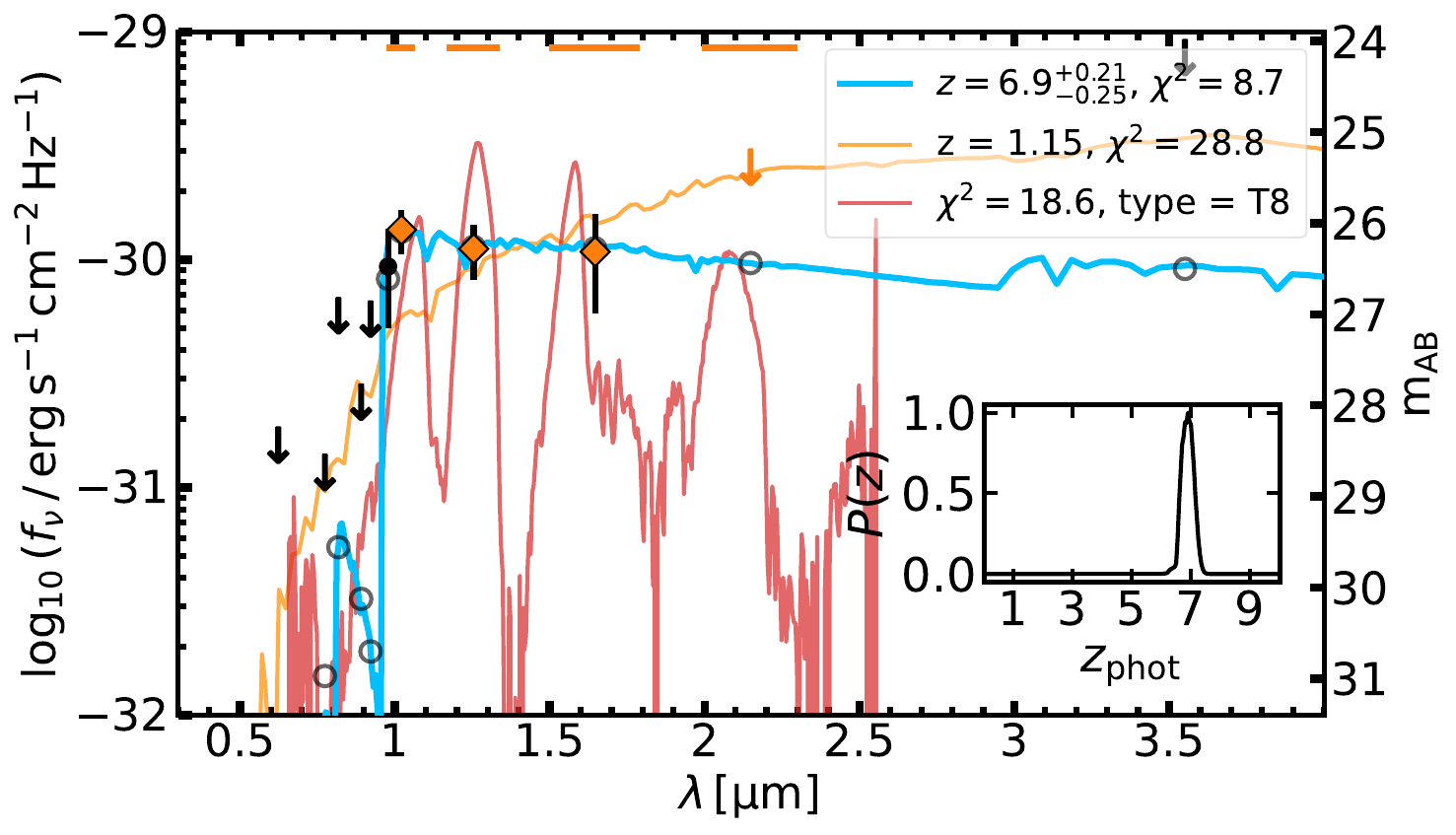}
    \includegraphics[width=0.49\linewidth]{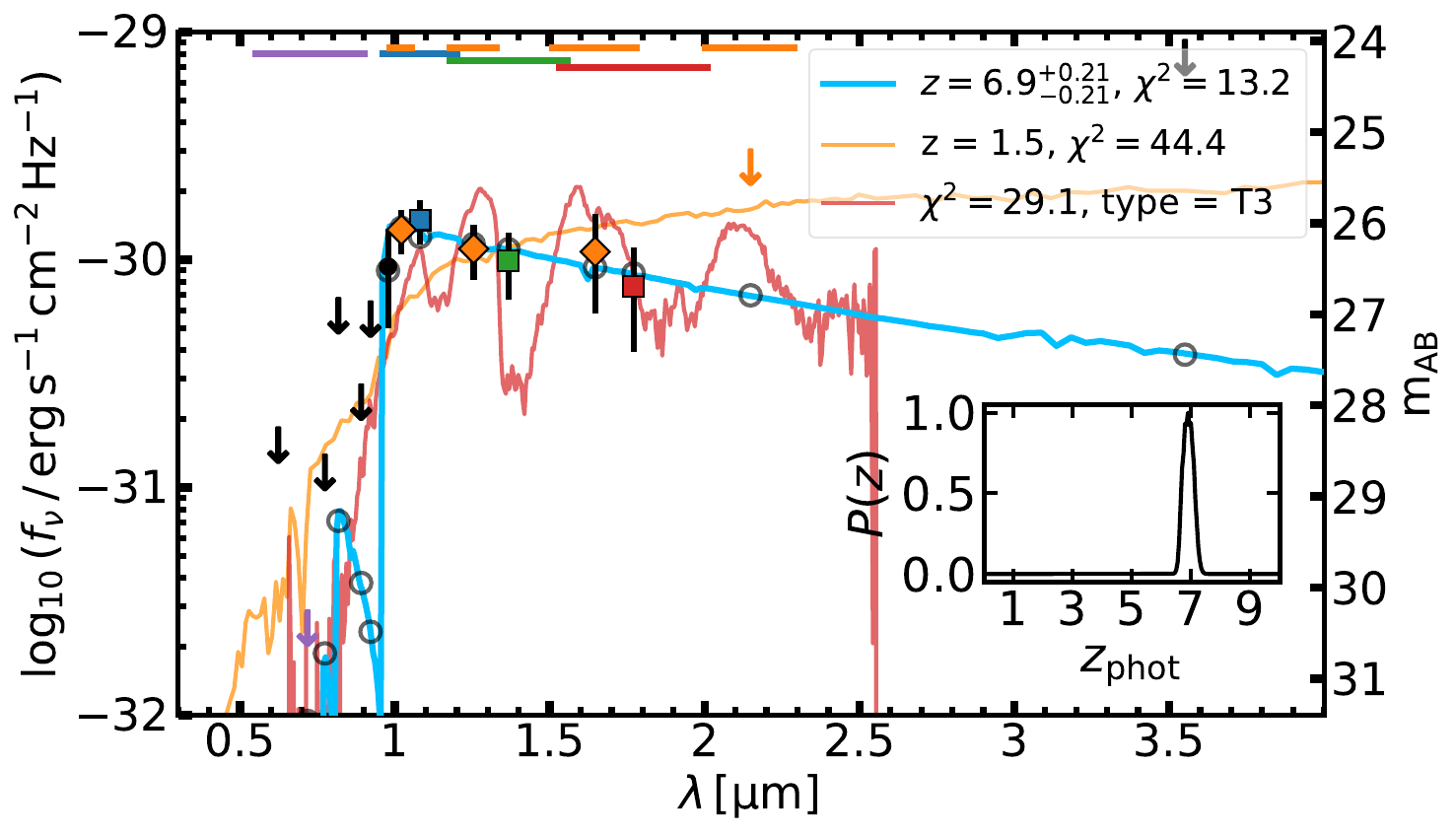}
    
    \caption{
    Spectral energy distribution fitting of a candidate LBG, EUCL\,J100041.40$+$020157.5, at $z=6.90$.
    Additional candidates are presented in Appendix \ref{sec: candidate seds and stamps}.
    \emph{Left}: SED fitting without \Euclid data as part of the U-only selection (see Sect. \ref{sec: candidate selection}).
    \emph{Right}: SED fitting including \Euclid data as part of the U+E selection.
    The HSC, VISTA, and \Euclid photometry are shown by the points, diamonds, and squares, respectively.
    The photometry is coloured following Fig. \ref{fig:filters}.
    Non-detections are replaced with $2\,\sigma$ upper limits.
    We also show the filter widths for VISTA and \Euclid at the top of the plot.
    The \spitzer model photometry and upper limit in the $3.6\,\micron$ filter is also shown.
    The blue curve shows the best high-redshift solution, and the grey open circles are its expected model photometry. 
    The orange curve shows the best dusty low-redshift solution, and the red curve shows the best UCD solution. 
    The legend in the top right shows the redshift and $\chi^2$ of the galaxy solutions, and the $\chi^2$ and spectral type of the UCD solution. 
    The inset panel shows the redshift probability distribution for this source.
    The inclusion of \Euclid data reinforces the exclusion of the UCD solution.
    We also note that the inclusion of \Euclid data prefers a bluer slope.}
    
    \label{fig:UVISTA euclid SED}
\end{figure*}

Objects were first selected by requiring that they are sufficiently bright in the detection filters. 
We imposed a $5\,\sigma$ detection threshold in $Y+J$, which removed roughly one third of the sources in the catalogue (see Table \ref{tab:selection steps}).
We then required non-detections ($<2\,\sigma$ significance) in the HSC $g$, $r$, and $i$ bands.
No condition was imposed on HSC $z$ since the Lyman-break can enter this filter towards the lower end of the redshift range $6.5 \leq z \leq 7.5$.
These non-detection conditions reduced the catalogue from $1\,051\,995$ to 4849 objects. 
We again stress that this first selection is for a base UltraVISTA sample without using \Euclid data.

\subsection{SED fitting}
\label{sec: SED fitting}

We used all ground-based filters available to conduct an SED fitting analysis in order to identify LBG candidates at $z\simeq7$.
In general, SED fitting is more complete than a colour-colour selection \citep{Adams20}, although it can introduce a more complex selection function {because it depends on the adopted templates and on degeneracies between them, particularly at a low signal-to-noise ratio.}
We used the \texttt{LePhare} SED fitting code \citep{Arnouts99, Ilbert06}, which minimises $\chi^{2}$, to find the best-fitting photometric redshifts and SEDs.
Following \citetalias{Varadaraj23}, we used \citet{BC03} stellar population models with metallicities of $Z \in \{0.2, 0.4, 1.0\} \, Z_{\odot}$. 
The SFHs we used are constant, instantaneous bursts, and exponentially declining; for the latter, we adopted timescales from $\tau=0.05$--$10 \, \text{Gyr}$.
Uniform priors were placed on the following parameters.
The redshift was allowed to vary between $z=0$--$9$.
Stellar population ages were allowed to vary between $10 \, \text{Myr}$ and $13.8 \, \text{Gyr}$, limited by the age of the Universe at a given redshift.
We used the \citet{calzetti} attenuation law, allowing $A_{V}=0.0$--$4.0$,
a \citet{chabrier} initial mass function is assumed, and intergalactic medium absorption is applied according to \citet{madau95}.
We initially only used HSC+VISTA to determine photometric redshifts, excluding \spitzer since the rate of non-detections increases in the faint end of the sample due to the shallower depth {and to make the completeness analysis more consistent.}
For example, in the final U+E sample, at $Y<25.5$, 17 out of 18 sources have an IRAC detection of at least $2\,\sigma$.
However, at $Y>25.5$ this drops to around half of the sources.
The main constraining power of \spitzer in this work is therefore to remove bright low-redshift dusty galaxy interlopers.
It is used later in Sect. \ref{sec: removing interlopers} in cases where there is no confusion.

Lyman-$\alpha$ emission can provide additional flux to the broad-band photometry, acting to increase the photometric redshifts of objects by up to $\Delta z \sim 0.5$ \citep{Bowler14}. 
To account for this, we also add Lyman-$\alpha$ emission lines to the \citet{BC03} templates with equivalent widths $0 \leq \mathrm{EW_{0}} \, / \AA \leq 240$ by measuring the continuum level in the range $\lambda=1250$--$1300 \, \AA$.

Based on the best-fit galaxy templates, candidates are first required to have their photometric redshift $z>6$.
The fits then had to be sufficiently good. 
Given the five degrees of freedom in the SED fitting, a $2\,\sigma$ significance threshold corresponds to $\chi^{2} < 11.3$.
Additionally, we require that the high-redshift solution is preferred to the low-redshift solution with $2\,\sigma$ significance, corresponding to $\Delta \chi ^{2} > 4$ between the two solutions. 
Single-band detections are also removed since they do not present robust candidates.
We show the model photometry of an example LBG in Fig. \ref{fig:filters}.

\subsection{Visual inspection}

\begin{table*}
    \centering

    \caption{Number of sources remaining in our catalogues after each selection step, beginning from the initial VISTA $Y+J$-selected catalogue.
    }
    \setlength{\tabcolsep}{2.9pt} 
    \begin{tabular}{ccc}
    \hline
    \hline
        Selection step & \multicolumn{2}{c}{Objects remaining} \\
        \noalign{\vskip 1pt}\hline
         Initial catalogue  & \multicolumn{2}{c}{1\,051\,995}\\
         $5\,\sigma \ Y+J$ VISTA cut & \multicolumn{2}{c}{706\,607}\\
         $<2\,\sigma$ in HSC $g$ & \multicolumn{2}{c}{32\,165}\\
         $<2\,\sigma$ in HSC $r$ & \multicolumn{2}{c}{13\,212}\\
         $<2\,\sigma$ in HSC $i$ & \multicolumn{2}{c}{4849}\\
         \noalign{\vskip 1pt}\hline
         \noalign{\vskip 1pt}SED fitting step & UltraVISTA-only & UltraVISTA+\Euclid\\
        \noalign{\vskip 1pt}\hline
         Overlap with \Euclid & --- &  1850\\
         Initial $z>6$ cut & 1872 & 403\\
         Visual selection & 751 & 341 \\
         Low-$z$: $\chi^2_{\rm{low-z}} < \chi^2_{\rm{high-z}}+4$& 718 & 333\\
         UCDs: $\chi^2_{\rm{UCD}} < \chi^2_{\rm{high-z}}$ & 656 & 315\\
         $6.5\leq z_{\rm{phot}} \leq 7.5$, with Ly$\alpha$ & 289 & 140 \\ 
         \hline
    \end{tabular}
    \tablefoot{
    The first column shows the selection step.
    The second column shows the number of objects remaining.
    For the SED fitting steps (and the visual selection), we conduct two different selections: the U-only selection, where \Euclid photometry is not included, and the U+E selection, where \Euclid photometry is included.
    In both selections, ancillary HSC and \textit{Spitzer} data are also used as required.
    For the U+E selection, we restrict the catalogue to the $0.65 \, \rm{deg}^2$ \Euclid footprint.
    The high-redshift ($z>6$) cut is defined as the best-fitting SED solution having its photometric redshift $z>6$, $\chi^2<11.3$ (U-only) or $17.5$ (U+E), and $\Delta \chi ^{2} > 4$ between the low- and high-redshift solutions.
    }
    \label{tab:selection steps}
\end{table*}

We carried out a visual selection of the remaining objects to remove artefacts.
These artefacts are typically diffraction spikes and crosstalk in the UltraVISTA images.
The latter are caused during the readout from the Vista InfraRed CAMera instrument, producing ghost images at regular pixel intervals from bright stars \citep[see][]{Kauffmann22}.
We used a catalogue of bright stars selected from the UltraVISTA $J$ band using \texttt{SExtractor} to flag potential crosstalk artefacts.
This was done by constructing a grid of positions at multiples of 128 pixels from these bright stars, then flagging sources within a conservative $\ang{;;6}$ radius of these positions.
At the magnitudes probed in this work, the crosstalk artefacts usually appear more diffuse and extended compared to real sources, and the visual presence in either HSC $z$ or \spitzer, which do not suffer from crosstalk, is a clear signpost for a real object.
We created an optical stack from HSC $g$, $r$, and $i$ to check for low-level flux indicative of a low-redshift galaxy.
We also smoothed the optical stack with a Gaussian filter with $\sigma = 2$ pixels, which helps when searching for low-level flux.
Crosstalk artefacts are numerous in the deep UltraVISTA data, leading to the removal of 60\% of sources.

\subsection{Removing interlopers}
\label{sec: removing interlopers}

There are two primary classes of low-redshift interloper objects that act to contaminate $z\simeq7$ LBG samples -- dusty low-redshift galaxies and UCDs.
Dusty galaxies at $z\sim1$--$2$ can have Balmer breaks that can be confused as Lyman breaks. 
However, these galaxies tend to have SEDs that increase in flux rapidly towards longer wavelengths \citep[see e.g.][]{Rodighiero10, LeBail24}.
Additionally, a Lyman break is usually stronger than the Balmer break, which tends to have shallower slope (see the dusty low-redshift galaxy model in Fig. \ref{fig:UVISTA euclid SED}).
If the SED fitting prefers a low-redshift solution to the $z>6$ solution when the \spitzer imaging is included and unconfused at $2\,\sigma$ significance ($\Delta\chi^2>4)$, the object is removed.

The UCDs of spectral type M, L, and T have SEDs featuring very little blue optical flux and heavy molecular absorption complexes (see e.g. \citealp{Burgasser24} and \citealp{Luhman24} for recent JWST spectroscopic observations).
These sources present a pressing challenge for ground-based searches since the peaks of the UCD SEDs usually coincide with the VISTA $YJHK_s$ filters, mimicking a flat NIR colour that can be confused with a blue $z\simeq7$ UV continuum. 
This occurs because the molecular species responsible for the absorption complexes (e.g. $\mathrm{CH_{4}, H_{2}O}$) are the same species responsible for making Earth's atmosphere opaque at certain wavelength ranges in the NIR \citep[e.g.][]{Bailey07}.
We use UCD templates taken from the SpeX prism library \citep{burgasser14} for the SED fitting, covering spectral types M4 through to T8, with one template per spectral type.
We exclude the HSC $g$ and $r$ bands from this fitting since the templates do not contain any information at these wavelengths.
We remove sources with $\chi^{2}_{\mathrm{UCD}}<\chi^{2}_{\mathrm{high-z}}$, meaning the UCD fit is preferred to the high-redshift solution.
In Fig. \ref{fig:filters}, we show the model photometry for an example UCD template, and compare it to the model photometry for an LBG to highlight the differences in the expected photometry.

We retain sources with a redshift $z<6.5$ when fitted without the Lyman-$\alpha$ emission line, but fall within our redshift range when the line is included and it provides the best fit (see Sect. \ref{sec: SED fitting}).
Finally, we restrict the redshift of candidates to $6.5\le z \le 7.5$, removing 55\% of sources, which lie at $6.0\le z < 6.5$.

\subsection{Redoing the SED fitting with \Euclid}
\label{sec: incorporating euclid photometry}

We then reran the SED fitting by combining \Euclid with VISTA on the remaining objects after the initial photometric selection (Sect. \ref{sec:initial selection}).
When the 4849 objects from this step are constrained to within the $0.65 \, \rm{deg}^2$ \Euclid footprint, there are 1850 remaining.
The additional four filters change the $\chi^2$ cut we impose for the initial high-redshift selection in Sect. \ref{sec: SED fitting}, increasing this to $\chi^{2} < 17.5$.
This initial high-redshift cut with \Euclid photometry removes many more objects than the same step without \Euclid photometry.
This is because VISTA artefacts have non-detections in \Euclid, leading to poor SED fits.
Following on from this, the \Euclid imaging is also extremely powerful for the visual check -- existence of a source in both \Euclid and VISTA immediately confirms it as real, and not an artefact.
Such a technique may also be used in reverse for future studies that use VISTA+\Euclid imaging -- selections based on \Euclid can use VISTA to rule out common artefacts such as ghosts and persistence.
Similarly, \cite{EROLensVISDropouts} leverage \spitzer detections to rule out artefacts.
The results of the selection steps after incorporating \Euclid photometry for the SED fitting and visual selection are also shown in Table \ref{tab:selection steps}.

\subsection{Expected number of ultra-cool dwarfs}

Using the model from \citet{Bowler15} of a single exponential disc with a scale height of $300 \, \text{pc}$, we expect that there will be around $800$ UCDs (of spectral type M4 through to T8) brighter than the initial $5\,\sigma$ cut in the full UltraVISTA field.
For the U-only selection, the UCD removal step removes 62 objects, which is fewer than expected.
However, prior to the initial high-redshift cut, of the 4849 objects there are 1372 objects with a best-fitting UCD SED.
The increased number with respect to the expected amount is likely to be caused by upscattering of UCDs into the selection, some faint galaxies being misidentified as UCDs, and crosstalk artefacts which had not yet been removed in a visual selection.

\begin{table*}[ht]
\centering
\caption{First five rows of the table containing the candidate properties of the U+E sample. }
\begin{tabular}{ccccccccccc}
\hline
\hline
ID & RA & Dec & $z_\mathrm{phot}$ & $\chi^2$ & $z_\mathrm{phot,sec}$ & $\chi^2_\mathrm{sec}$ & UCD & $\chi^2_{\rm UCD}$ & EW$_\mathrm{Ly\alpha}$ & $M_\mathrm{UV}$ \\
 & [deg] & [deg] &  &  &  &  & Model &  & [\AA] & \\
\hline
EUCL\,J100054.15$+$015048.3 & 150.226 & 1.846 & $6.65^{+0.11}_{-0.08}$ & 8.4 & 1.45 & 13.4 & M6 & 12.2 & 0.0 & $-20.43^{+0.08}_{-0.08}$ \\
EUCL\,J100105.05$+$015227.0 & 150.271 & 1.874 & $6.88^{+0.15}_{-0.31}$ & 6.5 & 1.50 & 27.6 & M6 & 21.4 & 0.0 & $-21.02^{+0.09}_{-0.13}$ \\
EUCL\,J100048.32$+$015330.8 & 150.201 & 1.892 & $6.56^{+0.06}_{-0.08}$ & 10.3 & 1.45 & 38.0 & T3 & 22.1 & 0.0 & $-20.92^{+0.07}_{-0.08}$ \\
EUCL\,J100028.55$+$015503.9 & 150.119 & 1.918 & $7.18^{+0.10}_{-0.21}$ & 14.1 & 1.55 & 42.4 & T8 & 21.2 & 0.0 & $-21.08^{+0.08}_{-0.10}$ \\
EUCL\,J100120.73$+$015542.5 & 150.336 & 1.928 & $7.22^{+0.14}_{-0.15}$ & 7.1 & 1.55 & 31.1 & T3 & 21.7 & 0.0 & $-21.23^{+0.08}_{-0.09}$ \\
 &  & &  & \vdots & &  &   &  & & \\
\hline
\end{tabular}
\tablefoot{
The fluxes within $\ang{;;1.8}$ diameter circular apertures, along with uncertainties, are also provided in the full version available at the CDS. 
We also provide the same table for the U-only sample.
The first three columns show the ID, RA, and Dec of the source.
The next six columns show photometric redshift and $\chi^2$ value of the LBG solution, as well as the low-redshift dusty galaxy solution, and the stellar type and $\chi^2$ of the UCD solution. 
The final two columns show the equivalent width of the Lyman-$\alpha$ emission line if a template with the emission line is preferred, and the absolute rest-frame UV magnitude $\Muv$.
}
\label{tab:sample}
\end{table*}

Restricting to the \Euclid footprint for the U+E selection, we expect around $250$ UCDs.
The UCD removal step cuts only 18 objects.
Prior to the initial high-redshift cut, there are 205 objects with a best-fitting UCD SED.
The total is very close to the expected number. 
The \Euclid photometry improves constraints via SED fitting (see Sect. \ref{sec: euclid sed fitting}), and \Euclid non-detections immediately rule out VISTA crosstalk artefacts, explaining the elevated number of objects with best-fit UCD SEDs in the U-only selection (relative to the prediction).

In \citetalias{Varadaraj23}, which used $8.2 \, \deg^2$ of imaging from the VISTA Deep Extragalactic Observations survey \citep[VIDEO,][]{Jarvis2013}, reaching $5\,\sigma$ depths in $Y$ of $25.2$, a large fraction (80\%) of high-redshift candidates were identified as possible UCDs, since sources sampled the magnitude range $24.1 \leq J \leq 25.2$ where the surface density of UCDs is much higher than that of LBGs \citep{Ryan11}.
UltraVISTA is a magnitude deeper than VIDEO, and the sample in this work probes down to $J = 26$ where the UCDs do not outnumber LBGs as substantially.
We note that the number of sources best fit as UCDs that make it through the initial high-redshift cut (62 for U-only, 18 for U+E) is significantly smaller than the final sample of LBGs (see Table \ref{tab:selection steps}), showing that UCD contamination is not as severe as in \citetalias{Varadaraj23}, but still a large issue.

\begin{figure}
    \centering
    \includegraphics[width=\linewidth]{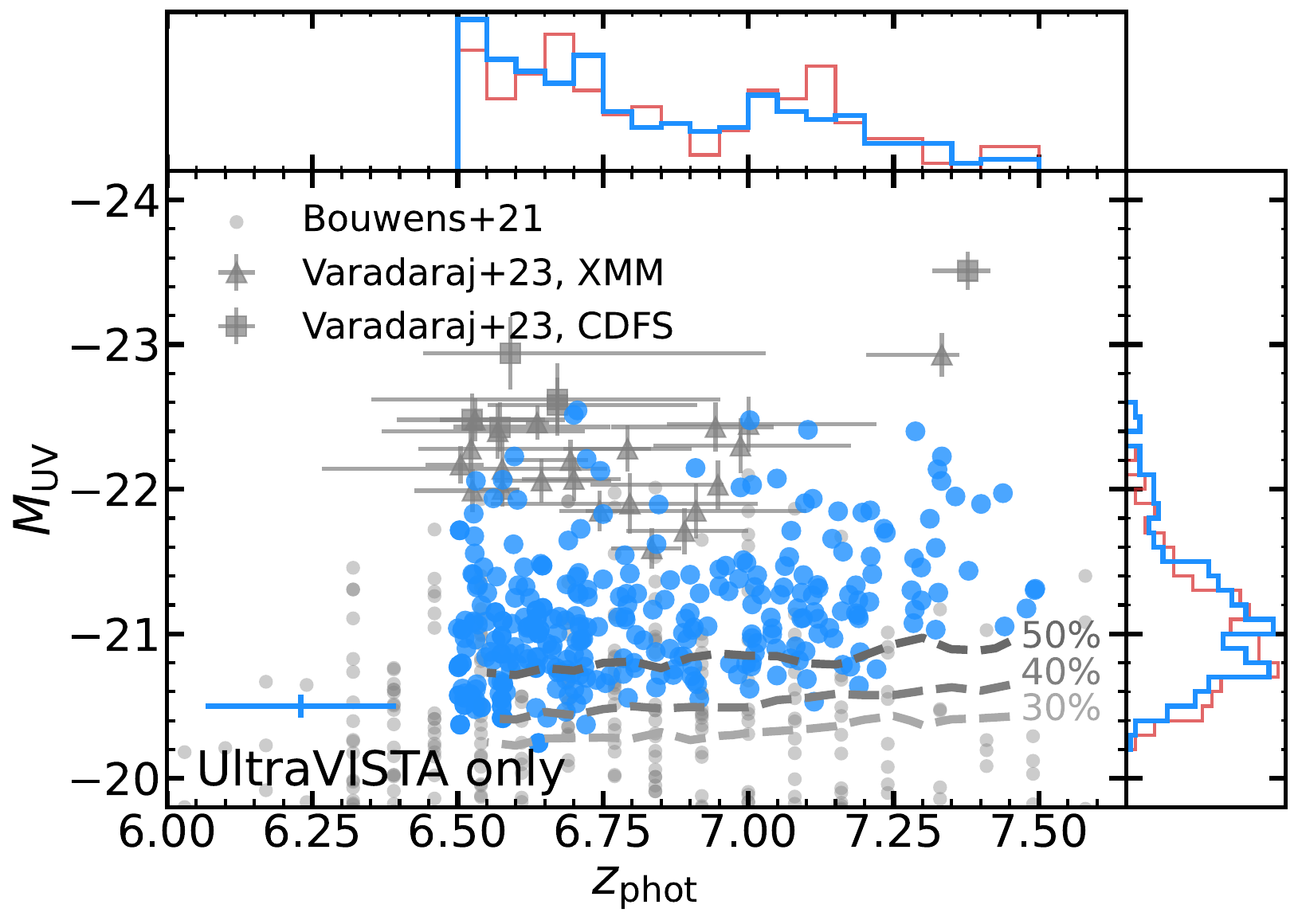}
    \includegraphics[width=\linewidth]{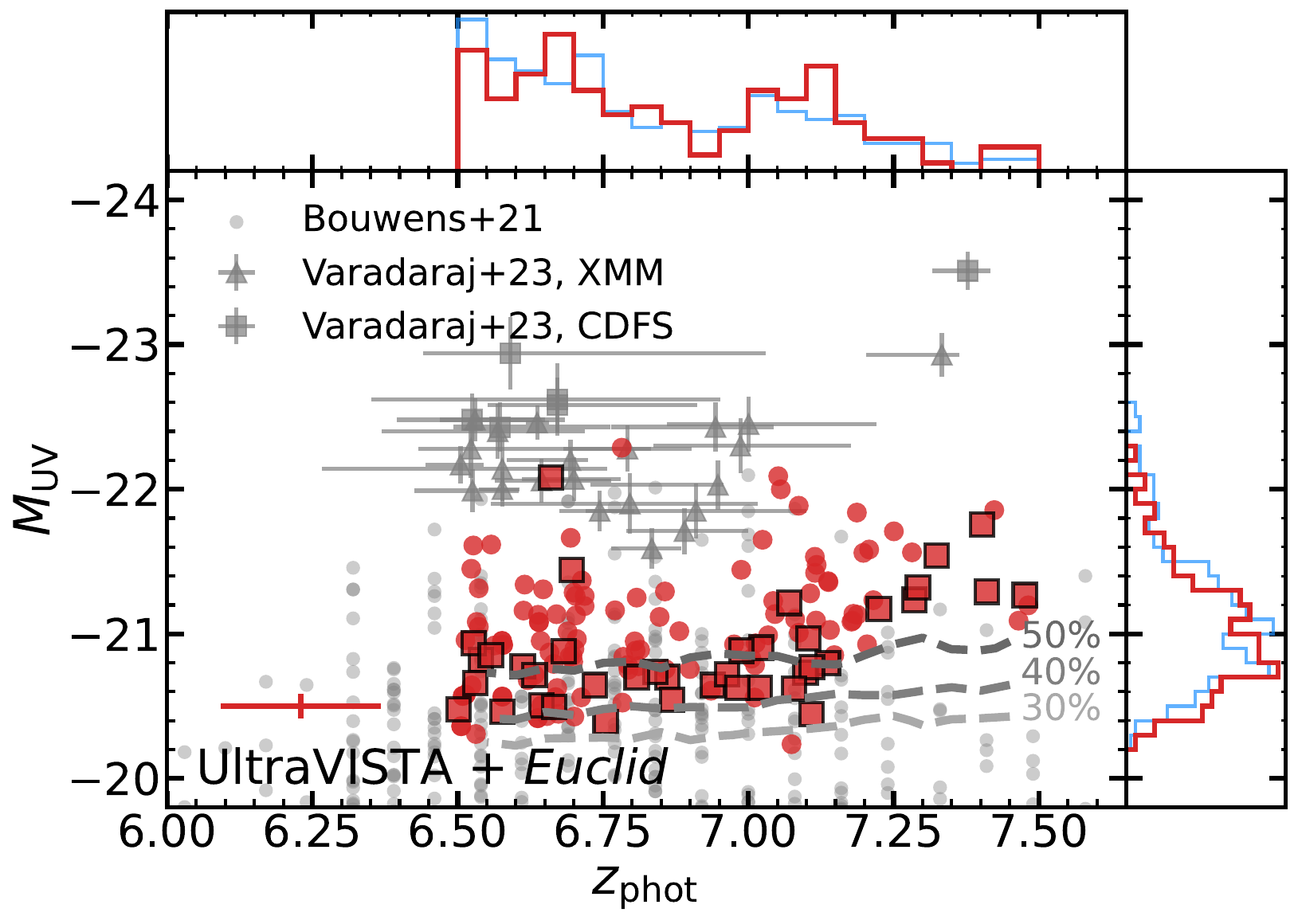}
    \caption{Sample of LBGs at $z\simeq7$ from this work plotted in photometric redshift ($z_{\rm{phot}}$) and absolute rest-frame UV magnitude ($\Muv$) space.
    \emph{Top}: U-only sample (blue).
    \emph{Bottom}: U+E sample (red).
    The red squares with a black outline indicate galaxies not recovered in the U-only sample and thus unique to the U+E sample.
    In both panels, we also show the candidates from \citetalias{Varadaraj23} in the XMM-LSS and ECDF-S fields, and the HST-selected candidates from \citetalias{Bouwens21}. 
    The dashed grey lines show the 30\%, 40\%, and 50\% completeness limits, as derived from the injection-recovery simulation (see Sect. \ref{sec:completeness}).
    We also show the marginalised distributions in $z_{\rm{phot}}$ and $\Muv$ as normalised histograms, and also overplot the distribution of the other sample as the thinner, fainter line for comparison. The mean uncertainties are shown on the bottom left.}
    \label{fig:z-Muv}
\end{figure}

\section{Candidate galaxies}
\label{sec: candidate galaxies}

The final U-only sample consists of 289 candidates selected from $1.72 \, \rm{deg}^2$ of UltraVISTA imaging.
We show the SED fitting of an example galaxy, EUCL\,J100041.40$+$020157.5 (hereafter LBG\,10004$+$02015), in Fig. \ref{fig:UVISTA euclid SED}.
Additional candidates are presented in Appendix \ref{sec: candidate seds and stamps}.
We measure the absolute rest-frame UV magnitude, $\Muv$, by placing a top-hat filter on the best-fit SED at $1500\,\AA$, with width $100\,\AA$.
The sample spans a range of over two magnitudes, $-22.5 \leq \Muv \leq -20.2$. Their distribution in $\Muv$ and photometric redshift $\zphot$ is shown in Fig. \ref{fig:z-Muv}.
The brightest candidates in our sample have similar $\Muv$ to the fainter sources in \citetalias{Varadaraj23}, where we used $8.2 \, \deg^2$ of shallower ($m_{\rm{AB}}\sim25$) imaging to select ultra-luminous $z\simeq7$ candidates.
Of course, due to the smaller area of UltraVISTA, our sample does not probe as bright as we could in the wider XMM-LSS and ECDF-S fields, but we are able to select a much fainter sample due to the improved depth.
Compared to the previous search for $z\simeq7$ LBGs in UltraVISTA by \citet{Bowler14} on DR2 imaging, we find an additional 257 galaxies thanks to the increase in depth by a magnitude, combined with an effective doubling in survey area since the `deep' stripes have been brought up to the same depth as the `ultradeep' stripes.  
UltraVISTA DR6 is also deep enough to reach magnitudes comparable to the bright end of the UV LF presented in \citetalias[][]{Bouwens21} (see Fig. \ref{fig:z-Muv}), the widest-area HST search for LBGs at $z\simeq7$, bridging the gap between space-based and ground-based observations of the UV LF at this redshift for the first time.

When we include \Euclid imaging over the $0.65 \, \rm{deg}^2$ which overlaps with UltraVISTA, for the U+E selection, our sample consists of 140 galaxies. 
Of these, 102 are also found in the U-only sample.
The remaining 38 galaxies are unique to the U+E sample.
In Fig. \ref{fig:UVISTA euclid SED}, we also show the SED fitting of LBG\,10004$+$02015 when the \Euclid photometry is included.
We show the distribution of the U+E sample in Fig. \ref{fig:z-Muv}, including the 38 galaxies that are not selected in the U-only sample.
These 38 galaxies have $\Muv$ largely corresponding to the 40--50\% completeness range, showing that the deeper \Euclid photometry recovers fainter galaxies in the sample.
Additionally, of these 38 galaxies, seven lie at $z>7.2$ with brighter magnitudes, $\Muv<-21$, where VISTA photometry is more susceptible to contamination by UCDs and crosstalk artefacts.
We present the SED fitting and postage stamp images of the brightest 30 galaxies in the U+E sample in Appendix \ref{sec: candidate seds and stamps}.
Tables of the U+E and U-only sources, along with \Euclid postage stamps of the U+E sources, is provided at the CDS.
We present the first five rows of the U+E sample in Table \ref{tab:sample}.

{In Appendix \ref{sec: Muv and zphot comparison} and Fig. \ref{fig:Muv comparison}, we compare the $\Muv$ and $\zphot$ derived with and without \Euclid.
This is done by comparing the measurements for sources which exist in both the U+E and U-only samples.
The photometric redshifts are in strong agreement across both samples.
The scatter appears to increase at $z>6.95$ (although all redshifts are consistent within errors).
This redshift corresponds to the blue side of the $\YE$ filter, suggesting that the increased scatter is driven by the improved determination of the position of the break when \Euclid data is included, driving shifts in the $\zphot$ by $\Delta z\sim0.1$-$0.2$.
Below this redshift, the wide $\IE$ filter does not provide significant information on the break position.
There is some scatter in the measurements of $\Muv$; however, the values only shift by up to 0.2 mag.
We briefly discuss the impact this has on the LF in Sect. \ref{sec: improved LF with euclid}.
}

\section{The UV LF with UltraVISTA and \Euclid}
\label{sec: rest-UV LF}

In this section we present the calculation of the UV LF at $z\simeq7$ using the U-only and U+E samples.
We then fitted a DPL and Schechter function using the U+E sample, before comparing our results to theoretical predictions at $z\simeq7$ and JWST determinations of the UV LF at $z>7$.

\subsection{Completeness}
\label{sec:completeness}

Incompleteness of the final galaxy sample must be accounted for before computing the UV LF.
Genuine high-redshift galaxies can become blended with other objects, and near the limiting magnitude of the detection images, photometric scattering can cause objects to drop in/out of the selection.
We therefore ran injection-recovery simulations to derive corrections for these effects in bins of $\Muv$ and $\zphot$.
First, we populated a redshift--absolute-magnitude grid with steps of $\Delta z = 0.05$ and $\Delta M = 0.1$, assuming the DPL LF derived in \citet{harikane24}. 
The absolute magnitude grid extends down to $\Muv = -19$, well below the limiting magnitude of UltraVISTA DR6, to account for the photometric up-scattering of faint sources.
The grid consists of $10^6$ sources.
Then, we generated mock $Y$ and $J$ photometry.
We drew rest-frame UV slopes, $\beta_{\rm{UV}}$, from a Gaussian distribution centred on $\beta_{\rm{UV}}=-2$ with standard deviation $\sigma=0.2$ \citep{Bowler14}.
We injecedt empirical PSF models (see Sect. \ref{sec:psf homogenisation}) into the imaging, assuming sources are generally unresolved in ground-based imaging \citep[see][]{Bowler14}.
We injected $10^3$ sources at a time into the $Y+J$ image so as to not artificially boost the number density too much.
Then, we ran \texttt{SExtractor} in the same manner as in Sect. \ref{sec:catalogues} and select sources using the same cuts.
We show the 30, 40, and 50\% completeness limits in Fig. \ref{fig:z-Muv}.
The faintest sources in our sample correspond roughly to the 40\% contour, below which the completeness drops rapidly.

\subsection{The rest-frame UV LF}
\label{sec: the rest-frame uv lf}
\begin{table*}
    \centering
    \caption{Values of UV LF at $z\simeq7$. }
    \begin{tabular}{c c c c c c}
        \hline
        \noalign{\vskip 1pt}\hline
         \noalign{\vskip 1pt}& \multicolumn{2}{c}{UltraVISTA-only} & \multicolumn{2}{c}{UltraVISTA+\Euclid} \\
         \noalign{\vskip 1pt}\hline

         $\Muv$ & $n_{\text{gal}}$ & $\phi$ & $n_{\text{gal}}$ & $\phi$ \\
         (mag) & & $(\rm{mag^{-1} \ Mpc^{-3}})$ & &$(\rm{mag^{-1} \ Mpc^{-3}})$  \\ 
        \noalign{\vskip 1pt}\hline
        $-22.6$ & $5$  & $(1.31\pm 0.60)\times10^{-6}$ & $0$  & $<1.83 \times 10^{-6}$ \\
        $-22.2$ & $12$ & $(3.33\pm 1.03)\times10^{-6}$ & $3$  & $(2.13\pm 1.28)\times10^{-6}$ \\
        $-21.9$ & $13$ & $(7.34\pm 2.18)\times10^{-6}$ & $4$  & $(6.03\pm 3.19)\times10^{-6}$ \\
        $-21.7$ & $11$ & $(6.28\pm 2.02)\times10^{-6}$ & $6$  & $(9.12\pm 4.01)\times10^{-6}$ \\
        $-21.5$ & $26$ & $(1.57\pm 0.35)\times10^{-5}$ & $10$ & $(1.59\pm 0.56)\times10^{-5}$ \\
        $-21.3$ & $43$ & $(2.74\pm 0.50)\times10^{-5}$ & $19$ & $(3.23\pm 0.88)\times10^{-5}$ \\
        $-21.1$ & $58$ & $(4.13\pm 0.68)\times10^{-5}$ & $25$ & $(4.42\pm 1.09)\times10^{-5}$ \\
        $-20.9$ & $47$ & $(3.83\pm 0.68)\times10^{-5}$ & $27$ & $(5.84\pm 1.42)\times10^{-5}$ \\
        \hline
        
    \end{tabular}
    \tablefoot{The first column shows the central absolute UV magnitude $\Muv$ of the bin.
    We then show the number of galaxies $n_{\text{gal}}$ and the UV LF value and uncertainty for two cases: firstly for the U-only sample, and secondly for the U+E. The brightest two bins centred at $\Muv=-22.6$ and $-22.2$ have a width of $\Delta\Muv=0.4$, and the remaining bins have width $\Delta\Muv=0.2$. 
    We present the $1\,\sigma$ upper limit for the U+E bin at $\Muv=-22.6$ \citep{Gehrels86}.}
    \label{tab:LF values}
\end{table*}

\begin{figure}
    \centering
    \includegraphics[width=\columnwidth]{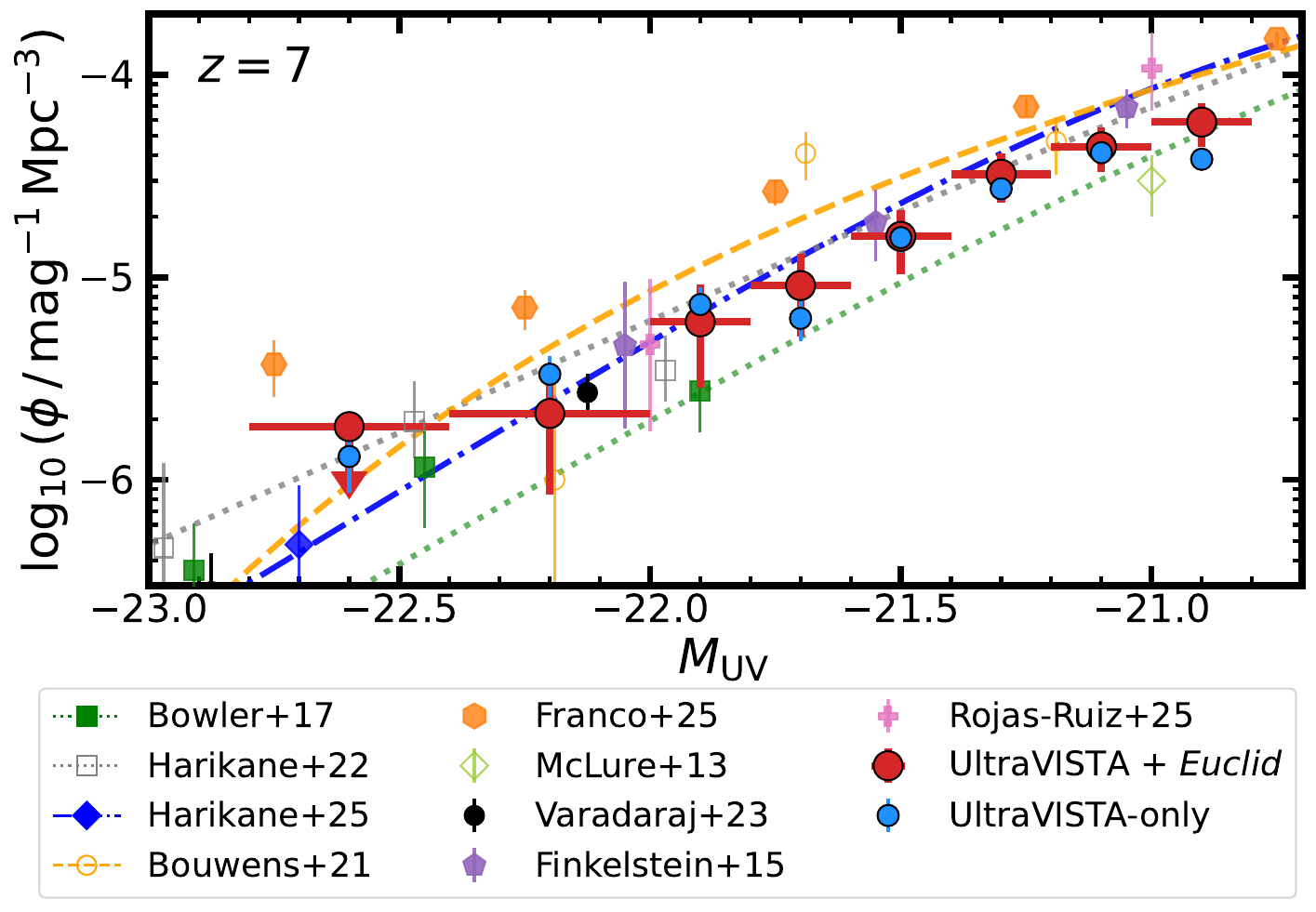}
    \caption{Comparison of the UV LF at $z\simeq7$ measured with the two samples presented in Fig. \ref{fig:z-Muv} and Sect. \ref{sec: candidate galaxies}.
    The smaller blue circles show the LF points calculated from the U-only sample.
    The larger red circles show the LF points calculated from the U+E sample.
    We used the same binning in both cases, and the LF values are presented in Table \ref{tab:LF values}.
    The brightest bin contains no galaxies from the U+E sample, so we show a $1\,\sigma$ upper limit whose value is noted in Table \ref{tab:LF values}.
    {We show results from \citet[][]{McLure13}, \citetalias{Finkelstein15}, \citetalias{Bowler17}, \citetalias{Bouwens21} ($6.3<z<7.3$), \citet[][$6.3<z<7.7$]{harikane22}, \citetalias{Varadaraj23}, \citet{harikane24}, and \citet[][$5.5<z<8.5$]{Franco25}.
    Unless otherwise stated, these LFs are measured at $6.5<z<7.5$.}
    Also shown are the best-fit Schechter and DPL fits found by \citetalias{Bouwens21} and \citet{harikane24} respectively.}
    \label{fig:LF UVISTA}
\end{figure}

We used {both samples of LBGs (U+E and U-only)} to determine the UV LF at $z\simeq7$ using the $1/V_{\rm{max}}$ method \citep{Schmidt68}. 
The term $V_{\rm{max}}$ is the {maximum comoving volume in which the galaxy could lie and still fulfil our selection criteria}.
The galaxy SEDs were redshifted in steps of $\Delta z = 0.01$ until they fell out of the $5\,\sigma$ detection threshold in $Y+J$, giving us its maximum redshift $z_{\rm{max}}$ at which it would still be detected in our selection. 
Then, $V_{\rm{max}}$ is the comoving volume between $z=6.5$ and $z=z_{\rm{max}}$. 
The value of $z_{\rm{max}}$ cannot exceed $z=7.5$, the maximum considered redshift.
The UV LF in a given bin of absolute magnitude, $\phi(M)$, in the redshift range $6.5<z<7.5$, is then given by
\begin{equation} \label{eq: lf}
    \phi(M) = \frac{1}{\Delta M} \, \sum^{N}_{i=1} \, \frac{1}{C(M_i,z_i)} \, \frac{1}{V_{\rm{max},i}}\;,
\end{equation}
where we summed over $N$ galaxies in the magnitude bin with width $\Delta M$.
The term $C(M_i,z_i)$ is the completeness value for the galaxy in its magnitude and redshift bin, calculated in Sect. \ref{sec:completeness}.
We assumed uncertainties are Poissonian, given by
\begin{equation} \label{eq: lf error}
    \delta \phi (M) = \frac{1}{\Delta M} \sqrt{\sum^{N}_{i=1} \left( \frac{1}{C(M_i, z_i)V_{\rm{max}, i}} \right) ^2} .
\end{equation}
\citep{Adams23, Adams24}.
We also accounted for the effect of cosmic variance.
Since galaxy surveys sample the wider large-scale structure of the Universe, over- and under-densities such as filaments and voids can bias measurements of the LF.
We used the \citet{Trenti08} calculator to estimate the uncertainty due to this effect.
We find that cosmic variance contributes no more than 12\% to the total error budget in our brightest bin, with Poissonian uncertainties dominating.
We added these cosmic variance contributions in quadrature to the uncertainties calculated with Eq. (\ref{eq: lf error}).
When computing the LF, we truncated to only include galaxies with $\Muv < -20.7$, corresponding to the 50\% completeness limit for the upper end of our redshift bin (see Fig. \ref{fig:z-Muv}), to ensure a robust and complete sample is used for the LF determination.
We chose $\Muv$ bins such that they span the magnitude range of the sample down to the 50\% completeness limit.
The brightest bins centred at $\Muv=-22.6$ and $-22.2$ have widths $\Delta \Muv = 0.4$ to ensure that at least five galaxies lie within them for the U-only sample. 
The remaining bins centred at $\Muv=-21.9$ down to $\Muv=-20.3$ have narrower widths of $\Delta \Muv = 0.2$.
This provides a finer binning than used for the HST results of \citetalias{Finkelstein15} and \citetalias{Bouwens21} whilst keeping at least ten galaxies in these narrower bins, benefiting both the fitting and the comparison with other studies.
We used the same binning scheme for the U+E sample to provide a direct comparison between the two samples.
The UV LFs from the two selections (U-only and U+E) are presented in Fig. \ref{fig:LF UVISTA}, and the LF values are presented in Table \ref{tab:LF values}.

\subsection{Improved LF measurement with \Euclid}
\label{sec: improved LF with euclid}

Comparing the two LFs computed with each sample in Fig. \ref{fig:LF UVISTA}, it is immediately clear that the U-only LF points show some scatter, whereas when \Euclid is included in the SED fitting, the points follow a smooth decline towards the bright end.
This is due to the additional \Euclid photometry providing a better characterisation of LBGs and low-redshift interlopers, and including \Euclid also allows for a more straightforward removal of VISTA crosstalk artefacts.
Additionally, the \Euclid photometry better probes the deep molecular absorption features seen in UCD SEDs, which are largely inaccessible with only ground-based NIR filters.
Therefore, it is also the case that the UCD removal step, namely removing objects with $\chi^2_{\rm{UCD}} < \chi^2_{\rm{gal}}$, performs better with \Euclid photometry since the degeneracy between a UCD SED and a flat UV continuum is broken.
Additionally, the LF value of the faintest U-only bin at $\Muv=-20.9$ decreases compared to the value at $\Muv=-21.1$ (although are consistent within the uncertainties).
When objects are selected in VISTA $Y+J$, the faintest sources may only have their detection in one or two of these filters, since the VISTA $H$ and $K_s$ filters drop in depth rapidly (see Table \ref{tab:Depths}). 
On the other hand, \Euclid provides uniform depth across all its NIR filters, leading to detections of the UV continuum across $\lambda_{\rm{obs}}= 1$--$2.5\,\micron$ for the faintest sources, leading to their robust characterisation as high-redshift LBGs. 
Finally, there is some scatter in the $\Muv$ determined with and without the \Euclid imaging. 
We show the differences in $\Muv$ in Fig. \ref{fig:Muv comparison}.
This leads to some objects lying in different bins after the inclusion of \Euclid data, providing an additional source of scatter in the LF bins.
As shown in Fig. \ref{fig:UVISTA euclid SED}, the inclusion of \Euclid photometry can alter the slope of the UV continuum slope, which contributes to differences in $\Muv$.
We further discuss the improved SED fitting constraints \Euclid provides in Sect. \ref{sec: euclid sed fitting}.

The $2\,\sigma$ upper limit provided for the $\Muv=-22.6$ bin for the U+E sample \citep[following][]{Gehrels86} is consistent with the LF value for the U-only sample in this bin.
However, we find that the sources from this brightest bin which overlap with the \Euclid footprint have their photometric redshifts shifted to $z>7.5$ when \Euclid photometry is included.

\subsection{Double power law and Schechter function fitting}
\label{sec: double-power law fitting}

\begin{figure}
    \centering
    \includegraphics[width=\linewidth]{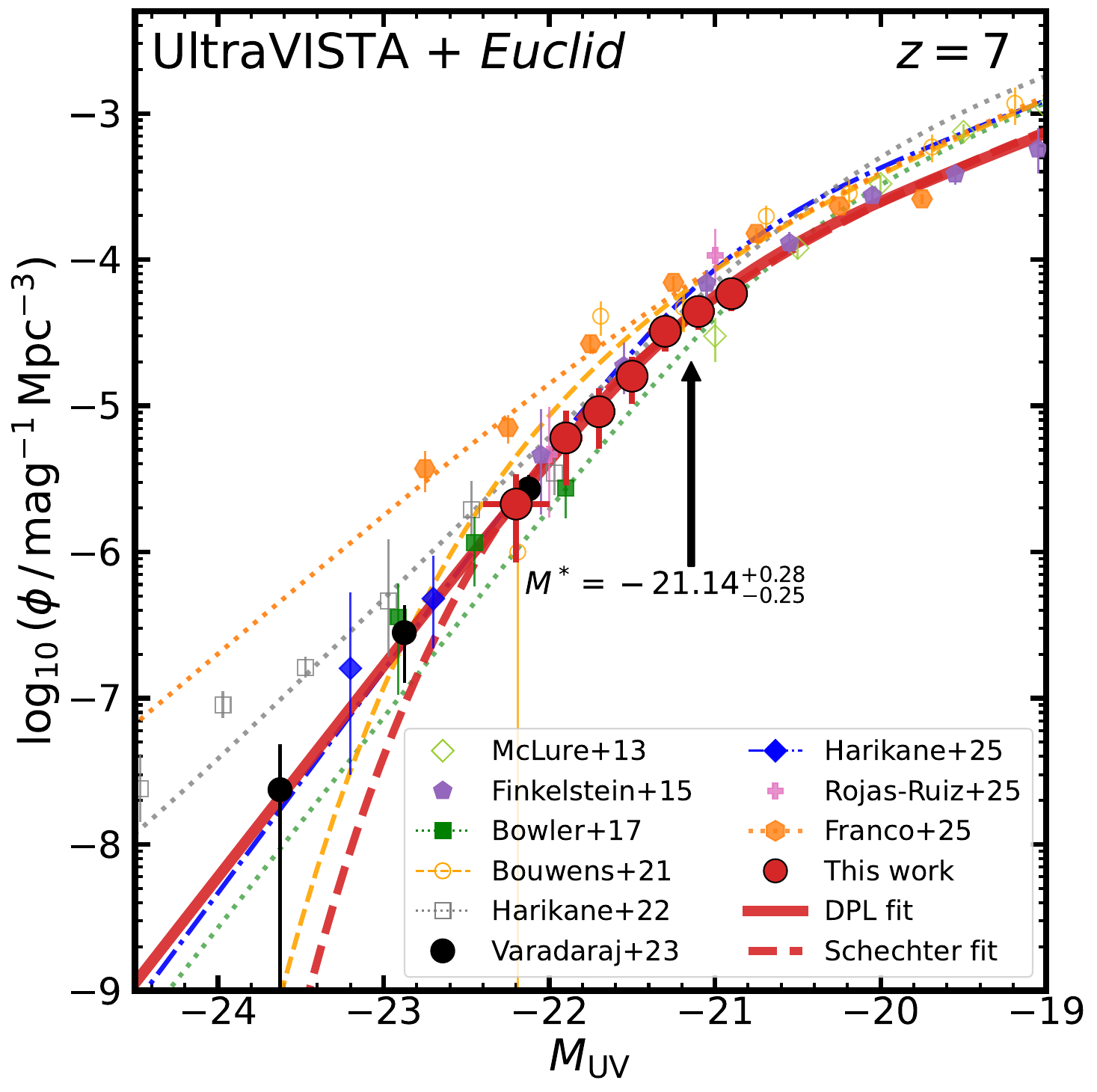}
    \caption{Best-fit DPL and Schechter function for the UV LF at $z\simeq7$ shown by the solid and dashed red lines, respectively.
    For the fitting, we used results from this work (red points, the U+E sample), bright-end results from \citetalias{Varadaraj23} (black points), and faint-end results from \citetalias{Finkelstein15} (purple pentagons).
    {We also show results at $6.5<z<7.5$ (unless otherwise stated) from \citet{McLure13}, \citetalias{Finkelstein15}, \citetalias{Bowler17}, \citetalias{Bouwens21} ($6.3<z<7.3$), \citet[$6.3<z<7.7$]{harikane22}, \citet{harikane24}, \citet[$5.5<z<8.5$]{Franco25}, and \citet[$7<z\leq8.4$]{RojasRuiz25}.
    For comparison to our fit, we plot the best-fit DPLs from \citetalias{Bowler17}, \citet{harikane22}, and \citet{harikane24}.}
    }
    \label{fig:DPL fit LF}
\end{figure}

Numerous studies have shown that a DPL provides a better fit to the UV LF at $z\simeq7$ than a Schechter function, due to an excess of bright galaxies at $L > L^*$ (\citealt{Bowler14}; \citetalias{Bowler17}; \citealt{harikane22}, \citetalias{Varadaraj23}; \citealt{harikane24}).
We also note that gravitational lensing may play an important role in shaping the bright end, particularly at $z\geq10$ with \Euclid \citep{Mason15}.
The DPL has the functional form
\begin{equation} \label{eq: dpl}
    \phi(M) = \frac{\phi^*}{ 10^{\,0.4\,(\alpha+1)\,(M-M^*)} +10^{\,0.4\,(\beta+1)\,(M-M^*)}}\;,
\end{equation}
where $\phi^*$ is the normalisation, $M^*$ is the characteristic magnitude, $\alpha$ is the faint-end slope, and $\beta$ is the bright-end slope.
We fitted a DPL to our U+E sample, incorporating bright-end results from \citetalias{Varadaraj23} and faint-end results from \citetalias{Finkelstein15}. 
{These two studies are chosen because they carefully remove UCD interlopers, providing natural extensions of the sample in this work to the bright and faint end.}
The best-fit parameters are presented in Table \ref{tab:dpl fitting}, and the resulting LF is shown in Fig. \ref{fig:DPL fit LF}. Our fit yields bright- and faint-end slopes $\beta=-4.63^{+0.34}_{-0.39}$, $\alpha=-2.10^{+0.21}_{-0.17}$, with the knee given by $M^*=-21.14^{+0.28}_{-0.25}$ and $\phi^*=0.91^{+0.67}_{-0.38}\times10^{-4} \, \rm{mag}^{-1} \, \rm{Mpc}^{-3}$.
By means of the deep degree-scale imaging used in this work, this is the first time ground-based imaging has robustly probed fainter than the knee of the UV LF at $z\geq6$. 
The constraints provided by combining UltraVISTA and \Euclid also enable reliable sampling around $-22 \leq \Muv \leq -21$, crucial for measuring $M^*$ and $\phi^*$.
We also fitted a Schechter function of the form
\begin{equation}
\phi(M) = 0.4\,\ln(10)\,\phi^*\,10^{\,0.4\,(M-M^*)\,(\alpha+1)}\,\exp\left(10^{\,-0.4\,(M-M^*)}\right)\;.
\end{equation}
The gradual decline seen in the LF points at $-22 < \Muv < -20.7$ is consistent with the Schechter function.
In fact, the DPL and Schechter function are indistinguishable at $\Muv > -22.5$
However, at $\Muv < -22.5$, the Schechter function diverges from the best-fit DPL, and is unable to account for the brightest LF points of \citetalias{Bowler17}, \citetalias{Varadaraj23}, and \citet{harikane24}.

Our derived slopes, $\alpha$ and $\beta$, are consistent with previous DPL measurements from \citetalias{Bowler17} and \citet[][see our Fig. \ref{fig:LF param evolution} for a comparison of the LF parameter measurements]{harikane24}.
The value of $\alpha$ also agrees with faint-end Schechter fits from \citetalias{Finkelstein15} and \citetalias{Bouwens21}. 
Our $M^*$ is consistent with \citetalias{Bowler17}, \citet{harikane24}, and the Schechter fit from \citetalias{Finkelstein15}, although note that the value of $M^*$ derived from a Schechter fit by \citetalias{Finkelstein15} is different from that for a DPL fit by the remaining studies.
The value we derive for $\phi^*$ is also in agreement with these studies, but with smaller errors.
Overall, our results agree with those of \citet{harikane24}, who use a spectroscopically confirmed sample, eliminating contamination. 
This further suggests that we have effectively mitigated the encroachment of low-redshift interlopers into our $z\simeq7$ U+E sample. 
Likewise, our DPL is consistent with \citetalias{Bowler17}, although with a slight shift to higher $M^*$ and lower $\phi^*$, but consistent within uncertainties. 
This implies that the constraining power of ground-based imaging for the LF, prior to sizeable overlap with space-based studies, was already quite strong.
However, our results show that the gentle DPL decline continues down to the knee of the UV LF, providing a definitive measurement of the bright-end slope.
This is best shown by the LF values of our faintest bins, which use deeper data and similar selection steps as \citetalias{Bowler17}, yet show no significant discrepancy with \citetalias{Finkelstein15}, indicating that there are no large systematic differences. 
The largest source of uncertainty in bright end of the LF thus remains the contamination by low-redshift interlopers and UCDs, which is reflected by the relative excesses seen in \citet[][as discussed in \citetalias{Varadaraj23}]{harikane22} and \citetalias{Bouwens21}.
This is discussed further in the next section.

In the EAFs, \Euclid will have substantial overlap with VIDEO.
A selection similar to \citetalias{Varadaraj23}, but incorporating \Euclid data to mitigate contamination, would be highly beneficial for confirming the bright-end shape.
Additionally, future work in the EDFs utilising the DAWN survey \citep{EP-McPartland} will provide unprecedented constraints on the bright end with $53 \deg^2$ of imaging.

\begin{table}
    \centering
    \caption{Best-fit parameters for the UV LF.}
    \begin{tabular}{cccc} 
    \hline
    \hline
    $\phi^*$ & $M^*$ & $\alpha$ & $\beta$ \\
    $\rm{mag}^{-1} \, \rm{Mpc}^{-3}$ & mag &  \\
    \noalign{\vskip 1pt}\hline
    \noalign{\vskip 1pt}$0.91^{+0.67}_{-0.38}\times10^{-4}$ & $-21.14^{+0.28}_{-0.25}$ & $-2.10^{+0.21}_{-0.17}$ & $-4.63^{+0.34}_{-0.39}$ \\
    \noalign{\vskip 1pt}$1.63^{+0.83}_{-0.61}\times10^{-4}$ & $-20.98^{+0.20}_{-0.21}$ & $-1.98^{+0.18}_{-0.16}$ & --- \\
    \noalign{\vskip 1pt}\hline
    \end{tabular}
    \tablefoot{The parameters for the DPL fit are shown in the top row, and for the Schechter fit in the bottom row.
    We performed our fit using our UV LF points, those from \citetalias{Varadaraj23}, and faint-end results from \citetalias{Finkelstein15}.}

    \label{tab:dpl fitting}
\end{table}

\subsection{Comparison with other studies}

In Fig. \ref{fig:LF UVISTA} we compare our LF points to other studies.
Our results are consistent with the bright-end studies of \citetalias{Bowler17}, \citet{harikane22}, \citetalias{Varadaraj23}, and \citet{harikane24}, confirming a gradual decline in the LF at $\Muv<-22$.
The constraining power of UltraVISTA DR6 is best seen at $\Muv>-22$.
For this, we focus on the U+E results.
As discussed in the previous section, our LF points are remarkably consistent with the HST results of \citetalias{Finkelstein15}, indicating that any differences in methodology do not cause major systematic offsets.
For example, UCD contaminants were removed by \citetalias{Finkelstein15} using a combination of the \texttt{SExtractor} FWHM and their colours, whereas we use SED fitting.
As we show in Sect. \ref{sec:UCDsize}, it is not possible to remove faint UCDs as point sources at the magnitudes probed in this work with \Euclid imaging.
Additionally, \citetalias{Finkelstein15} calculate the effective volume for a galaxy using their injection-recovery simulations, whereas we directly redshift the SED of each galaxy iteratively to determine its maximum redshift, and then its maximum occupied volume.
In their simulations, they also allowed the galaxy size to vary.
\citetalias{Finkelstein15} note that had they fixed their galaxy sizes to $R_{\rm{e}} = 1\,\rm{kpc}$, they would have derived similar effective volumes, mirroring results from \citet{Grazian12}.
\citetalias{Finkelstein15} also determine their LF points using a non-parametric stepwise maximum likelihood calculation, although this approach produces equivalent results to the $1/V_{\rm{max}}$ method for bins with high enough number counts.
The consistency with \citetalias{Finkelstein15}, despite slight differing strategies for removing low-redshift interlopers, suggests that both of our approaches are successful at mitigating contamination.

Also, as discussed in the previous section, the consistency with the spectroscopic sample of \citet{harikane24} suggests little contamination in our sample.
Similarly, we also find that our results are consistent with the findings of \citet{RojasRuiz25}, who used the BoRG-JWST survey to spectroscopically confirm candidates identified in pure-parallel HST fields.
In Fig. \ref{fig:LF UVISTA} we only compare to the sample from their GO 2426 programme since it covers a similar redshift range, $7.0 < z \leq 8.4$, although extends to higher redshifts.
However, there is little evolution in the bright end of the LF between $z=7$--$8$ \citepalias{Varadaraj23}.

Our four brightest U+E bins are in some tension with the brightest bins of \citetalias{Bouwens21}, who see a drop of $1.6$ dex between the two bins from $\Muv=-21.7$ to $\Muv=-22.2$. 
Since we are able to use narrower binning than the $\Delta M = 0.5$ mag bins used by \citetalias{Bouwens21} in this range, we can probe the finer evolution in number density across this magnitude range. 
We observe a gentle decline in the number density across our four brightest bins, dropping by only $0.9\,\text{dex}$, which is more consistent with the behaviour of the underlying DPL distribution of \citet{harikane24} compared to the Schechter function found by \citetalias{Bouwens21}.
Whilst the discovery of numerous $\Muv < -22.5$ galaxies at $z\simeq7$ has convincingly ruled out the Schechter function form for the LF (\citealt{Bowler14}; \citealt{harikane22}; \citetalias{Varadaraj23}), our results show this gradual decline also occurs from the knee of the LF.
As discussed by \citetalias{Finkelstein15} and \citetalias{Bowler17}, the discrepancy with \citetalias{Bouwens21} may be due to a lack of deep $Y$ band imaging in the majority of the CANDELS fields, critical for determining the strength of a break to rule out UCDs and low-redshift galaxy interlopers.

Our LF values, when combined with \citetalias{Varadaraj23}, are significantly lower than the results of \citet{Franco25} at $\Muv \leq -22$, who use the COSMOS2025 catalogue \citep{Shuntov25} to measure the UV LF in a redshift bin $z=5.5$--$8.5$ from JWST COSMOS-Web imaging \citep{CWEB} by selecting sources which drop out of the HST F814W filter and are detected in JWST F115W (along with detections in the redder filters).
For this sample they require $\rm{F814W} - \rm{F115W} > 0.5$, selecting the Lyman break.
In this work, our reddest dropout filter, HSC $i$, is 0.8--1 mag deeper than our detection filters, $Y$ and $J$. 
Additionally, \IE is 1.5 mag deeper than our detection filters.
\citet{Franco25} are likely robust against UCD contamination, since the high-resolution JWST imaging provides the means to remove them as point sources.
However, the shallower depth of their dropout filter, relative to their detection filter, may introduce contamination by low-redshift dusty galaxies.
A large magnitude difference between the dropout and detection filter is useful for distinguishing between a Balmer break and a Lyman break.
As discussed in \citetalias{Varadaraj23}, imposing a brighter selection in the detection filter provides one method for combatting a relatively shallow dropout filter.

\subsection{Comparison with theory}
\label{sec: comparison with theory}

\begin{figure}
    \centering
    \includegraphics[width=\linewidth]{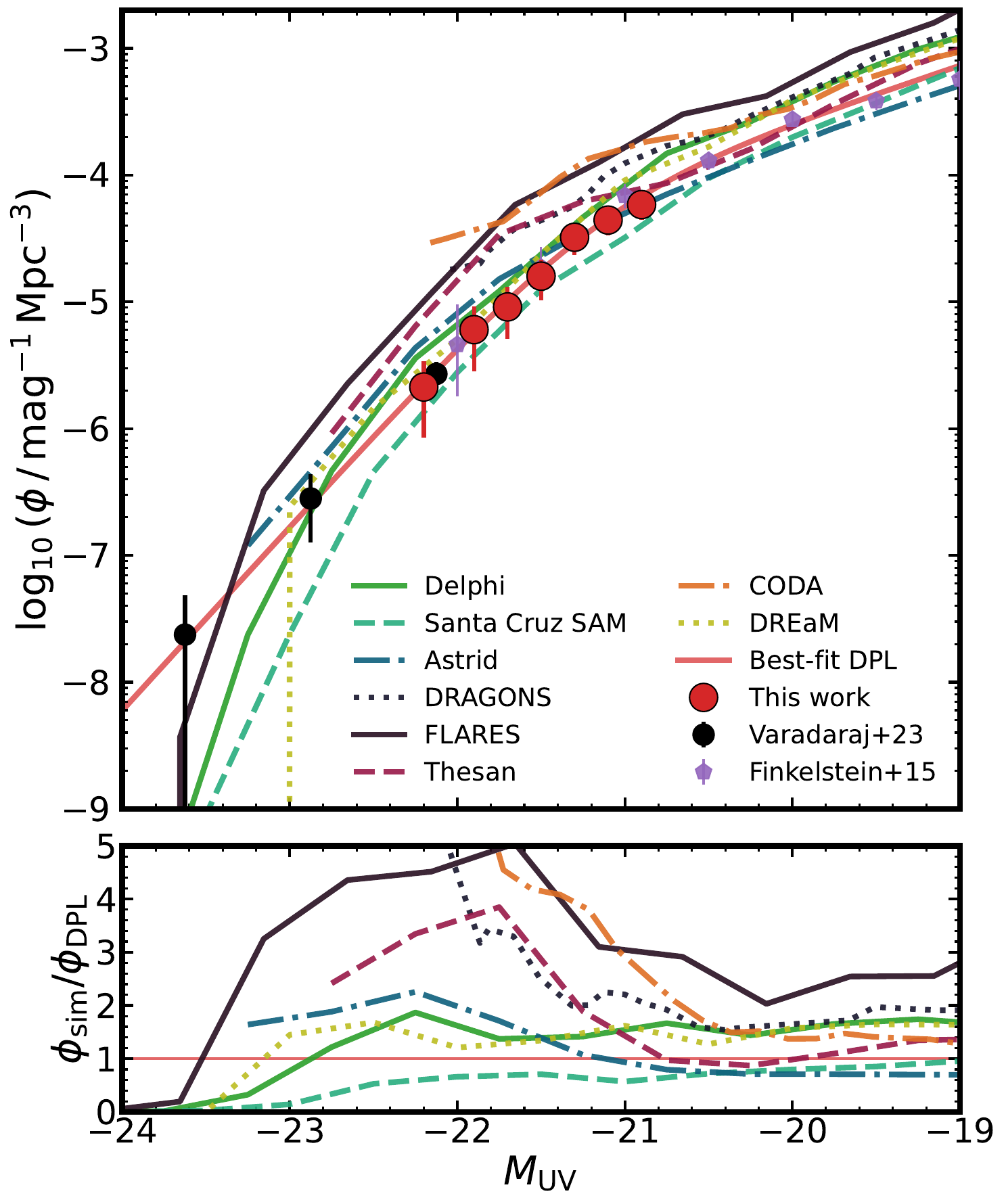}
    \caption{Comparison of the UV LF measured in this work with predictions from semi-analytic models and hydrodynamic simulations.
    (See Sect. \ref{sec: comparison with theory} for an outline of the studies.)
    We also show LF results from \citetalias{Varadaraj23} and \citetalias{Finkelstein15}.
    We show our best-fit DPL with the red line.
    The bottom panel shows the ratio of the LF predictions to our best-fit DPL, as a function of $\Muv$.}
    \label{fig:LF sim comparison}
\end{figure}

In Fig. \ref{fig:LF sim comparison} we compare our UV LF results (along with results from \citetalias{Varadaraj23}) to predictions from various simulations and theoretical models.
These include Delphi \citep[][]{Delphi, Dayal22}, the Santa Cruz semi-analytical model \citep[SAM,][]{SantaCruzSAM}, Astrid \citep{ASTRID}, DRAGONS \citep{DRAGONS}, FLARES \citep{Lovell21, Vijayan21, Wilkins23}, Thesan \citep{Kannan22}, CoDa \citep{CODA}, and DREaM \citep{DREAM}.
We compare to these studies since they have a large enough volume to extend to at least $\Muv = -22$ in their predictions of the UV LF at $z=7$.
Broadly speaking, there are two features that can be drawn from this comparison. 
{Firstly, most of the predictions reproduce the faint end  of the LF ($\Muv \gtrsim-21$, although FLARES is in excess of our LF points by at least factor of two, and CoDa begins to over-predict significantly towards the knee of the LF).}
Secondly, it appears that there is a split in the bright end - some predictions are in significant excess of our results, whereas others agree with our results out to $\Muv = -23$.
Specifically, Delphi, Astrid and DREaM are consistent with our results.
{The Santa Cruz SAM is consistent out to $\Muv = -22$, beyond which it begins to under-predict the LF.}
The agreement is likely because of the calibration of some of these simulations to relevant observations, in particular to observations of dust. 
For example, Delphi includes fully coupled treatment of metal and dust enrichment in order to explain the dust masses of REBELS galaxies \citep{REBELS, Inami22}, which are luminous LBGs at $z=7$ selected from ground-based imaging, analogous to (and overlaps with) the sample presented in this work. 
The Santa Cruz SAM tunes the dust extinction optical depth to match the UV LF at $z=4$--$10$. 
Astrid calibrates dust extinction at $z=4$, and assumes no evolution of dust extinction across $z=3$--$10$.
DREaM calibrates its $\Muv$-stellar mass relation on results at $z\leq4$, and extrapolations at $z>4$ are consistent with observational results from \citet{Stark13}.
On the other hand, DRAGONS and FLARES both calibrate their dust extinction on results from \citep{Bouwens14, Bouwens15}, which seems to cause a large excess compared to our results at $\Muv=-22$.
Both Thesan and CoDa discuss they do not produce enough dust in the highest mass haloes and at $\Muv<-21$ respectively, explaining the excess.

Clearly, the shape of the bright end is highly sensitive to dust obscuration, and the prescription of dust is sensitive to the observations used to tune it. 
However, most predictions agree that significant reduction and steepening of the intrinsic UV LF occurs due to dust obscuration, producing the observed bright-end slope of $\beta\sim4.5$--$4.6$, and galaxies at $z=7$ with $\Muv < -22$ are able to experience dust attenuation of up to 2 mag in the rest-frame UV.
A suggestion by \citet{Dayal22} is that the dust and star-forming regions are spatially offset or perhaps a significant fraction of the total dust mass diffuses into the ISM and therefore no longer contributes to the attenuation of UV light.
Indeed, recent studies have found significant dust build-up and offsets between dust and UV emission in $z\simeq7$ LBGs \citep[e.g.][]{Bowler22, Inami22, Lines24, Algera25}.
There is still uncertainty in the $z=7$ UV LF beyond $\Muv<-23$, so pinning down the precise shape of the ultra-bright end with \Euclid in the EDFs, whilst simultaneously making attempts to measure the dust content of the sample \citep[akin to][]{REBELS} is critical for further comparison with simulations.

\subsection{Comparison with JWST: A gradual evolution in the bright-end slope}
\label{sec: lf discussion}

\begin{figure*}
    \centering
    \includegraphics[width=\linewidth]{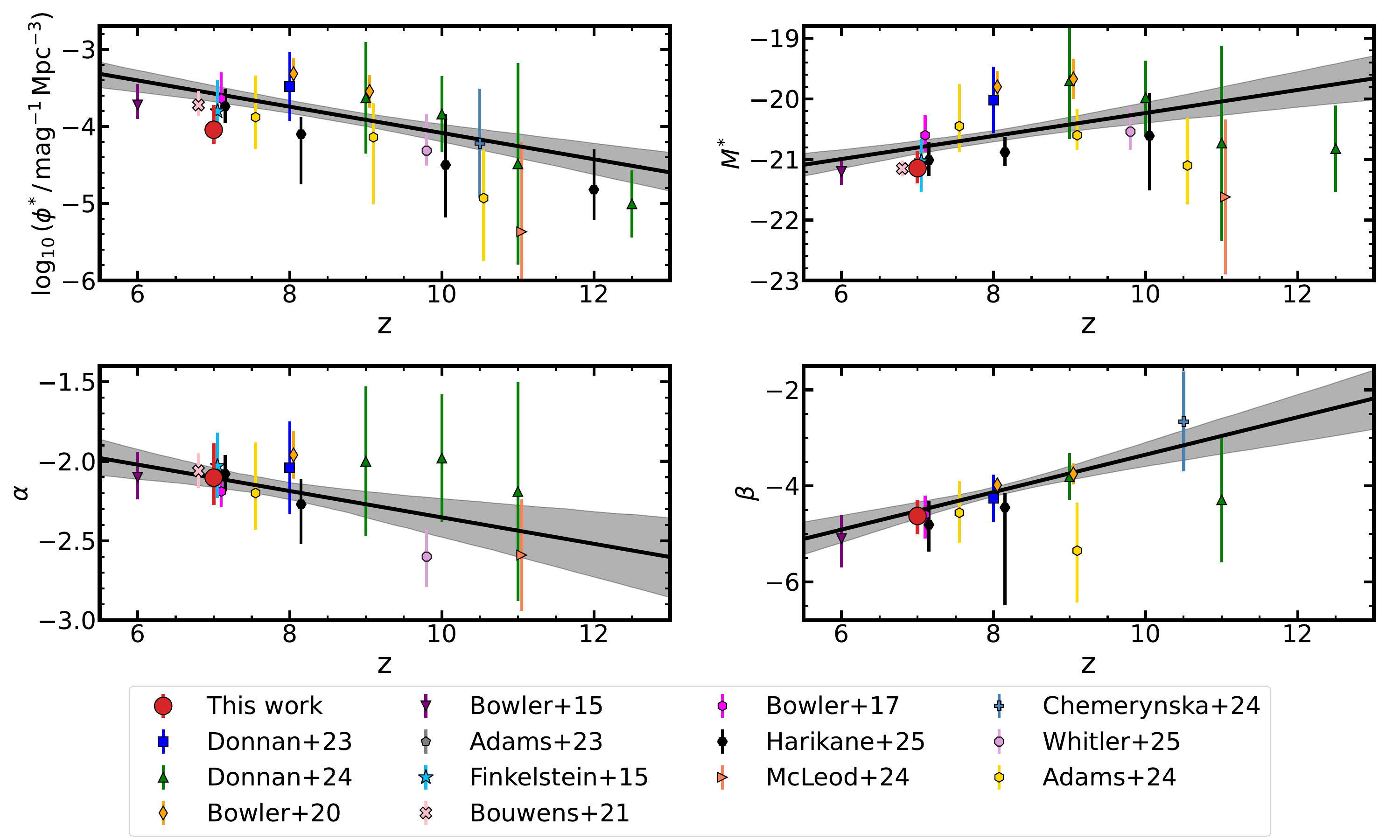}
    \caption{Evolution of the DPL parameters with redshift, namely the normalisation, $\phi^*$; the characteristic magnitude, $M^*$; the faint-end slope, $\alpha$; and the bright-end slope, $\beta$ (see Eq. \ref{eq: dpl}).
    The red points show the parameters derived from the DPL fitting in this work. 
    We show a compilation of results at $z=6$--$12.5$ from \citet{Bowler15, Bowler20}, \citetalias{Bowler17}, \citet{Adams23, Adams24}, \citet{Donnan23, Donnan24}, \citet{Chemerynska24}, \citet{McLeod24}, \citet{harikane24}, and \citet{Whitler25} with some slight offsets in redshift for clarity.
    We do not show points from these studies that were fixed during the DPL fitting.
    For $\beta$, we only show studies which have at least two LF points brighter than their $M^*$, meaning they have some constraints on the bright-end slope.
    The black line indicates the best-fit straight-line to the $z\geq6$ data.
    {We also show the results of Schechter fitting from \citetalias{Finkelstein15} and \citetalias{Bouwens21} for ease of comparison in Sect. \ref{sec: double-power law fitting}.}
    }
    \label{fig:LF param evolution}
\end{figure*}

By enabling the discovery of luminous sources at early cosmic time, JWST has revolutionised our understanding of the $z>7$ Universe \citep[e.g.][]{Naidu22, Finkelstein22, Castellano22}.
However, ground-based imaging is still highly important for probing the bright end of the LF at $z>6$.
Currently, JWST lacks the volume to probe the bright end.
Additionally, at $z=6$--$7$, without deep ancillary optical HST imaging, JWST does not have enough dropout filters bluewards of the expected position of the Lyman break at $\lambda_{\rm{obs}}\sim1\,\micron$.
We also note that the luminous sources discovered by JWST are fainter than the brightest ground-based candidates presented in this work and in \citetalias{Varadaraj23} \citep[GNz-11 is currently the brightest spectroscopically confirmed galaxy at $z>10$ with $\Muv = -21.5$,][]{Bunker23}.
Our $z\simeq7$ LBGs may be linked to these early sources.
It is likely that $\Muv<-22$ sources abundant in ground-based imaging at $z=7$ occupy similar dark matter haloes to that of the brightest JWST sources such as JADES-GS-z14-0 and JADES-GS-z14-1 \citep{Carniani24}, GHZ2 \citep{Castellano24}, and GNz-11 \citep{Bunker23}.
It is thus natural to compare our LF results to $z>7$ JWST results to understand the evolution of luminous LBGs in the first Gyr of cosmic time.

In Fig. \ref{fig:LF param evolution} we show the evolution of the LF parameters across $z=6$--$13$ from a range of ground-based, HST and JWST studies.
We only show results from DPL fitting, and for $\beta$ we only show results which have at least two LF points at $\Muv<M^*$, such that they sufficiently probe the bright end.
Note that JWST bright-end determinations have large error bars at $z>9$ due to the limited volume available.
We fitted straight lines to these studies at $z\geq6$ to probe the linear evolution, and find that, relative to $z=6$, these can be expressed as
\begin{equation}
\begin{aligned}
\log_{10}(\phi^*/\,\mathrm{mag}^{-1}\,\mathrm{Mpc}^{-3}) 
    = (-3.40 \pm 0.15) \\
    + (-0.17 \pm 0.05)\,(z-6)\;, \\
M^* = (-20.99 \pm 0.15) + (0.19 \pm 0.07)\,(z-6)\;, \\
\alpha = (-2.02 \pm 0.09) + (-0.08 \pm 0.05)\,(z-6)\;, \\
\beta  = (-4.91\pm 0.29) + (0.39 \pm 0.13)\,(z-6)\;,
\end{aligned}
\end{equation}
for the DPL LF parameters. 
Overall, our results are consistent with an evolution in the LF driven by a shallower $\beta$ at higher redshifts, a mild evolution in the position of the knee, $\phi^*$ and $M^*$, and weak evolution in $\alpha$.
These results differ slightly from those of \citet{Bowler20}, who found that the evolution of the LF was dominated by $\beta$ and $M^*$, with $\phi^*$ remaining relatively constant over $z=7$--$10$.
\citet{Donnan24} use JWST multi-field imaging to show that $\phi^*$ has a stronger evolution than $M^*$, although they are required to fix their $M^*$ at some redshifts.
However, their evolution in $\phi^*$ is consistent with the luminous sources found at $z>7$, with a gentle evolution in $M^*$ allowing for the existence of these sources.

We note that a major caveat is that the LF parameters are degenerate during the fitting, limiting discussion regarding the evolution of $\phi^*$ and $M^*$.
However, ground-based studies are able to probe sufficiently bright LBGs to determine the bright-end slope and allow for a meaningful discussion of the bright-end evolution.
When fitting the evolution of $\beta$, we again note that we only used JWST studies that had at least two LF bins brightwards of their $M^*$, such that they sufficiently probe the bright end of the LF.
Comparing to the number of studies in the plot above for $M^*$, most JWST studies do not have the dynamic range to provide reliable measurements of the bright end of the LF at $z>8$, because large areas are needed to find the rarest sources.
This is reflected in much smaller uncertainties on $\beta$ from ground-based studies (\citealt{Bowler15}; \citetalias{Bowler17}; \citealt{Bowler20}; \citealt{Donnan23}, and this work).
A lack of bright-end measurements also results in weaker constraints on the knee of the LF.
This can be seen in the large amounts of scatter in the values of $M^*$ found by JWST studies. 
In fact, the increasing trend we see towards higher redshift is driven by the small uncertainties (relative to the JWST studies) on $M^*$ found by the ground-based study of \citet{Bowler20}.
Their small errors are caused by their sample lying entirely at $\Muv < M^*$ (with DPL fits determined by combining with faint-end results from \citealt{McLure13} and \citealt{McLeod16}).
It is therefore not entirely clear whether $M^*$ increases at higher redshift (as suggested by ground-based studies) or remains fairly constant with redshift (as suggested by JWST studies, but with large uncertainty).
This demands a determination of the UV LF with \Euclid at $z\geq8$.

The DPL form of the LF at $z\simeq7$ is more akin to the functional form of the halo mass function than a Schechter function, indicating that quenching of star formation and/or dust obscuration has not begun to dominate luminous LBGs at this epoch \citep{Bowler14}.
However, numerous studies have also shown that $z\simeq7$ LBGs can host large dust reservoirs and exhibit signs of substantial dust obscuration \citep[e.g.][]{Bowler24, Algera25}.
The discrepancy arises due to unobscured star formation being probed by optical+NIR studies such as this work, and obscured star formation requiring sub-millimetre observations to measure the dust emission.
Interestingly, we observe a gradual steepening in $\beta$ from $z=9$ down to $z=6$ based on ground-based studies.
The results of \citet{Chemerynska24} and \citet{Donnan24} are consistent with this gradual evolution in $\beta$, but with larger uncertainties.
Note that \citet{Chemerynska24} use a large redshift bin of $9<z<12$, a useful strategy for boosting number counts over a redshift range corresponding to only $180 \, \text{Myr}$.
The bright-end slope is sensitive to dust obscuration \citep[e.g.][]{Cai14} and quenching of star formation \citep{Peng10}, so a gradual steepening at these epochs suggests a steady and gradual physical mechanism driving the evolution.
\citet{Donnan25} find tentative evidence that at fixed stellar mass, dust attenuation increases with decreasing redshift, in line with our steepening of $\beta$.
We stress that \Euclid studies are necessary to complement JWST for constraining the bright end of the UV LF at $z\simeq8$--$10$, which will reduce the roughly two magnitudes of scatter in $M^*$ at these redshifts.
\Euclid studies will do this by identifying thousands of $\Muv<-22$ sources in $53 \, \deg^2$ of imaging in the EDFs to depths comparable to UltraVISTA.
However, based on the results of this work, it may be difficult to decontaminate UCDs from LFs determined from \Euclid without some correction factors, due to the lack of ancillary NIR data available.
Indeed, EDF-North is expected to have severe contamination due to a lower area of coverage by \spitzer, and since this field is closer to the Galactic plane (Allen et al. in prep.).

\vspace{-5pt}
\section{Outlook for \Euclid}
\label{sec: Outlook with euclid}

In this section we explore the additional information provided by \Euclid for our sample of $z\simeq7$ LBGs, in terms of photometry and morphology.
We also discuss approaches for high-redshift galaxy selections in the EDFs.

\subsection{SED fitting with \Euclid}
\label{sec: euclid sed fitting}

In Sect. \ref{sec: rest-UV LF} we have shown that adding \Euclid photometry to SED fitting of our UltraVISTA-selected sample eliminates scatter in the LF points due to contamination and loss of genuine high-redshift galaxies. In Sect. \ref{sec: candidate galaxies} we have shown that \Euclid recovers faint galaxies that are removed in the U-only selection.
In this section we explore in further detail the additional information provided by the \Euclid photometry with some example SED fits.

\begin{figure*}
    \centering
    \includegraphics[angle=0, trim=0 0 0 0, clip, width=0.49\linewidth]{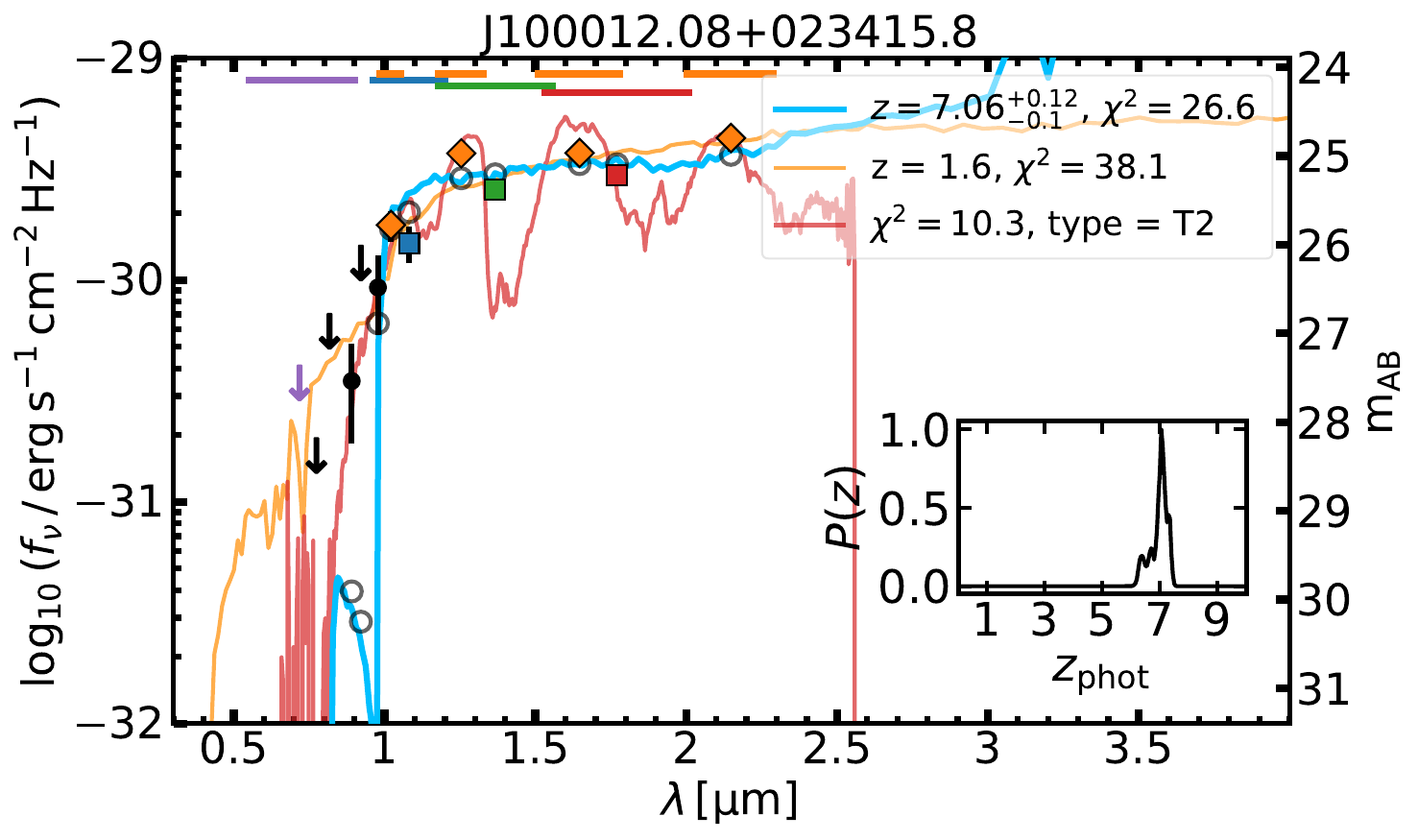}
    \includegraphics[angle=0, trim=0 0 0 0, clip, width=0.49\linewidth]{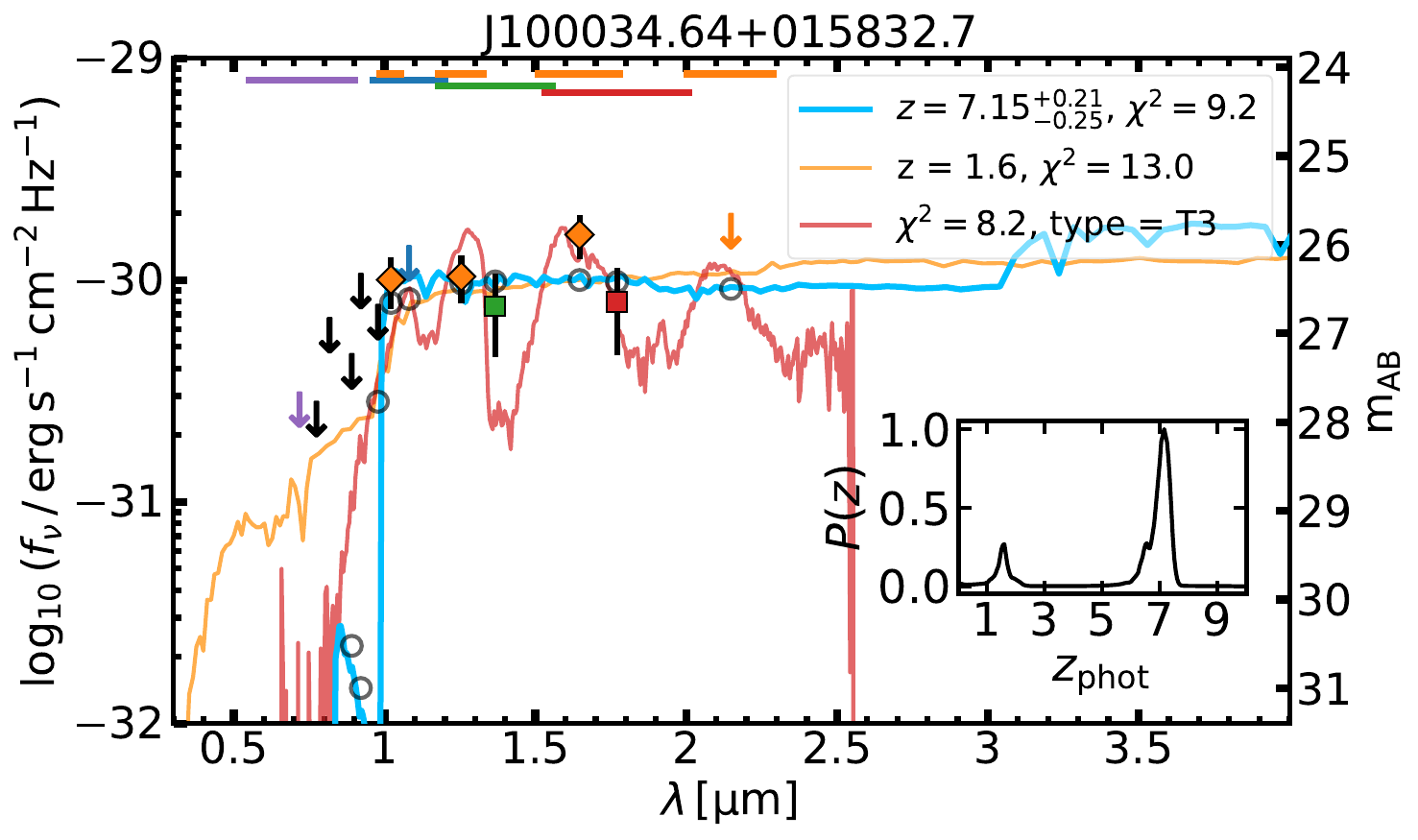}

    \includegraphics[width=0.45\linewidth]{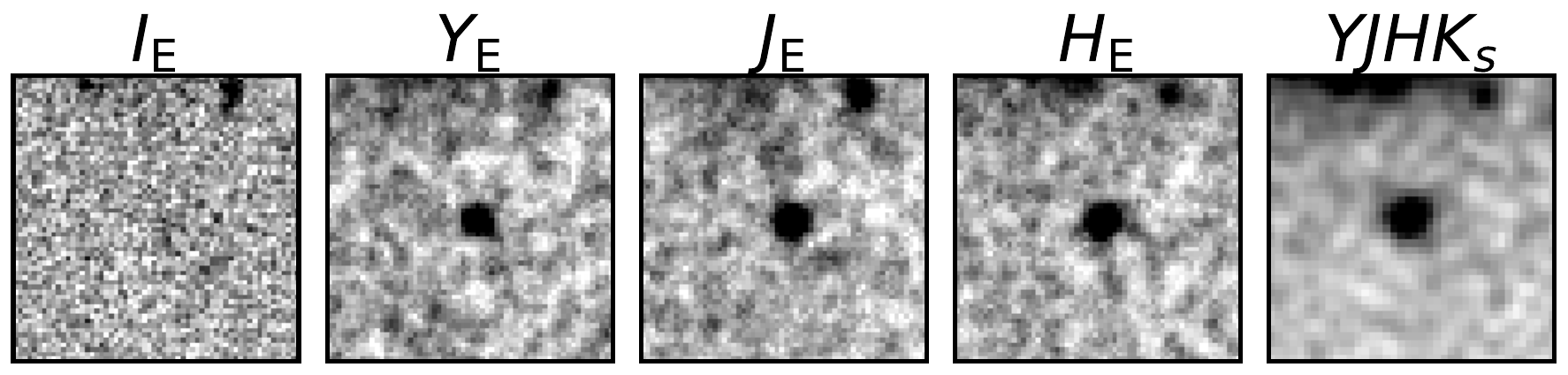}
    \hspace{10pt}
    \includegraphics[width=0.45\linewidth]{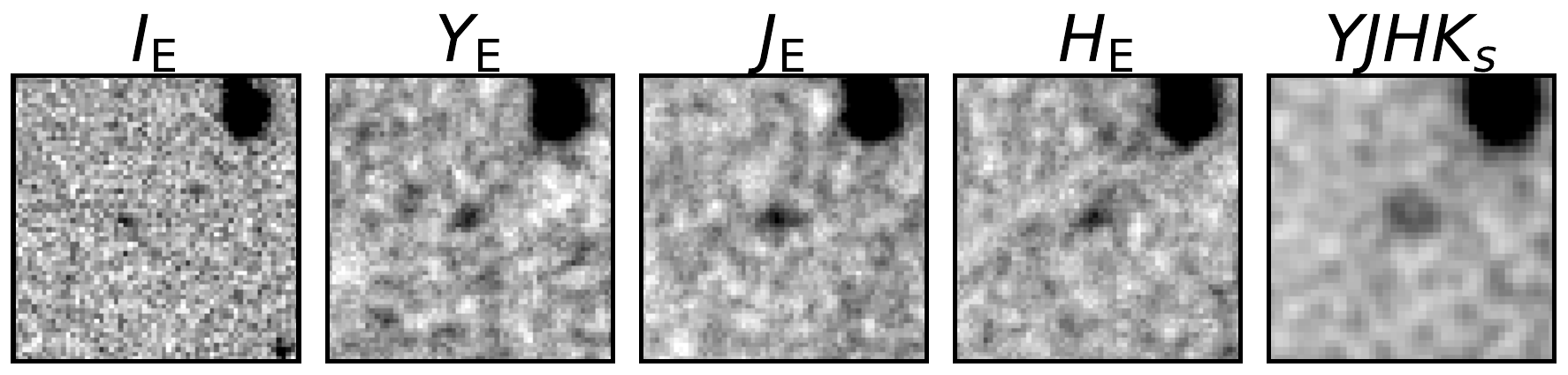}
    \caption{Spectral energy distribution fitting and postage stamp cutouts (in \Euclid and a VISTA $YJHK_s$ stack) of two candidate T-type UCDs.
    \emph{Left}: Brighter UCD with $m_{\rm{AB}}=25$. 
    \emph{Right}: Fainter UCD with $m_{\rm{AB}}=26$. 
    {As in Fig. \ref{fig:UVISTA euclid SED}, the orange diamonds represent the UltraVISTA photometry.
    The coloured squares show the \Euclid photometry, with the \IE non-detection replaced with a $2\,\sigma$ upper limit.
    The remaining HSC photometry is shown by the black points.
    The coloured horizontal lines indicate the \Euclid ad VISTA filter widths.
    The blue, orange, and red curves represent the high-redshift LBG, low-redshift dusty galaxy, and UCD solutions, respectively.
    }
    The postage stamp cutouts are $6\arcsec\times6\arcsec$ and are scaled to saturate at $2\,\sigma$ below and $5\,\sigma$ above the noise level.
    Note the clear molecular absorption features seen by the VISTA+\Euclid NIR photometry at $\lambda>1\,\micron$.
    Also note the PSF-like morphology for the bright source, but a more irregular morphology for the faint source driven by noise.
   }
    \label{fig:euclid BDs}
\end{figure*}

Returning to Fig. \ref{fig:UVISTA euclid SED}, where we show the SED fitting of a candidate LBG, LBG\,10004$+$02015, we see that the \Euclid photometry is powerful for identifying flat UV continuum slopes, and the deep $\IE$ imaging strongly rules out low-redshift galaxy and M-type dwarf solutions, which do not exhibit as strong a break as LBGs.
Although beyond the scope of this work, combined VISTA+\Euclid photometry will be powerful for UV slope measurements. 
In this case, the addition of \Euclid suggests a bluer UV slope than for the SED found with UltraVISTA NIR data alone.
Note that the VISTA filters have shallower depths at longer wavelengths (see Fig. \ref{fig:filters} and Table \ref{tab:Depths}).
The \Euclid NISP filters have uniform depths, so the deep $\HE$ imaging can allow for the discovery of very blue galaxies which would be undetected in VISTA $H$ and $K_s$.

In Fig. \ref{fig:euclid BDs} we show the SED fitting and postage stamp cutouts of two candidate T-type UCDs -- one bright with $m_{\rm{AB}}=25$, and one faint with $m_{\rm{AB}}=26$.
We first note that the brighter UCD has a clear point source morphology, which would allow for its removal from an LBG sample due to being a point source.
Such a removal would require an additional completeness calculation to account for the potential removal of active galactic nuclei and compact star-forming galaxies but would aid in improving the purity of $z=7$ samples.
However, the visual morphology of the fainter source is more ambiguous, with noise spikes beginning to contribute to the shape of the object, giving it a `fuzzier' morphology. 
We discuss the prospect of removing UCDs based on size further in Sect. \ref{sec:UCDsize}.
Now looking to the SEDs for both candidate UCDs, we can see the deep absorption features with \Euclid which are inaccessible with VISTA alone.
The SED fitting for the brighter source also benefits from the pre-existing HSC optical photometry being able to detect the gentler blue slope relative to a Lyman break.
Prior to \Euclid, this gentle blue slope, combined with deviations from flat NIR photometry, was the main discriminant used to remove UCDs in ground-based data. 
However, for the fainter UCD, the blue slope is not visible. As a result, it is more challenging to extend LBG searches down to the signal-to-noise limit without \Euclid.
This was a challenge encountered in \citetalias{Varadaraj23} in the ECDF-S field due to a lack of deep optical imaging, where we selected objects in $Y+J$ at a brighter $8\,\sigma$ significance to account for this (compared to $5\,\sigma$ in this work), resulting in only six luminous galaxy candidates.
VISTA imaging is available in the EAFs and in EDF Fornax as part of VIDEO imaging in ECDF-S.
This overlap will be powerful for providing additional photometry on $z\geq7$ sources found in ground-based imaging \citep[e.g.][]{Stefanon19, Bowler20, Donnan23}, as well as identifying new ones.
A clear next step is to repeat this experiment with VISTA $J$- and $H$-band dropouts, and then adding in the \Euclid photometry.

A natural experiment to conduct with the U+E sample is to test what fraction of these sources are recovered as high-redshift LBGs when only using \Euclid photometry, i.e. when we remove the VISTA photometry (hereafter called E-only). 
\citet{vanMierlo-EP21} used real UltraVISTA galaxies at $z=1$--$8$ and simulated \Euclid photometry to predict the recoverability of UltraVISTA galaxies with \Euclid.
They found that $91\,\%$ of bright galaxies at $z>6$ are recoverable, with a contamination rate of $20$--$40\,\%$ depending on magnitude.
We repeated the SED fitting on the 108 galaxies in the U+E sample which are also present in the U-only sample.
We recover $96\,\%$ of the sources with an E-only photometric redshift $z>6$, and we recover $87\,\%$ with $|z_{\rm E-only}-z_{\rm U+E}|\, / \, (1+z_{\rm U+E}) < 0.15$, consistent with the findings of \citet{vanMierlo-EP21}.
The $z_{\rm E-only}$ span from $z=6$--$8.3$ since the wide \Euclid filters are unable to precisely constrain the position of the Lyman break.
An analysis of the contamination rate would require a reselection based on the \Euclid imaging, which is beyond the scope of this work.

\begin{figure}
    \centering
    \includegraphics[width=\linewidth]{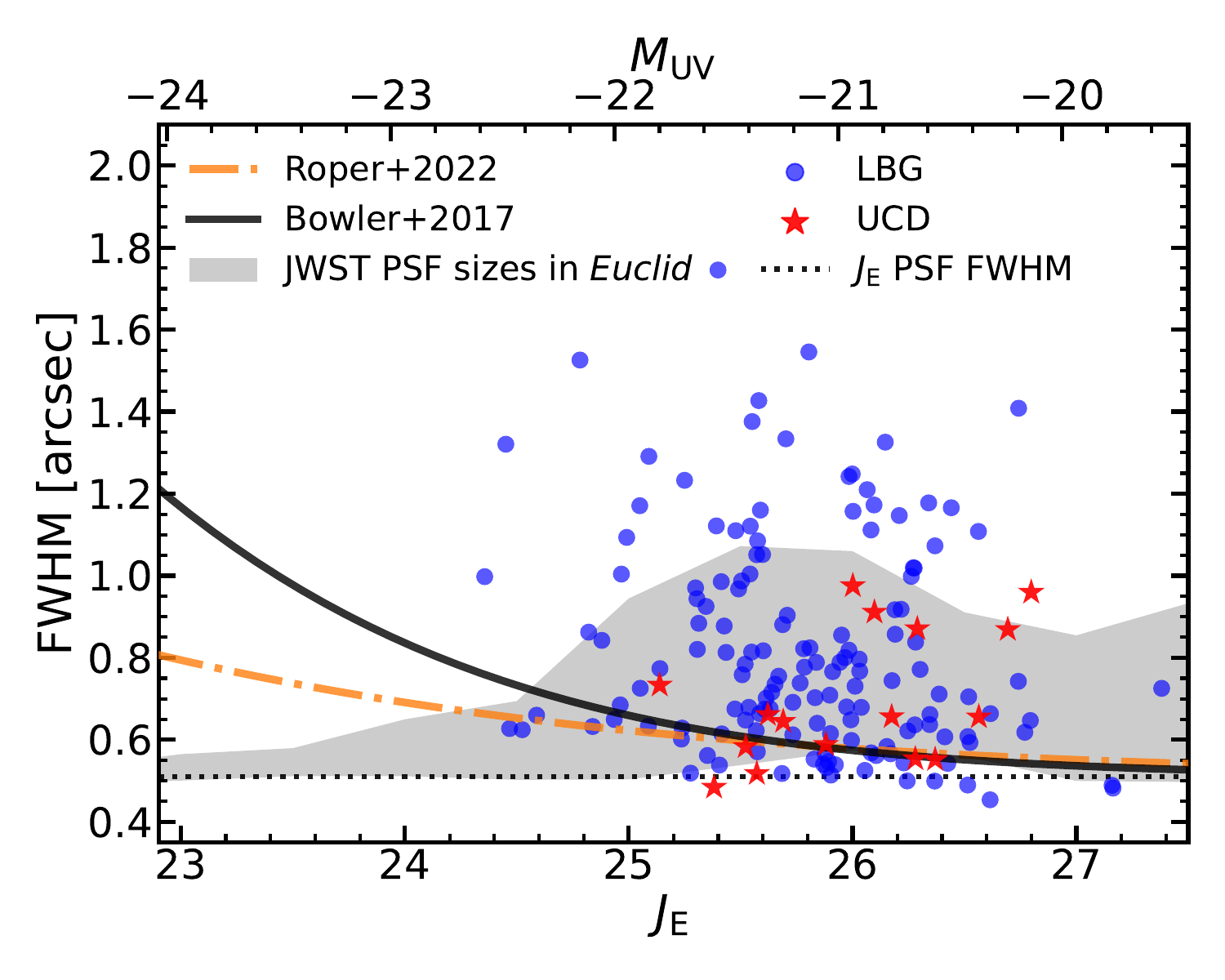}
    \caption{\Euclid \JE FWHMs for sources identified as LBGs (blue dots) and UCDs (red stars) using SED fitting on HSC+VISTA+\Euclid photometry. 
    Also shown is the size-luminosity relation from HST imaging of a bright ground-selected sample \citepalias{Bowler17}, and the prediction from the FLARES simulation \citep{Roper22}.
    The dotted grey line is the {median} \Euclid PSF FWHM.
    The grey shaded region shows the FWHMs of PSFs selected from JWST F444W imaging when they are observed in \Euclid \JE.
    }
    \label{fig:size-mag}
    \vspace{-5pt}
\end{figure}

\subsection{Removal of ultra-cool dwarfs as point sources with \Euclid}
\label{sec:UCDsize}

It may be expected that space-based imaging with \Euclid will allow for the removal of UCDs from LBG samples as unresolved sources.
The first results from the Early Release Observations (EROs) were based on data that had a pixel scale of $\ang{;;0.3}\,\text{pix}^{-1}$, which is larger than ground-based surveys such as VIDEO \citep[$\ang{;;0.2}\,\text{pix}^{-1}$,][]{Jarvis2013}.
This allowed for the identification of \IE dropout sources as $z>6$ LBG candidates, but objects were generally undersampled \citep{EROLensData, EROLensVISDropouts}.
\citet{EROLensVISDropouts} injected PSF, clumpy, and disc models into the imaging and found that galaxies and PSFs are indistinguishable at this pixel scale, so no attempt was made to separate bright sources based on morphology.

We used \texttt{SExtractor} FWHMs to conduct a first-order analysis of LBG and UCD sizes in the \Euclid performance verification imaging.
This is possible thanks to the higher resolution of the \Euclid imaging ($\ang{;;0.1}\,\text{pix}^{-1}$ compared to $\ang{;;0.15}\,\text{pix}^{-1}$ for UltraVISTA, with a PSF FWHM of $\ang{;;0.5}$ in NISP compared to $\ang{;;0.85}$ in VISTA $Y$).
We first investigated the sizes of PSFs in the \Euclid imaging. 
We took point sources from the JWST COSMOS-Web survey \citep[][]{CWEB},\footnote{The data described here may be obtained from \url{https://doi.org/10.17909/g3fb-ez59}.} selected from the F444W FWHM-MAG\_AUTO diagram using a catalogue produced with \texttt{SExtractor}.
We then crossmatch these point sources with \Euclid.
F444W is thus used to provide the `ground-truth' since the objects are unambiguously PSFs.
Since F444W is much deeper than \JE ($5\,\sigma$ depth of 26.9 following the methodology in Sect. \ref{sec:depths}), we are able to select these `ground-truth' PSFs far below the limiting magnitude of the \Euclid imaging.
In Fig. \ref{fig:size-mag}, the grey region show the \texttt{SExtractor} FWHM in \Euclid \JE imaging of these JWST PSFs as a function of their \JE magnitude.
This region is determined as the interquartile range around the median in magnitude bins of width 0.5 mag.
For very bright ($\JE \lesssim 24$) sources, PSFs in \Euclid are consistent with the FWHM measured in Sect. \ref{sec:psf homogenisation}, as expected.
However, as we approach fainter magnitudes, the mean and standard deviation both begin to increase, before turning over near the limiting magnitude of the \JE imaging.
This behaviour is expected as noise begins to boost the FWHM measurements of faint sources.
Additionally, the turnover is caused by biases in the selection of faint PSFs in the \Euclid imaging, since only those with an associated large positive noise will be detected.
\citet{Bowler14} conduct a similar investigation by injecting PSFs into UltraVISTA and UDS imaging, and find similar behaviour.
In Fig. \ref{fig:size-mag} we also show the FWHM-\JE magnitude distribution of LBGs and UCDs identified through SED fitting with combined HSC+VISTA+\Euclid photometry for the U+E sample.
The FWHMs of the UCDs are consistent with the shaded region.
Additionally, none of the UCDs identified with SED fitting exceed a FWHM of $1\arcsec$.
Whilst a large fraction of the LBG sample has FWHMs above $1\arcsec$, the majority are consistent with the scatter in sizes measured for point sources.
{Some LBGs have FWHMs consistent with that of the PSF. 
These could be compact sources with high star-formation surface densities \citep{Morishita24, harikane24}.}
We also show the size-luminosity relations determined by \citetalias{Bowler17} from HST imaging of a ground-based sample, and the prediction from the FLARES simulation \citep{Roper22}.
The size-luminosity relation can be written as
\begin{equation}
    R_{\rm{e}} = R_0 \left( \frac{L}{L_0} \right) ^ \gamma\;,
\end{equation}
where $L_0$ is the characteristic luminosity corresponding to an absolute magnitude $M = -21$, $R_0$ is the size at $L_0$ and $\gamma$ is the slope of the relation.
We convert their relations for the effective radius in terms of $R_{\rm{e}}$ in kpc to an FWHM by assuming the galaxies have a S\'ersic profile with index $n=1$, corresponding to an exponential disc.
Results from JWST show that galaxies at $z\simeq7$ are well-fit by such profiles \citep[e.g.][]{Kartaltepe23, Ormerod24, Westcott24}.
We then convolved the literature size-luminosity relations with the \Euclid PSF in order to simulate what would be seen with \Euclid in terms of FWHM.
For this, we followed \citet{Oesch10} and added the measured PSF size in quadrature.
We note that the UV size-luminosity relation at $z=7$ has been measured by HST and JWST \citep[e.g.][respectively]{Shibuya15, Yang22}.
However, due to the lack of area, these studies do not place strong constraints on the very bright end.
\citet{Yang22} base their results on lensed galaxies to place constraints on the ultra-faint end of this relation, but only have one source with $\Muv < -22$ at $z=6$--$7$.
\citet{Shibuya15} is based on a brighter sample, but only have two sources with $\Muv < -22$ at $z\simeq7$ (see their figure 9).
The relation determined by \citetalias{Bowler17} is currently the only relation at $z=7$ which samples galaxies at $\Muv<-22$ (with HST follow-up of ground-selected LBGs), once again highlighting the necessity for ground-selected samples prior to \Euclid.
They find a size-luminosity relation with slope $\gamma=0.50$.
This provides a positive forecast for the removal of UCDs as PSFs as point sources, since it suggests that galaxies will grow rapidly in size at $\Muv < -22$.
As shown in Fig. \ref{fig:size-mag}, the FWHMs of JWST PSFs, as measured from \Euclid \JE imaging, deviates from the \citetalias{Bowler17} size-luminosity relation at $\JE \lesssim 24.5$.
The prediction from the FLARES simulation \citep{Roper22} also shows a similar deviation in the size-magnitude relation from the point source sizes at bright magnitudes.
Comparison to these studies suggests that within the EDFs at these magnitudes, where VISTA imaging is not available to remove UCDs with SED fitting, it will be possible to remove UCDs as unresolved sources in bright samples (noting expected $5\,\sigma$ depths in the EDFs of around $26$).

The key VISTA fields (COSMOS for UltraVISTA, XMM-LSS and ECDF-S for VIDEO) cover around $ 10 \, \rm{deg}^2$ and will overlap with both the \Euclid Deep/Auxiliary Fields, also expected to reach depths of $26$.
This offers the unique opportunity to identify LBGs and UCDs with SED fitting and determine a size-luminosity relation for the most luminous $z\simeq7$ LBGs.
Additionally, at $z\simeq6$ and below, the UV continuum will enter \IE, which has a much smaller PSF FWHM than NISP.
\Euclid will thus provide some of first resolved measurements of ultra-luminous sources in the middle of, towards the end of, and after reionisation.

\subsection{Lessons learned for the Euclid Deep Fields}

Our results have shown that additional ground-based NIR photometry is powerful for removing L- and T-type UCDs via SED fitting, which \Euclid alone is unable to do due to its broad filter response curves (see Fig. \ref{fig:filters}).
We have also shown that in the range $25 < J < 27$, UCDs cannot be cleanly removed from an LBG sample as point sources because noise spikes contribute to the morphology of faint point sources, and the size-luminosity relation of LBGs corresponds to a small FWHM.
This is critical for the EDFs because they will eventually reach $5\,\sigma$ depths of around $26$, allowing for the discovery of many sources in this magnitude range.
At $J \lesssim 24.5$, this will be much easier since UCDs will have a clear PSF morphology and LBGs are expected to be larger \citepalias{Bowler17}.
At these redshifts, however, little red dots \citep[][]{Matthee24} may also exhibit a clear PSF morphology.
These are sources with a distinct V-shaped SED, compact morphology and a large subset of these have broad $\rm{H}\alpha$ lines, indicating that they may harbour supermassive black holes \citep[e.g.][]{Harikane23_agn, Greene24, Kocevski25}.
Additionally, at these magnitudes, one may expect to also find quasars that will have a PSF morphology. 
It may be possible to distinguish these from UCDs based on the strength of the $\IE-\YE$ break at $z>6$ \citep{Barnett-EP5}, although this is expected to be quite difficult \citep{Banados25}.
Full SED fitting and Bayesian classification methods are likely required to identify quasars and little red dots, along with spectroscopic follow-up.
Despite this, bright magnitudes ($J \lesssim 24.5$) are the strength of the EDFs, where there will be enough area to discover many of these rare, ultra-luminous LBGs.
However, the EDFs will not have ancillary NIR data to help remove UCDs via SED fitting -- although Rubin/LSST will provide imaging in EDF-Fornax and EDF-South in the optical $ugrizy$ filters in the southern sky, allowing for the UCD blue-end slope to be detected (see Fig. \ref{fig:euclid BDs}).
This means relatively faint samples at $J>25$ may still suffer from contamination by UCDs.
This magnitude range, however, is the strength of the key VISTA fields, namely XMM-LSS, ECDF-S, and COSMOS, where VISTA surveys such as VIDEO and UltraVISTA have/will have overlap with \Euclid imaging.
We suggest that $z\gtrsim7$ LBG searches in the EDFs should focus on bright selections at $J<25$ to mitigate contamination, and supplement this with selections in the VISTA fields where SED fitting can be used to construct clean samples.
There is the potential to use ancillary \spitzer to reduce contamination rates. However, at $J>25$, the available \spitzer imaging would be too shallow, as found for the faint end of the sample in this work, further motivating a selection at $J<25$.

\begin{figure}
    \centering
    \includegraphics[width=\linewidth]{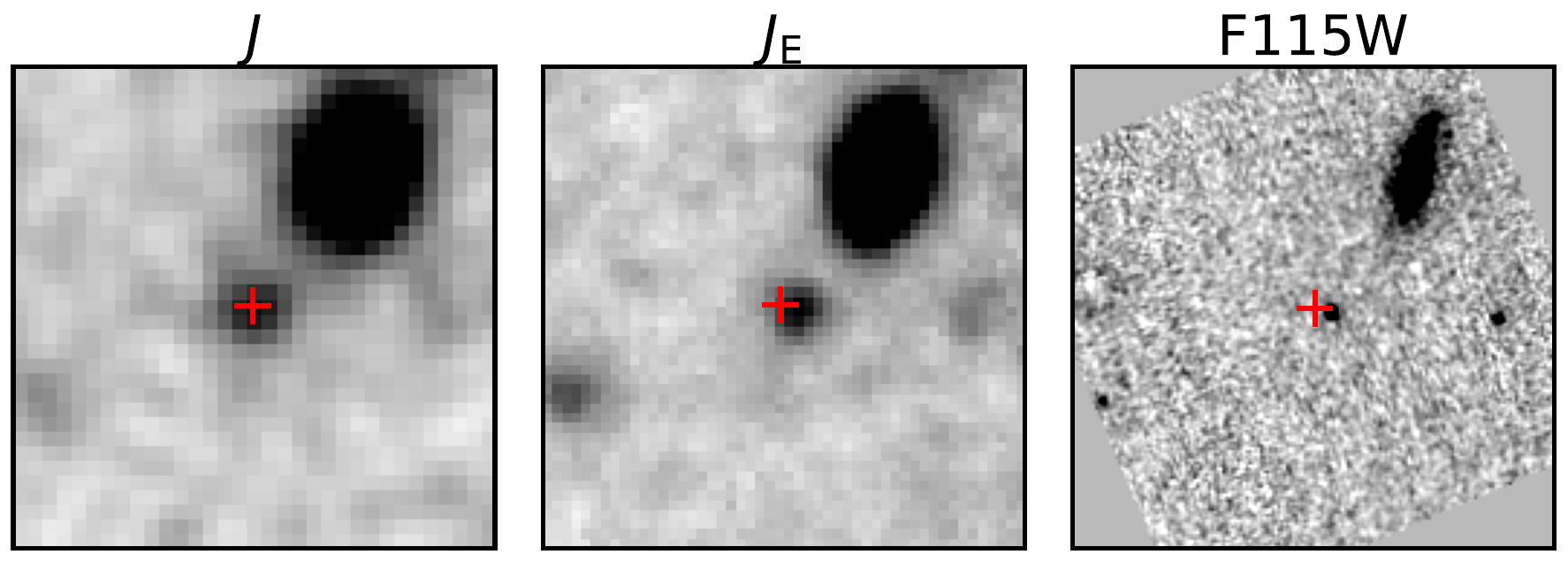}
    \caption{Proper motion of a candidate UCD. The postage stamp cutouts show the source in VISTA $J$, \JE, and F115W from JWST.
    The cutouts are $5\arcsec\times5\arcsec$ and saturate at $2\,\sigma$ and $5\,\sigma$ below and above the noise level. The red crosshair is placed at the centroid of the source as determined from the VISTA image to highlight the proper motion when observed with \Euclid and JWST.}
    \label{fig:proper motion}
\end{figure}

Not only is SED fitting powerful in these fields, but since \Euclid imaging has been taken over a decade after the beginning of VIDEO and UltraVISTA, some of the brightest candidate UCDs found in our selection show small amounts of proper motion.
In Fig. \ref{fig:proper motion} we show an example, with the offset relative to VISTA being much clearer when compared to imaging from JWST COSMOS-Web.
We note this source is removed from the sample based on SED fitting.
However, sources {with} large proper motion are removed as part of our selection, since they {will have} a low or non-detection in \Euclid, resulting in a poor SED fit. 
Additionally, we do not use proper motion as a screening step since this is complicated by the appearance of faint stars in \Euclid, as discussed in Sect. \ref{sec:UCDsize}.
Removal of UCDs by their proper motion will thus be more viable in the EDFs.

\subsection{Lyman$-\alpha$ emitters with pseudo-narrowbands}

\begin{figure}
    \centering
    \includegraphics[width=\linewidth]{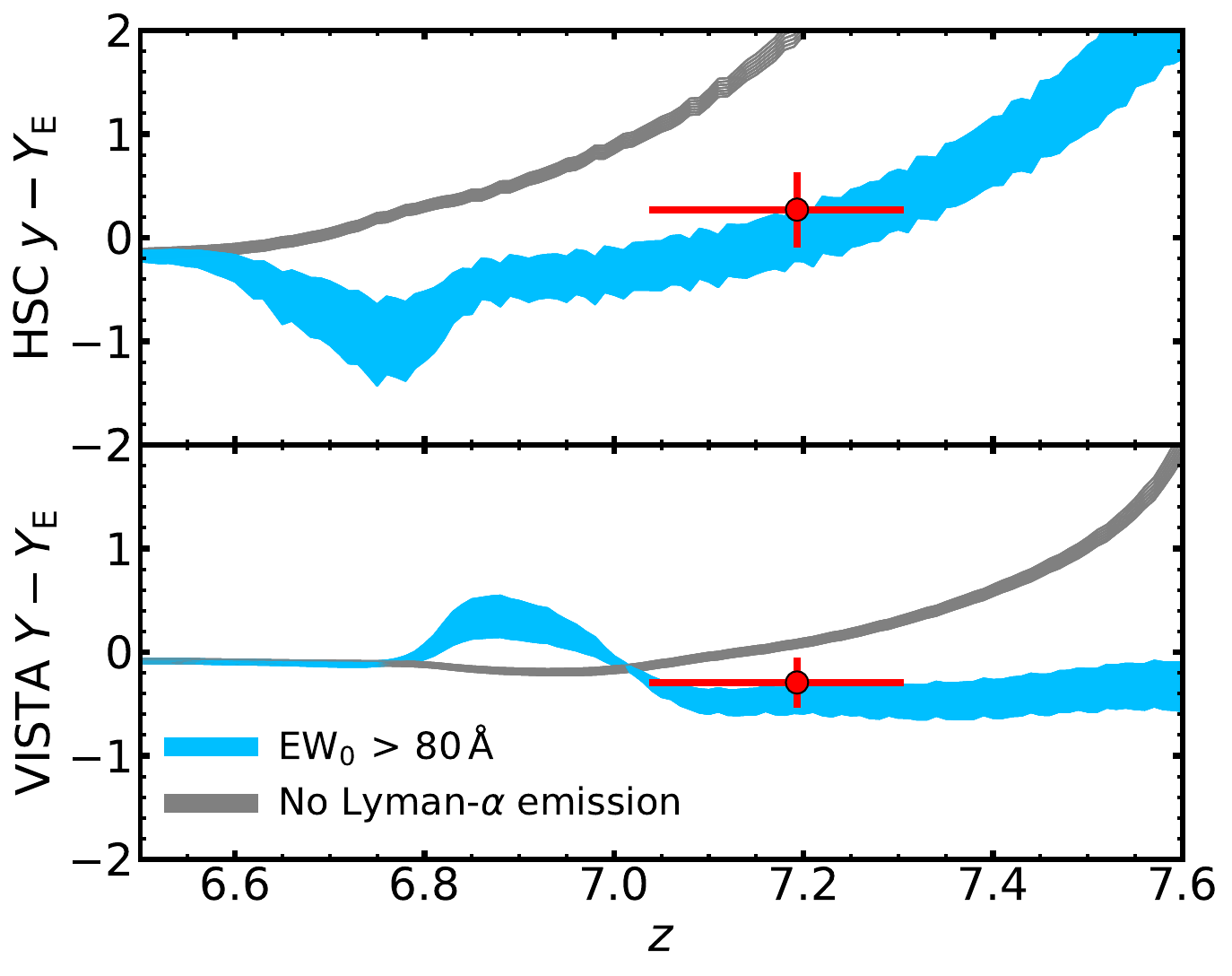}
    \caption{Expected $\mathrm{HSC}\ y - \YE$ (\emph{top}) and $\mathrm{VISTA} \ Y - \YE$ colours (\emph{bottom}) of LBGs (grey) and strong ($\rm{EW}_0 > 80\,\AA$) LAEs (blue) as a function of redshift, generated using mock \texttt{BAGPIPES} galaxies. 
    The red point shows the position of our LAE candidate (see Fig. \ref{fig:LAE}) in this colour-redshift space.
    }
    \label{fig:LAE colour simulation}
\end{figure}

\begin{figure}
    \centering
    \includegraphics[width=1.\linewidth]{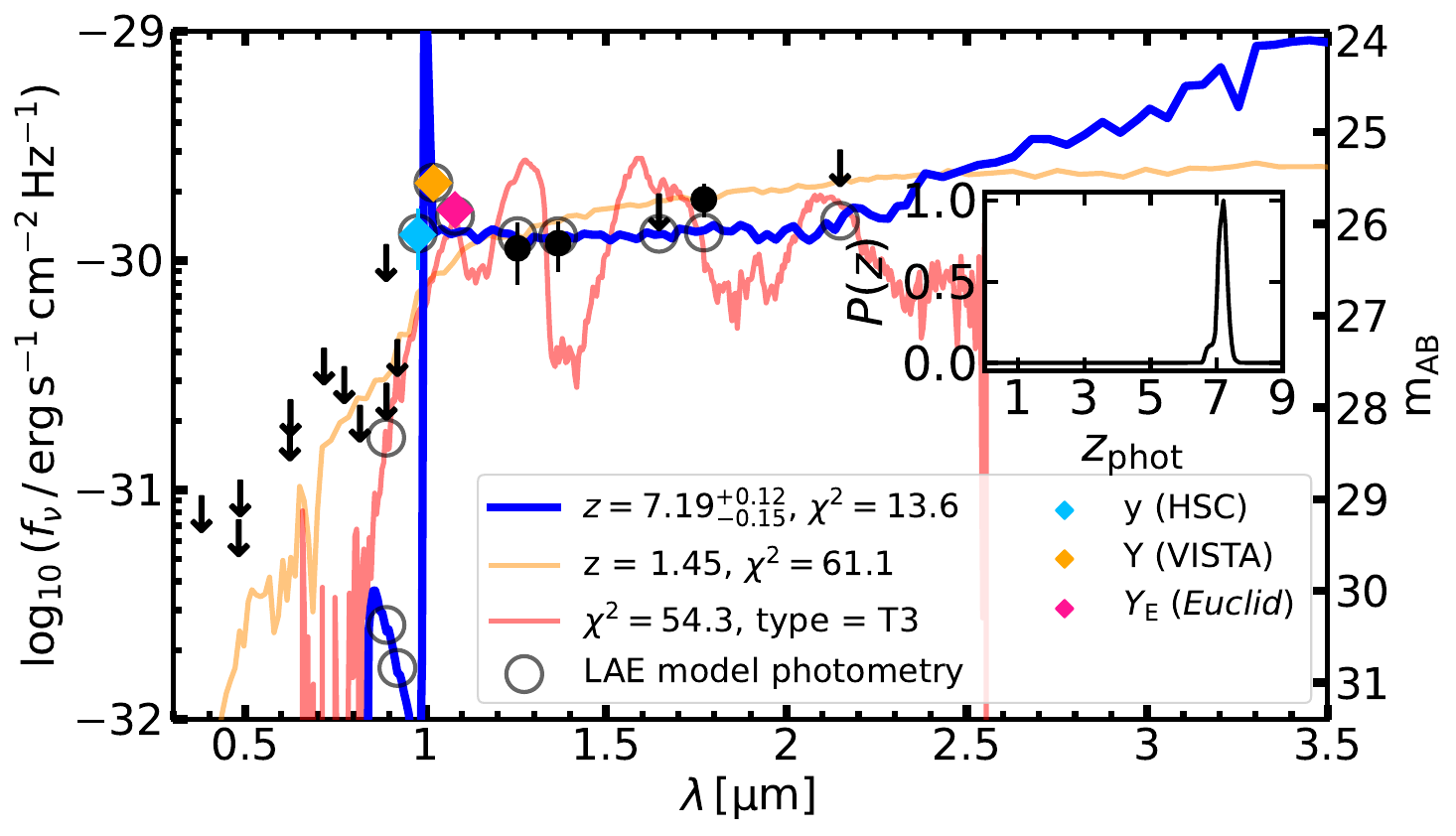}
    \includegraphics[width=1.\linewidth]{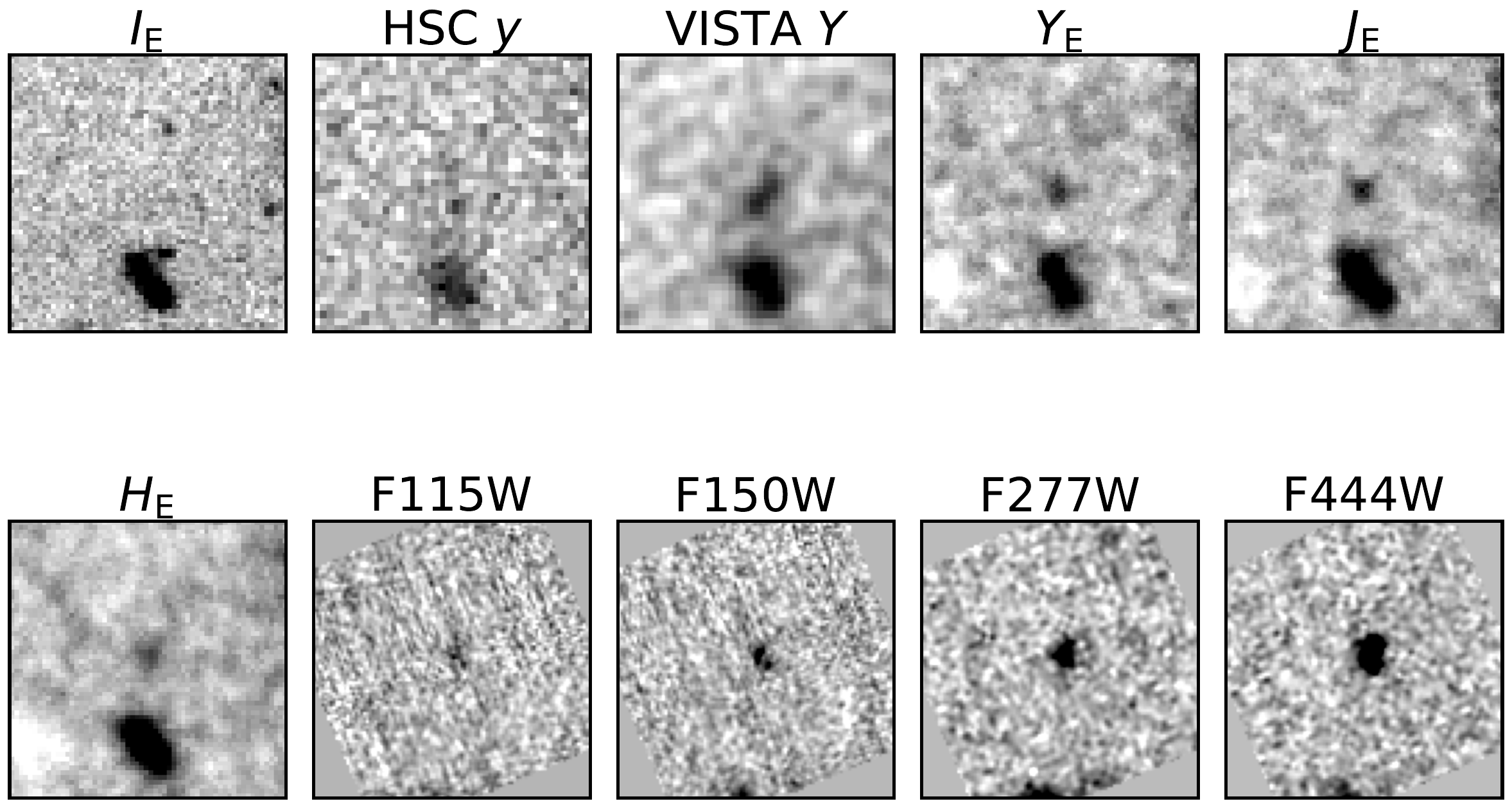}
    \caption{Spectral energy distribution fitting (\emph{top}) and postage stamp cutouts (\emph{bottom}) of our LAE candidate, EUCL\,J100028.39$+$021508.0.
    The SED fitting plot is similar to Fig. \ref{fig:UVISTA euclid SED}, but here we colour the HSC $y$, VISTA $Y$ and \Euclid $\YE$ filters differently to highlight them.
    The ground-based and \Euclid stamps are $6\arcsec\times6\arcsec$ and the JWST stamps are $3\arcsec\times3\arcsec$. The stamps saturate at $2\,\sigma$ below and $5\,\sigma$ above noise level.}
    \label{fig:LAE}
\end{figure}

\Euclid not only provides strong colour information on L- and T-type dwarfs when combined with VISTA, but can also reveal the nature of LBGs with strong emission lines.
The identification of emission lines with photometry has been done before using \textit{Spitzer}/IRAC colours to identify strong \textsc{[Oiii]}+H$\beta$ emitters \citep[e.g.][]{Smit15, Roberts-Borsani16} which have then been found to be LAEs through spectroscopic follow-up \citep[e.g.][]{Oesch15}.
Ground-based studies have used narrowbands to identify LAEs in narrow redshift ranges \citep[e.g.][]{Banados13, Endsley21nebular, Umeda24, Lambert24}, and JWST has used medium- and narrowbands to search for flux excess relative to the broadband photometry \citep[e.g. JELS,][]{Duncan24, Pirie24}.
The overlap of the HSC $y$, VISTA $Y$ and \Euclid $\YE$ filters around $\lambda = 1 \, \micron$ (see Fig. \ref{fig:filters}) provides a unique opportunity to identify strong Lyman-$\alpha$ emitters since the filters will behave differently as the line is redshifted through the filters relative to a normal Lyman-break.
We use \texttt{BAGPIPES} \citep{BAGPIPES} to generate the synthetic galaxy photometry with $A_V = 0$--$0.6$, allowing ages up to the age of the Universe at a given redshift and subsolar metallicity $Z = 0.2 \, Z_{\odot}$.
Since \texttt{BAGPIPES} does not model Ly$\alpha$ emission, we add the line with equivalent widths $\rm{EW}_0 = 80$--$240\, \AA$. 
This is done in the same manner as in Sect. \ref{sec: SED fitting}, by measuring the continuum level between $\lambda_{\rm{rest}}=1250$--$1300\, \AA$.
In Fig. \ref{fig:LAE colour simulation}, we show the $\mathrm{HSC} \ y - \YE$ and $\mathrm{VISTA} \ Y - \YE$ colours of LBGs and LAEs with rest-frame equivalent width $\rm{EW}_0 > 80\,\AA$.
A strong colour difference can be seen between the LBGs and LAEs at $z>7$.
We find one source at $z=7.19$ (EUCL\,J100028.39$+$021508.0) which lies within the LAE colour region.
The SED fitting for this galaxy, along with postage stamp cutouts, is shown in Fig. \ref{fig:LAE}.
A clear excess in flux is seen in VISTA $Y$ relative to HSC $y$ and $\YE$, boosted by a potential strong Ly$\alpha$ emission line.
The morphology in VISTA $Y$ also differs from $\YE$ with an additional clump to the north-east.
Physical offsets between the UV continuum and Ly$\alpha$ emission are often seen in LAEs targeted by MUSE \citep[e.g.][]{Claeyssens22}, thought to be caused by star-forming substructures, merging galaxies or scattering effects inside the circumgalactic medium.
Follow-up imaging with the F090W NIRCam filter would confirm the Lyman-$\alpha$ morphology.
This source also overlaps with JWST COSMOS-Web, and these stamps are also shown in Fig. \ref{fig:LAE}.
We use \texttt{BAGPIPES} to conduct SED fitting using the additional JWST photometry to determine the physical properties of this galaxy.
We fix the redshift to that found by \texttt{LePhare}, $z=7.19$.
We use a delayed-$\tau$ SFH and allow the time since star formation began to vary between $10 \, \text{Myr}$ and the age of the Universe at this redshift, and allow the characteristic timescale, $\tau$, to vary between $50 \, \text{Myr}$ and $10 \, \text{Gyr}$.
We fix metallicity to $Z = 0.2 \, Z_{\odot}$.
We find that this galaxy has a mass of $\mathrm{log}_{10}(M_*/M_{\odot})=9.81^{+0.13}_{-0.20}$ and a star-formation rate of around $26 \, M_{\odot} \, \si{\year}^{-1}$.
This source was previously identified in the COSMOS2020 Farmer catalogue \citep{COSMOS2020} as a $z=7$ galaxy with Farmer ID 586756.
Additionally, based on a continuum level of $\sim1.3 \times10^{-30} \, \rm{erg} \, \rm{s}^{-1} \, \rm{cm}^{-2} \, \rm{Hz}^{-1}$ and an equivalent width $\rm{EW}_0 = 240 \,\AA$, we estimate a Ly$\alpha$ line flux of $7.7 \times10^{-17} \, \rm{erg} \, \rm{s}^{-1} \, \rm{cm}^{-2} \, \rm{Hz}^{-1}$ and a Ly$\alpha$ luminosity $L_{\rm{Ly}\alpha}=5.5\times10^{43}\,\rm{erg}\,\rm{s}^{-1}$.

This galaxy is an ideal candidate for follow-up observations with JWST and ALMA, since similar sources have shown signs of primordial ISM conditions, Ly$\alpha$ haloes, and complex gas exchange mechanisms between components, hinting at merging activity \citep[e.g. `Himiko' and `COSMOS Redshift 7'; see][]{Ouchi13, Sobral15, Marconcini24, Kiyota25}.
Based on the estimated line flux, this source would also be detectable with ground-based observatories such as Keck \citep{Schenker12}.
Looking forward, the three filters used to identify this source will also be available in the XMM-LSS field as part of the EAFs and Euclid Wide Survey, providing additional opportunities to identify strong LAEs embedded in the epoch of reionisation.
In the Euclid Wide Survey, overlap with the LSST $y$ filter can also be used to identify such sources.
LAEs at lower redshifts can be identified in the EDFs by comparing HSC $r$, $i$, and $z$ to the \Euclid $\IE$ filter, although this will be more difficult due to overlap between only two filters at a time and the large width of $\IE$.
Finally, the EDFs and EAFs will be observed with the blue and red grism ($BG_{\rm E}$ and $RG_{\rm E}$) available with NISP-S, covering $\lambda=0.9$--$1.9\,\micron$ \citep{EuclidSkyNISP, EuclidSkyOverview}.
Current observations with NISP-S in the EAFs are not deep enough to detect Lyman-$\alpha$ emission at $z=7$, since the SIR pipeline only does spectral extraction with $\HE<22.5$. 
However, in future data releases, this threshold is planned to be brought down to $\HE<24$.
$BG_{\rm E}$ will be able to detect Lyman-$\alpha$ emission at $z=6.5$--$10$, providing a method for redshift confirmation of LAE candidates presented in this work, as well as a sample of extreme line emitters over degree scale imaging.

\section{Summary}
\label{sec: conclusions}

We have conducted a search for $z\simeq7$ LBGs in the final data release (DR6) of the UltraVISTA survey covering $1.72 \, \rm{deg}^2$. 
We combined this NIR imaging with deep optical imaging from HSC and infrared imaging from \textit{Spitzer}/IRAC to conduct a full SED fitting analysis using \texttt{LePhare}.
We conducted two different selections: one where \Euclid photometry is included for the SED fitting (the U+E sample) and one without the use of \Euclid (the U-only sample).
The U-only sample consists of 
289 candidate galaxies at $6.5\leq z_{\rm{phot}}\leq7.5$ with $-22.5\leq \Muv \leq -20.2$.
With the improved depth of UltraVISTA DR6, this sample reaches magnitudes faint enough to overlap with the HST-selected samples of \citetalias{Finkelstein15} and \citetalias{Bouwens21}.
The U+E sample consists of 140 galaxies, with 38 not identified in the U-only sample.
\Euclid acts to recover very faint galaxies that lack robust SED constraints in VISTA photometry alone.

We find that \Euclid comprehensively addresses contamination by UCDs and artefacts as well as the unintended removal of genuine galaxy candidates, which are seen as scatter in the UV LF points computed from the U-only sample.
Correspondingly, the UV LF based on the U+E sample shows a smooth decline in number density towards brighter magnitudes and is in excellent agreement with \citetalias{Finkelstein15} as well as previous bright-end studies (\citetalias{Bowler17}; \citetalias{Varadaraj23}; \citealt{harikane24}).
Our DPL fitting reveals that this ground-selected sample probes fainter magnitudes than the LF knee for the first time at $z>6$.
We compared our UV LF at $z\simeq7$ to JWST results at $z>7$.
We found some evidence of a gentle evolution in the bright-end slope, although this is limited by a lack of robust measurements of the bright end at $z>9$, which will require robust samples from current and upcoming degree-scale JWST programmes as well as the EDFs.

We then explored in detail the additional information provided by \Euclid for this UltraVISTA-selected sample.
We showed that whilst combined VISTA+\Euclid photometry is powerful for removing faint UCDs via SED fitting, at the magnitudes probed in this work, UCDs cannot be separated morphologically from LBG samples as point sources.
This is because faint point sources have their FWHMs boosted by positive noise spikes.
At $J\lesssim24.5$, since the size-luminosity relation of galaxies rapidly increases, we forecast that it will be straightforward to remove UCDs based on their morphology.
This is particularly crucial for the EDFs where ancillary NIR data are lacking.

Finally, we also presented an extreme Lyman-$\alpha$ emitter candidate at $z=7.2$ identified via its strong colours in HSC $y-\YE$ and VISTA $Y - \YE$, which differ significantly from the expected colours of a normal LBG.
The differences in morphology between VISTA $Y$ and $\YE$ may indicate a Ly$\alpha$-emitting clump physically offset from the UV continuum.
These three slightly different $y$ filters will be available in XMM-LSS, providing further opportunity for the identification of extreme sources during the epoch of reionisation.
Such sources are prime candidates for follow-up with JWST.

\section*{Data availability}

Table \ref{tab:sample}, which provides the photometry and properties of the U+E sample, along with a similar table for the U-only sample, and $10\arcsec\times10\arcsec$ postage stamp cutouts of the U+E sample in the \Euclid \IE, \YE, \JE, and \HE filters, are available in electronic form at the CDS via anonymous ftp to cdsarc.u-strasbg.fr (130.79.128.5) or via http://cdsweb.u-strasbg.fr/cgi-bin/qcat?J/A+A/. 

\begin{acknowledgements}
RGV acknowledges funding from the Science and Technology
Facilities Council (STFC; grant code ST/W507726/1). 
RAAB acknowledges support from an STFC Ernest Rutherford Fellowship (grant number ST/T003596/1). 
MJJ acknowledges support of the
STFC consolidated grants (ST/S000488/1 and ST/W000903/1) and
from a UKRI Frontiers Research Grant (EP/X026639/1). 
MJJ also
acknowledges support from the Oxford Hintze Centre for Astrophysical Surveys that is funded through generous support from the
Hintze Family Charitable Foundation. 
The Cosmic Dawn Center (DAWN) is funded by the Danish National Research Foundation under grant DNRF140.

\AckEC  

This work is based on observations collected at the European Southern Observatory under ESO
programmes 179.A-2005, 198.A-2003, 1104.A-0643, 110.25A2 and 284.A-5026 and
on data obtained from the ESO Science Archive Facility with DOI https://doi.
org/10.18727/archive/52, and on data products produced by CANDIDE and the
Cambridge Astronomy Survey Unit on behalf of the UltraVISTA consortium.
The Hyper Suprime-Cam (HSC) collaboration includes the astronomical communities of Japan and Taiwan, and Princeton University.
The HSC instrumentation and software were developed by the
National Astronomical Observatory of Japan (NAOJ), the Kavli
Institute for the Physics and Mathematics of the Universe (Kavli
IPMU), the University of Tokyo, the High Energy Accelerator
Research Organization (KEK), the Academia Sinica Institute for
Astronomy and Astrophysics in Taiwan (ASIAA), and Princeton
University. Funding was contributed by the FIRST program from
the Japanese Cabinet Office, the Ministry of Education, Culture,
Sports, Science and Technology (MEXT), the Japan Society for
the Promotion of Science (JSPS), Japan Science and Technology
Agency (JST), the Toray Science Foundation, NAOJ, Kavli IPMU,
KEK, ASIAA, and Princeton University.
This paper is based on data collected at the Subaru Telescope and
retrieved from the HSC data archive system, which is operated by
the Subaru Telescope and Astronomy Data Center (ADC) at NAOJ.
Data analysis was in part carried out with the cooperation of Center
for Computational Astrophysics (CfCA), NAOJ.

This work is based in part on observations made with the NASA/ESA/CSA \textit{James Webb} Space Telescope. The data were obtained from the Mikulski Archive for Space Telescopes at the Space Telescope Science Institute, which is operated by the Association of Universities for Research in Astronomy, Inc., under NASA contract NAS 5-03127 for JWST. These observations are associated with program \#1727.

\end{acknowledgements}

%
%

\bibliography{bibliography} 

\begin{appendix}

\appendix

\section{\Euclid photometry check}
\label{sec: euclid photometry check}

\begin{figure}
    \centering
    \includegraphics[width=\linewidth]{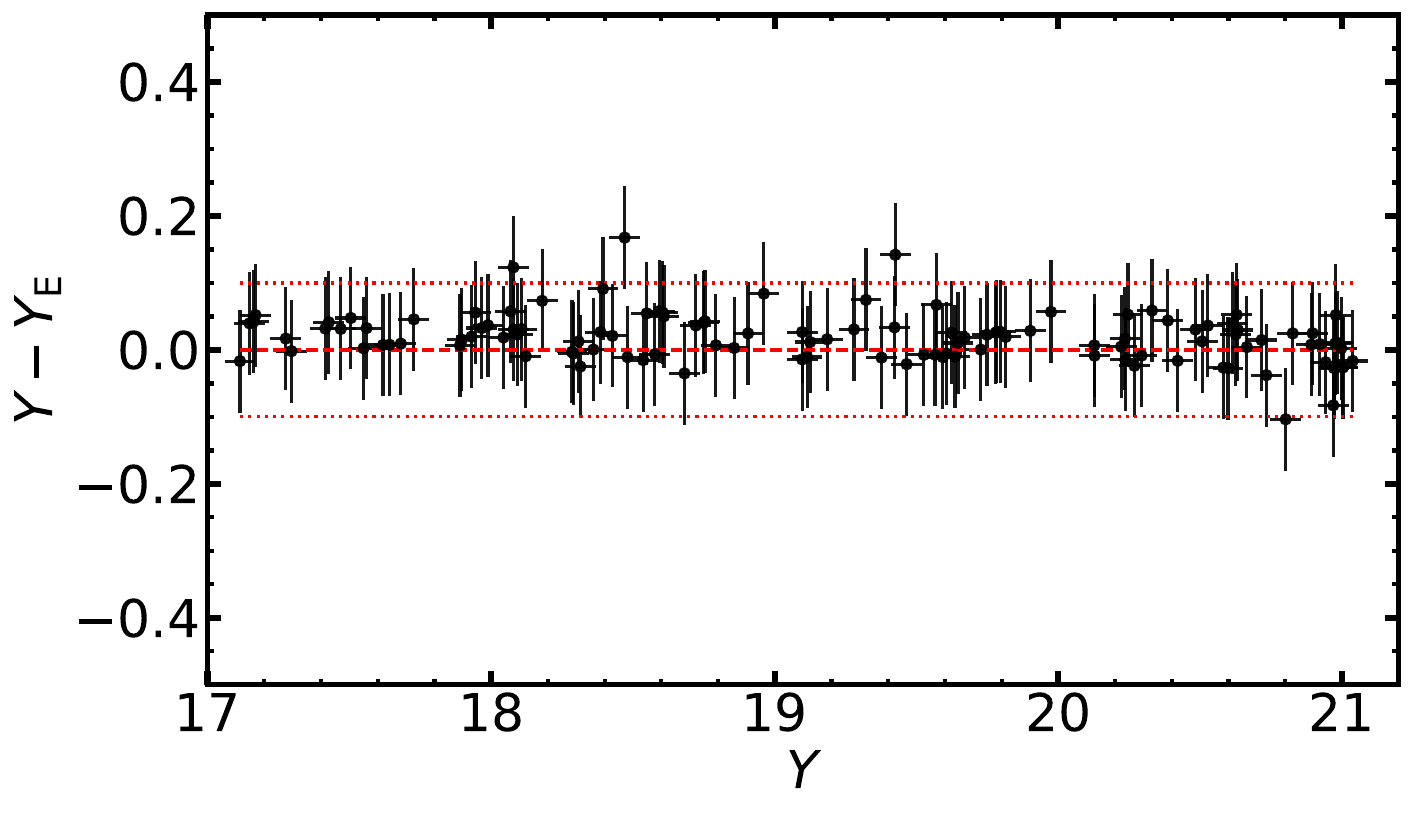}
    \includegraphics[width=\linewidth]{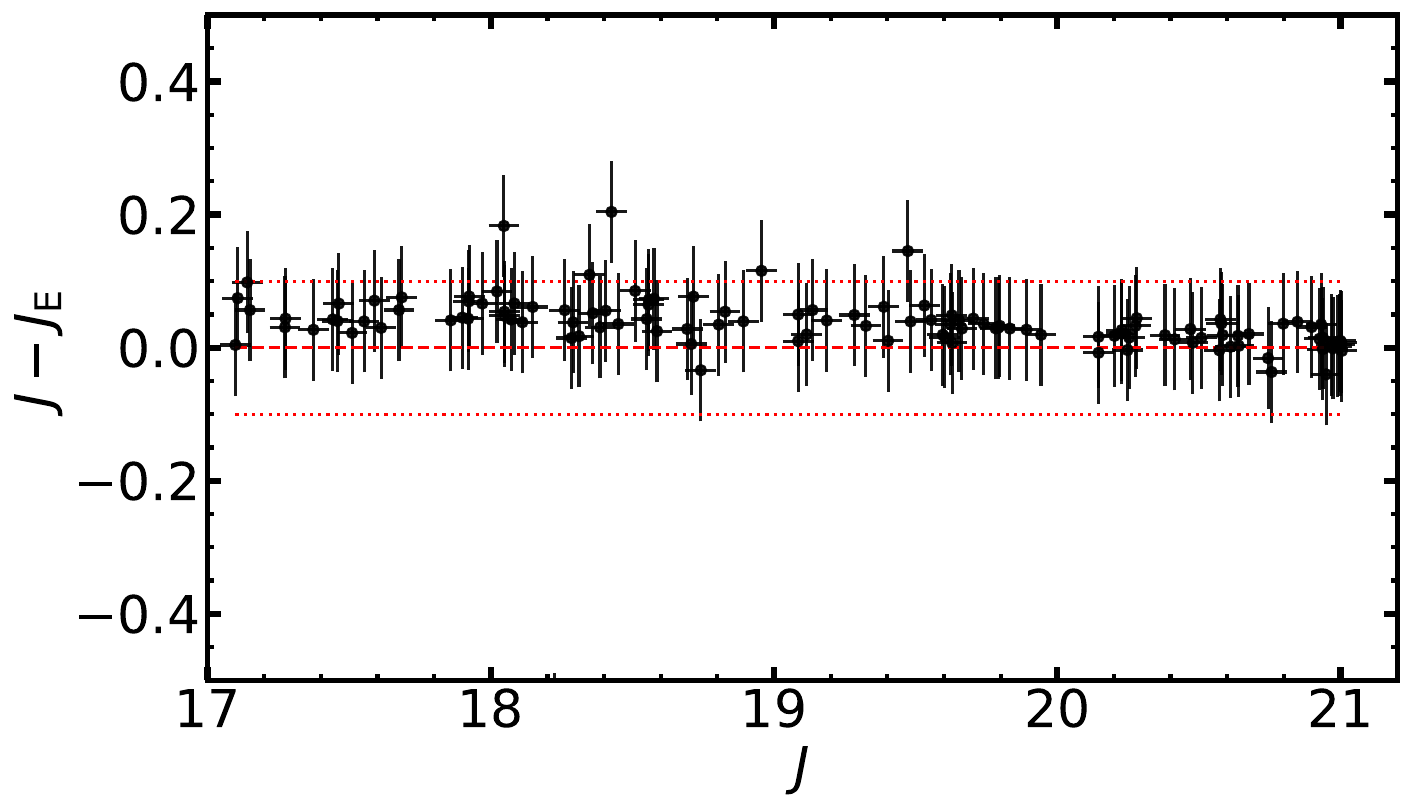}
    \includegraphics[width=\linewidth]{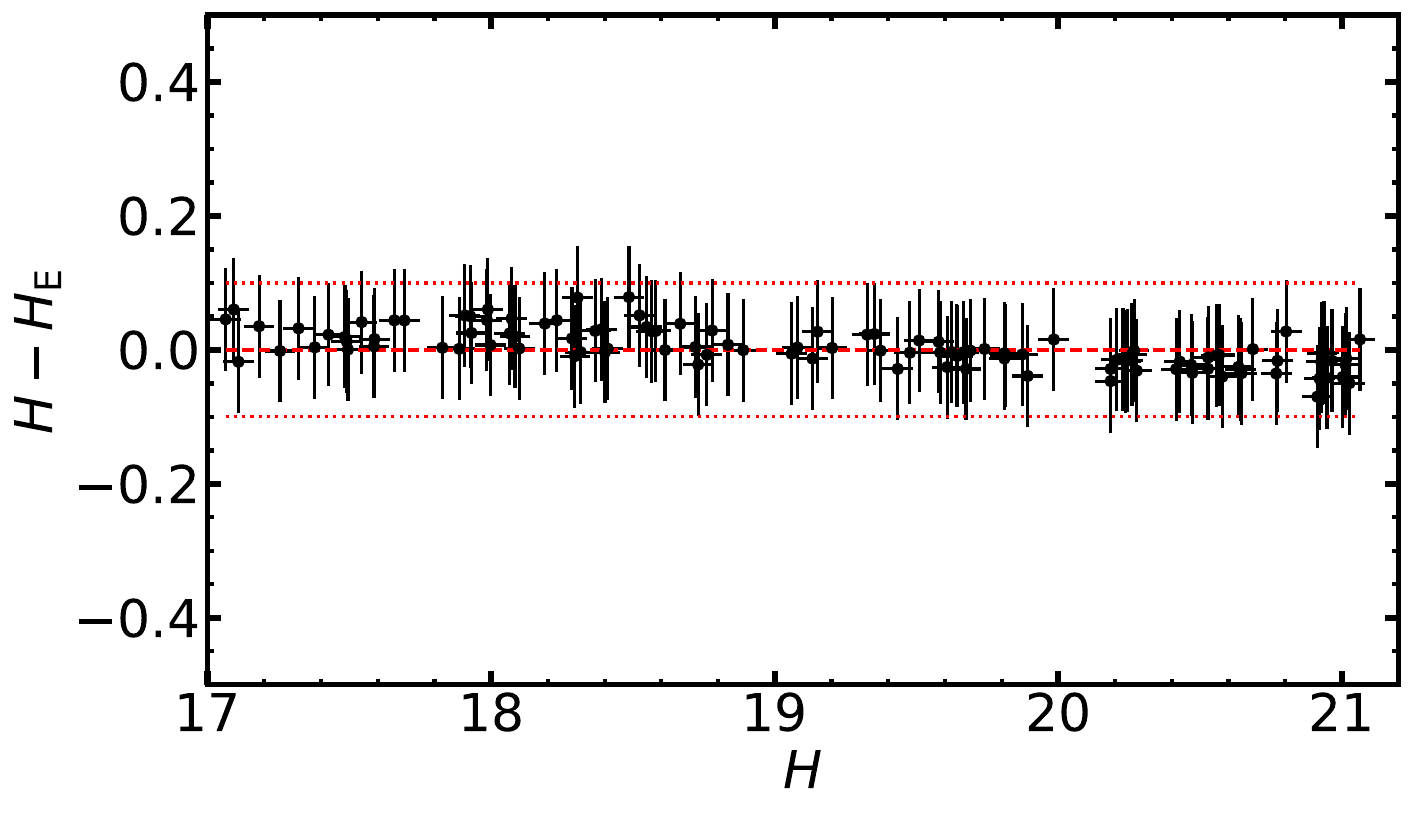}
    \caption{Difference in magnitude between VISTA $Y$ and \Euclid \YE (\emph{top}), VISTA $J$ and \Euclid \JE (\emph{middle}), and VISTA $H$ and \Euclid \HE (\emph{bottom}) and for bright stars with flat colours in VISTA, selected as $| Y - J| < 0.05 \wedge |J - H| < 0.05$. The red dashed line indicates no colour difference, and the red dotted lines indicate a difference in magnitude of 0.1. A minimum uncertainty of 5\% is imposed on the photometry.}
    \label{fig:flat stars}
\end{figure}

In Fig. \ref{fig:flat stars} we present the colours in comparable \Euclid and VISTA filters of stars with flat $YJH$ photometry. 
This provides a check that the \Euclid photometry, after PSF homogenisation, is consistent with VISTA.

\section{Comparison of $\Muv$ and $\zphot$ between the U-only and U+E samples}
\label{sec: Muv and zphot comparison}

\begin{figure}
    \centering
    \includegraphics[width=\linewidth]{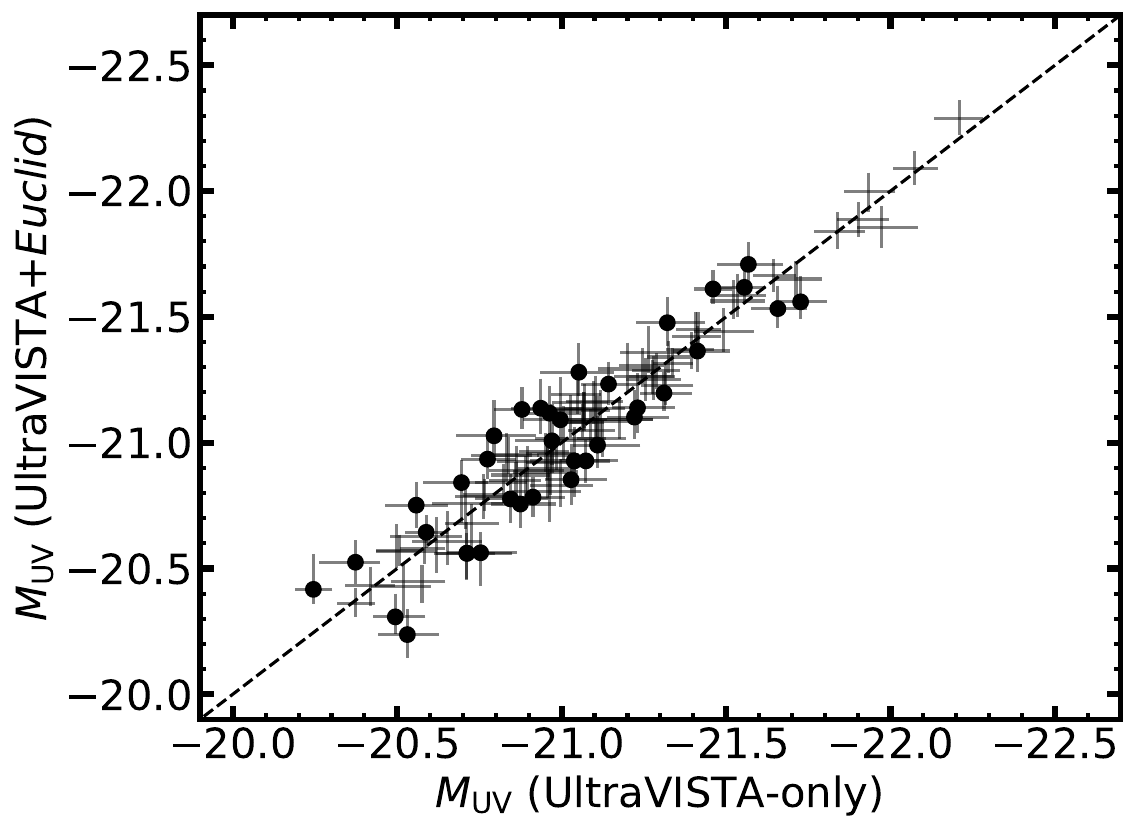}
    \includegraphics[width=\linewidth]{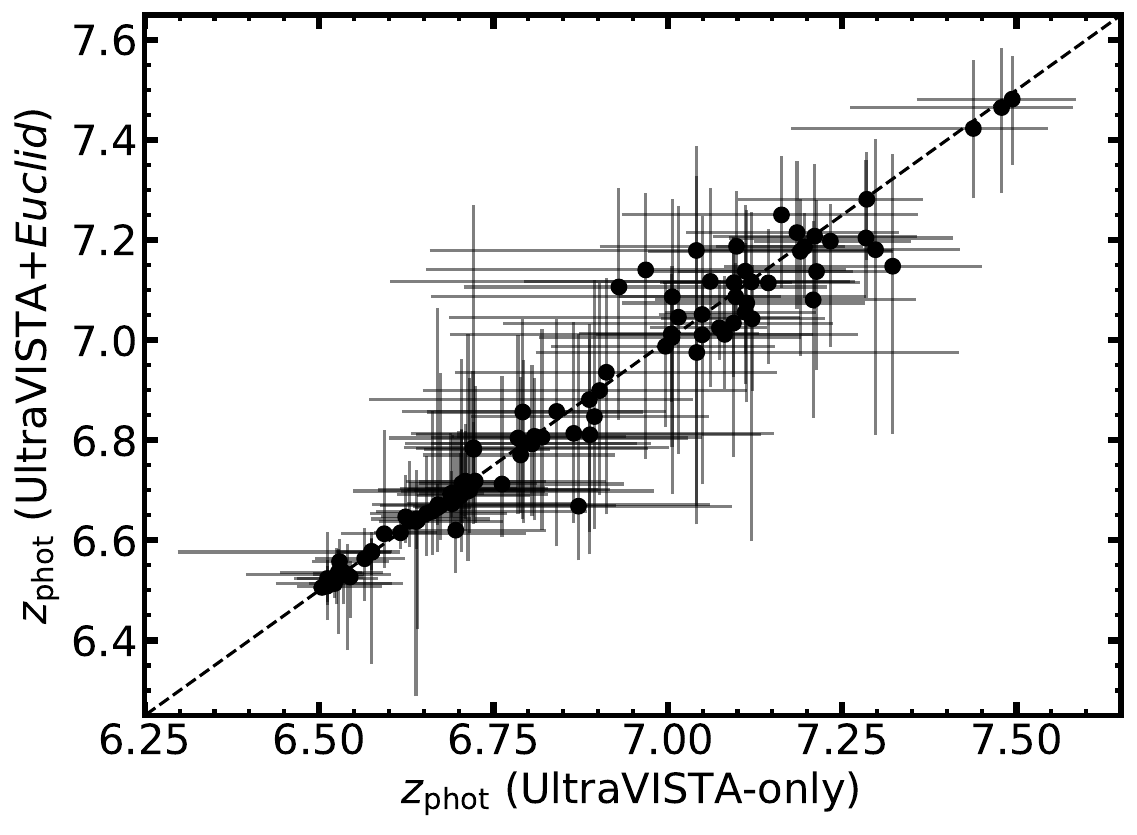}
    \caption{Comparison between the $\Muv$ {(\emph{top}) and $\zphot$ (\emph{bottom})} derived with and without \Euclid photometry. 
    The sample here is based on sources which are present in both the U-only and U+E samples. 
    {In the top panel,} the large black points indicate sources that have their $\Muv$ change enough after the inclusion of \Euclid photometry that they move into a different LF $\Muv$ bin. 
    {In both panels,} the dashed line shows the one-to-one relation.}
    \label{fig:Muv comparison}
\end{figure}

In Fig. \ref{fig:Muv comparison} we show the differences in the determination of the absolute rest-frame UV magnitude $\Muv$ with and without \Euclid photometry.
We also show a similar plot for the photometric redshifts $\zphot$.

\section{SED fits and postage stamp images}
\label{sec: candidate seds and stamps}

In Fig. \ref{fig:seds and stamps} we present the results of the SED fitting and postage stamp images for the 30 most luminous galaxies in the U+E sample, spanning $-22.3<\Muv<-21.3$.
They are ordered from brightest to faintest.

\begin{figure*}

    \includegraphics[width=0.47\linewidth]{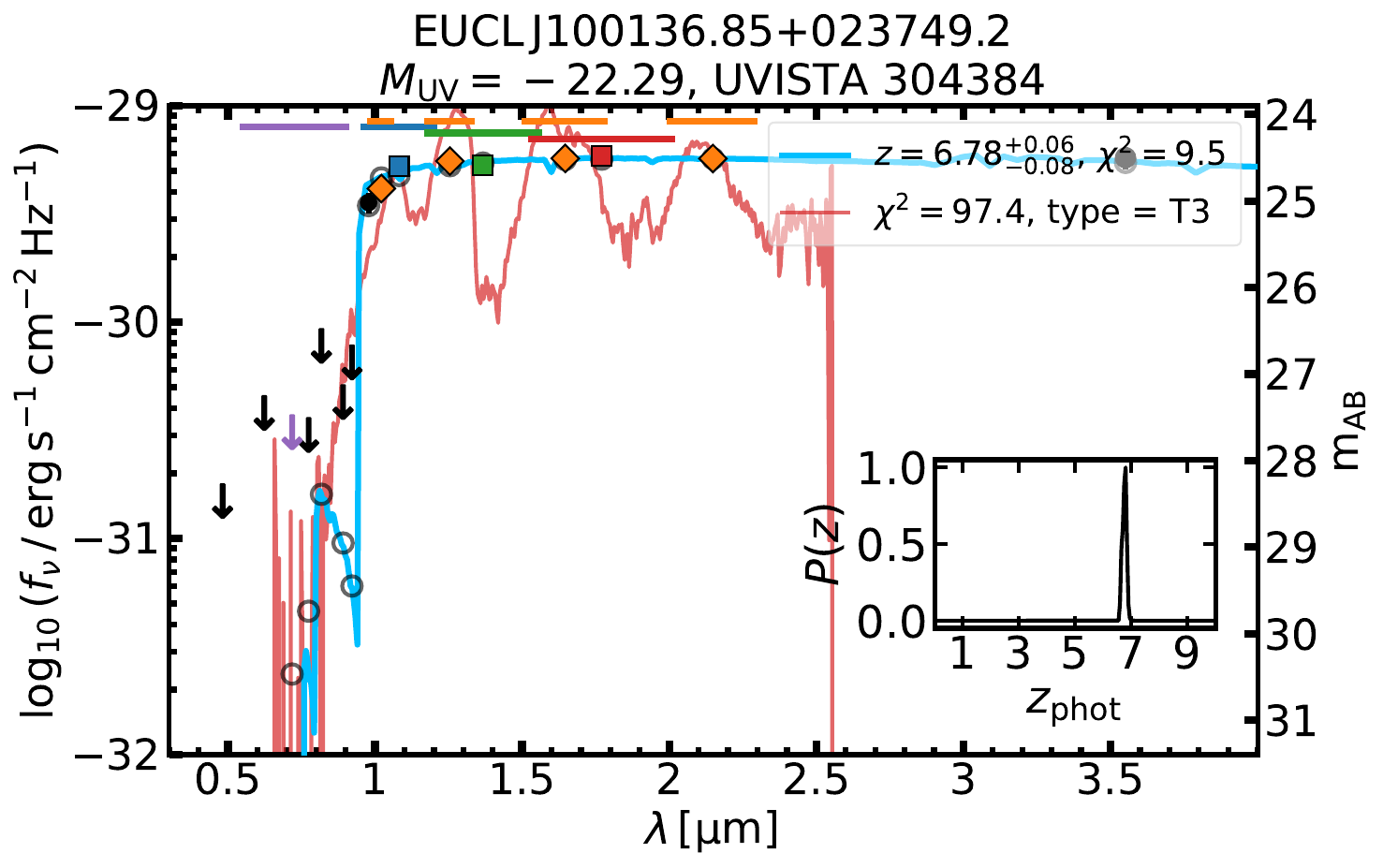}
    \includegraphics[width=0.47\linewidth]{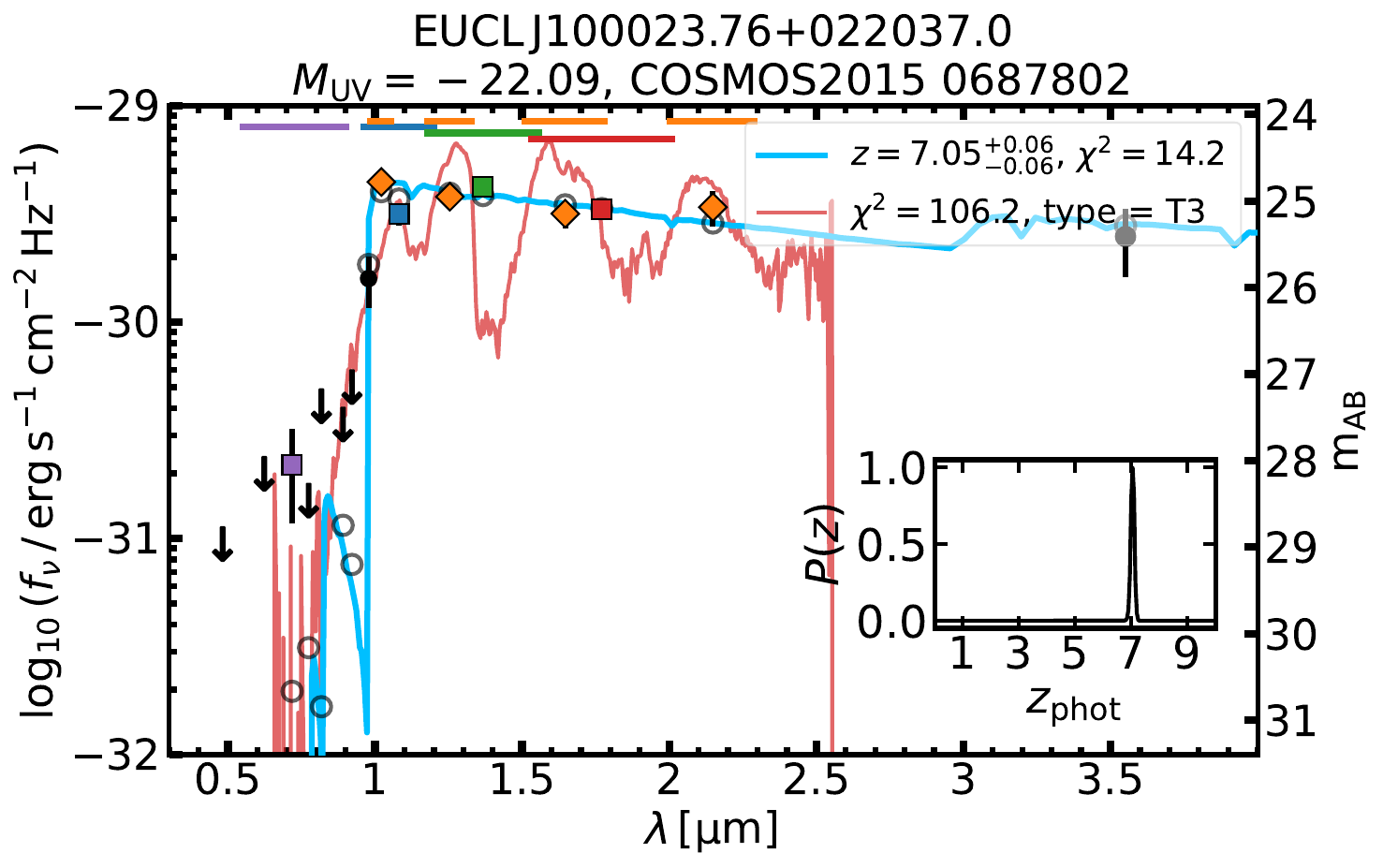}
    
    \includegraphics[width=0.45\linewidth]{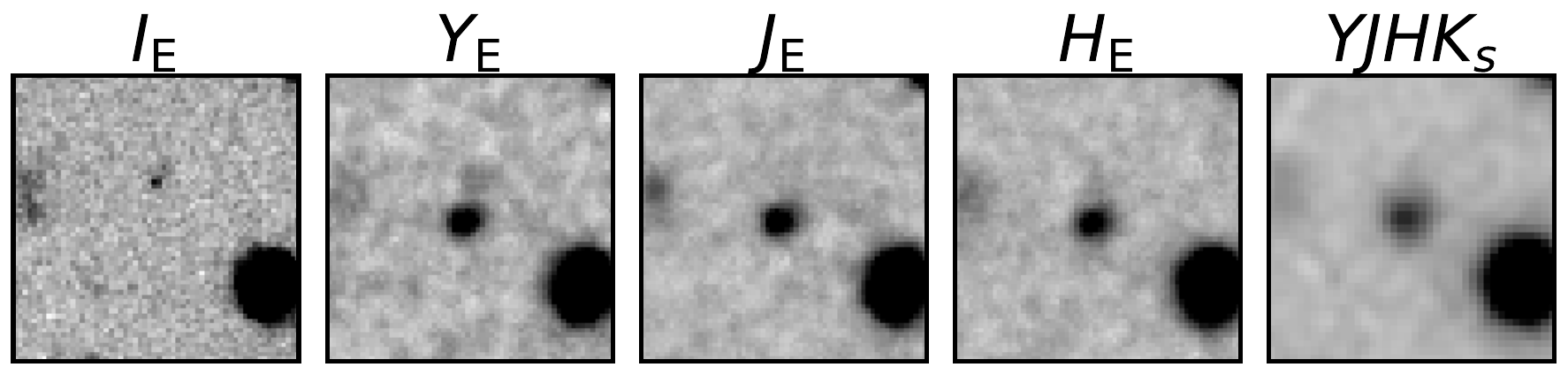}
    \hspace{12pt}
    \includegraphics[width=0.45\linewidth]{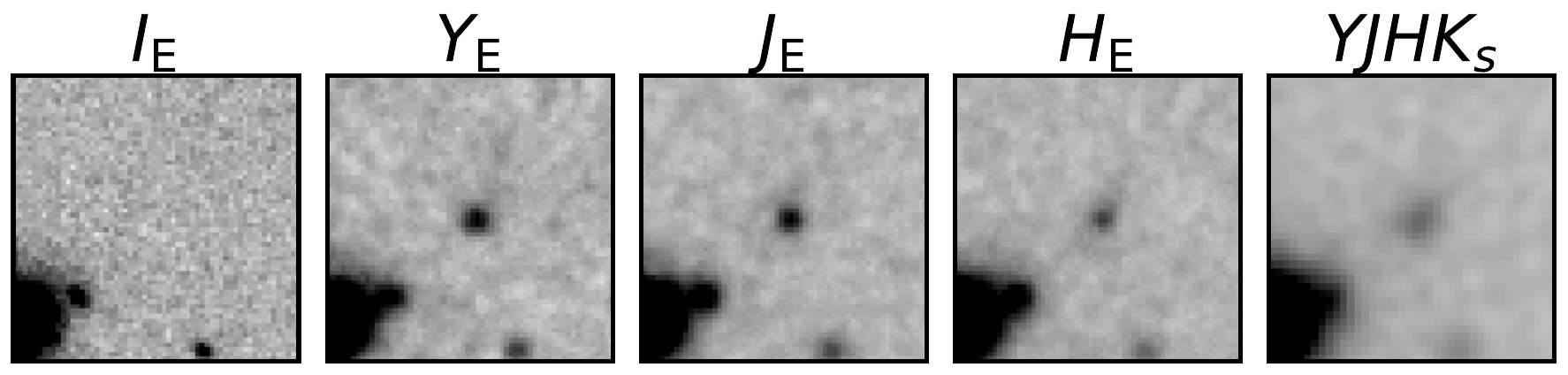}

    \vspace{12pt}

    \includegraphics[width=0.47\linewidth]{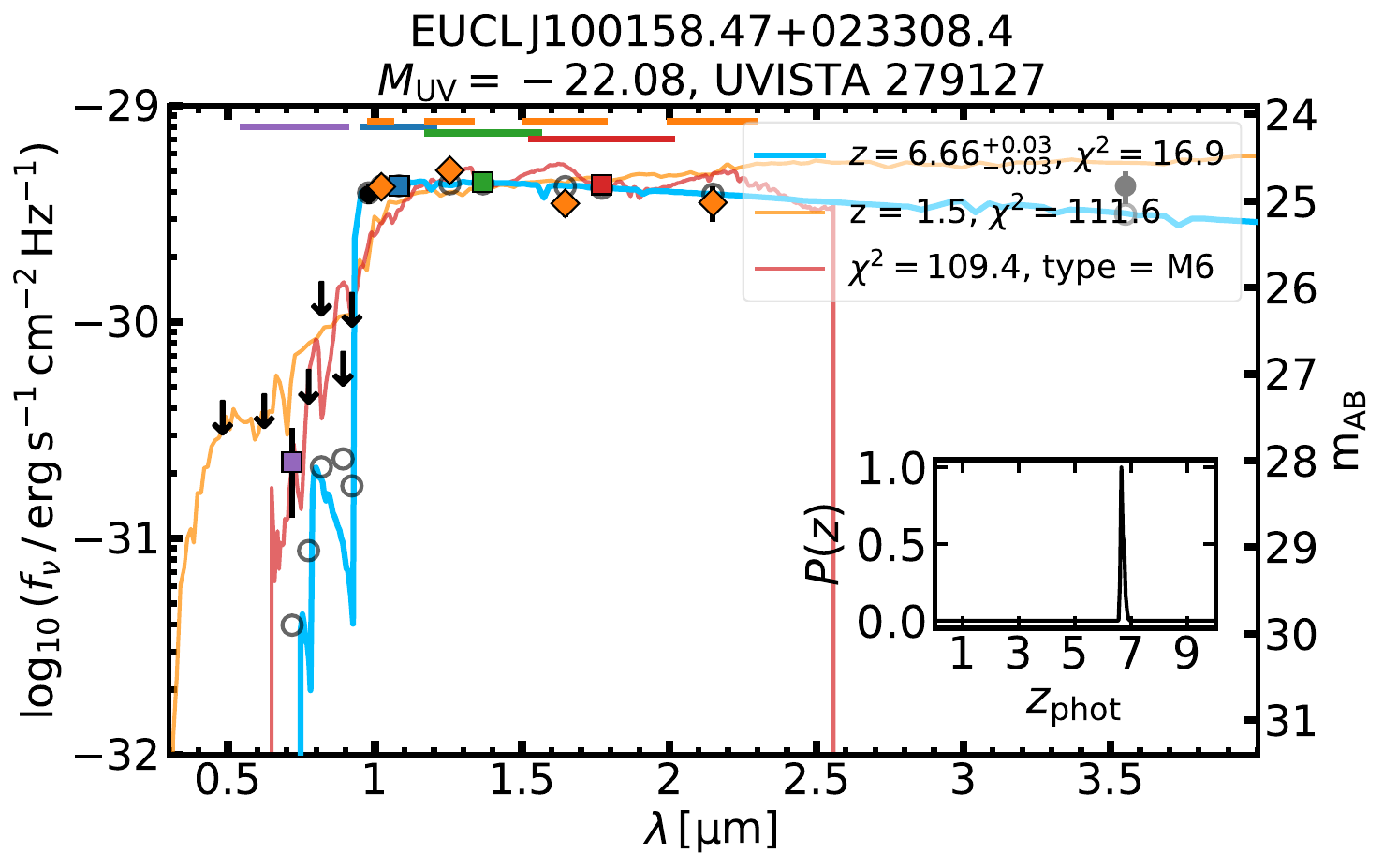}
    \includegraphics[width=0.47\linewidth]{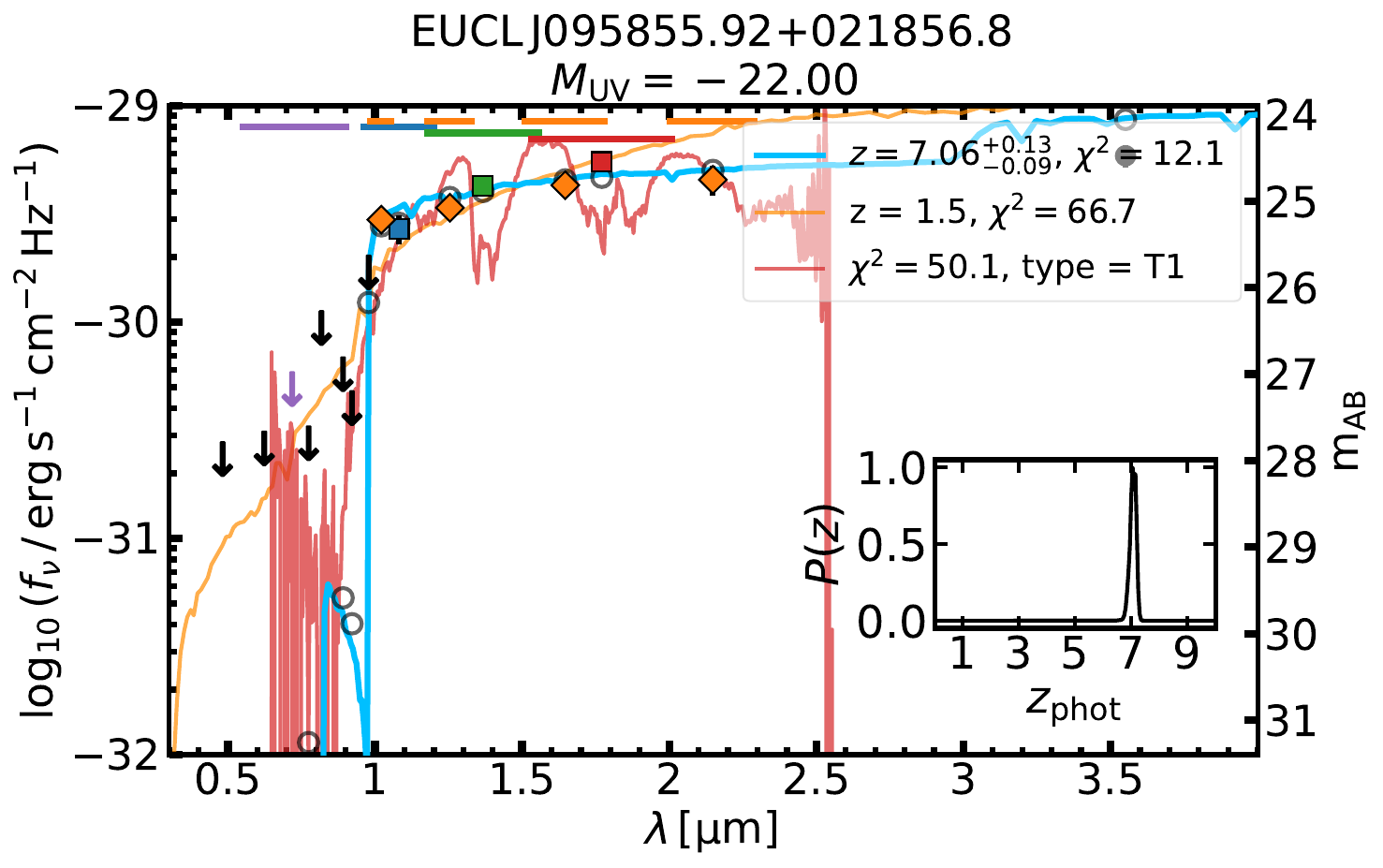}

    \includegraphics[width=0.45\linewidth]{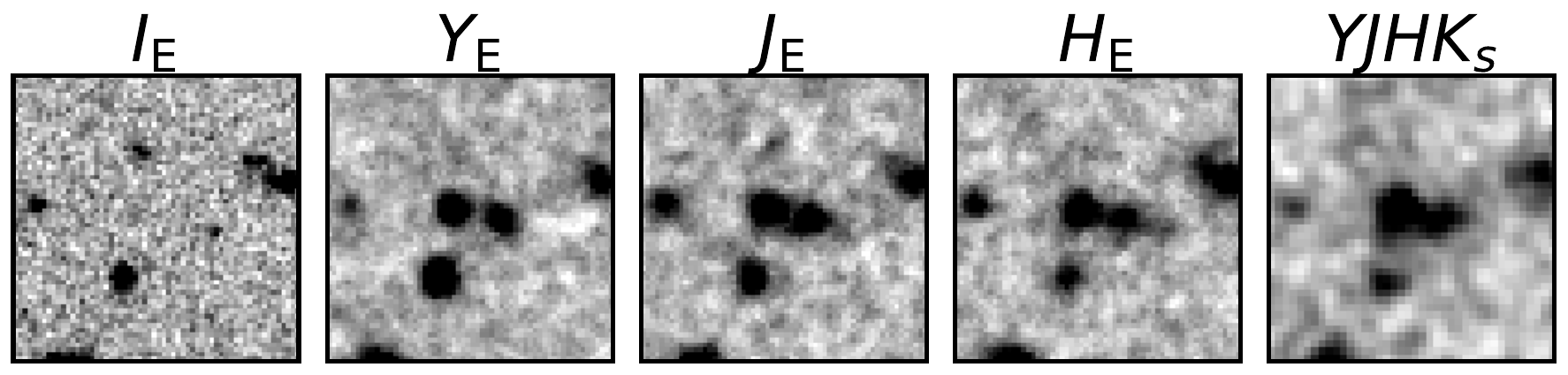}
    \hspace{12pt}
    \includegraphics[width=0.45\linewidth]{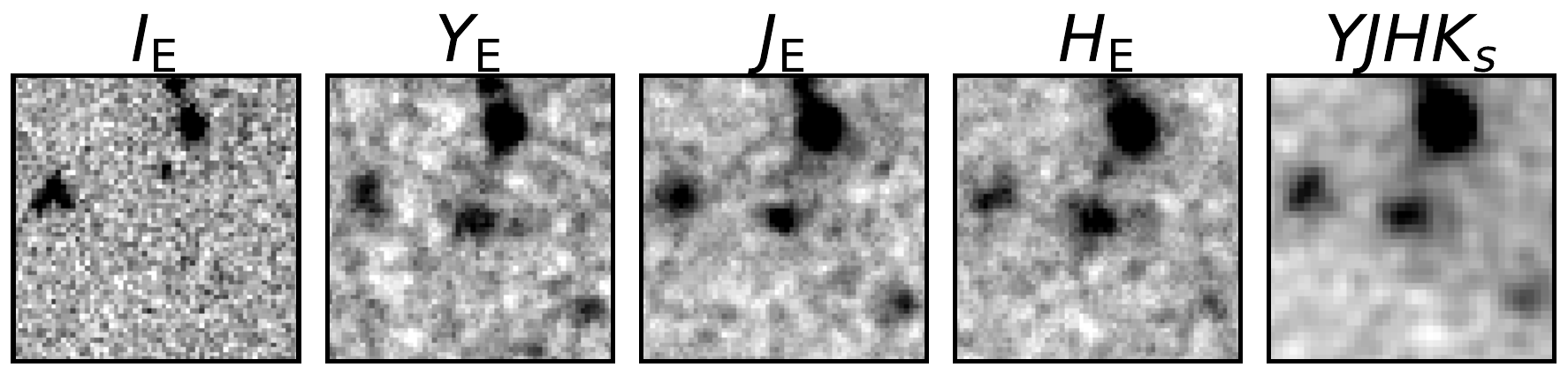}

    \caption{Spectral energy distribution fits (\emph{top}) and postage stamp cutouts (\emph{bottom}) of the 30 most luminous galaxies in the U+E sample.
    (See Fig. \ref{fig:UVISTA euclid SED} for a description of the SED plots.)
    The title of the SED indicates the object ID, absolute magnitude $\Muv$, and the literature ID if it has been previously identified.
    The postage stamps, from left to right, are from \IE, \YE, \JE, \HE, and a VISTA $YJHK_s$ stack. The stamps are $6\arcsec\times6\arcsec$ and are scaled to saturate at $5\,\sigma$ above and $2\,\sigma$ below the noise level.
    Objects with names beginning with UVISTA were identified in \citet{Bowler14}. 
    Objects with names labelled as [ESC2021] and [ESC2021a] were first identified in \citet{Endsley21nebular} and \citet{Endsley21MMT}, respectively. 
    [DMD23] 892014 is from \citet{Donnan23}.
    COSMOS2015 and COSMOS2020 objects were identified in \citet{COSMOS2015} and \citet{COSMOS2020}, respectively.
    S-CANDELS objects are galaxies identified in HST imaging \citep{SCANDELS}, and LAE-17 was identified in \citet{Hu17}.
    The SSTSL source is also identified as REBELS-21 \citep{REBELS}.}
\label{fig:seds and stamps}
\end{figure*}

\begin{figure*}
    \ContinuedFloat

    \includegraphics[width=0.47\linewidth]{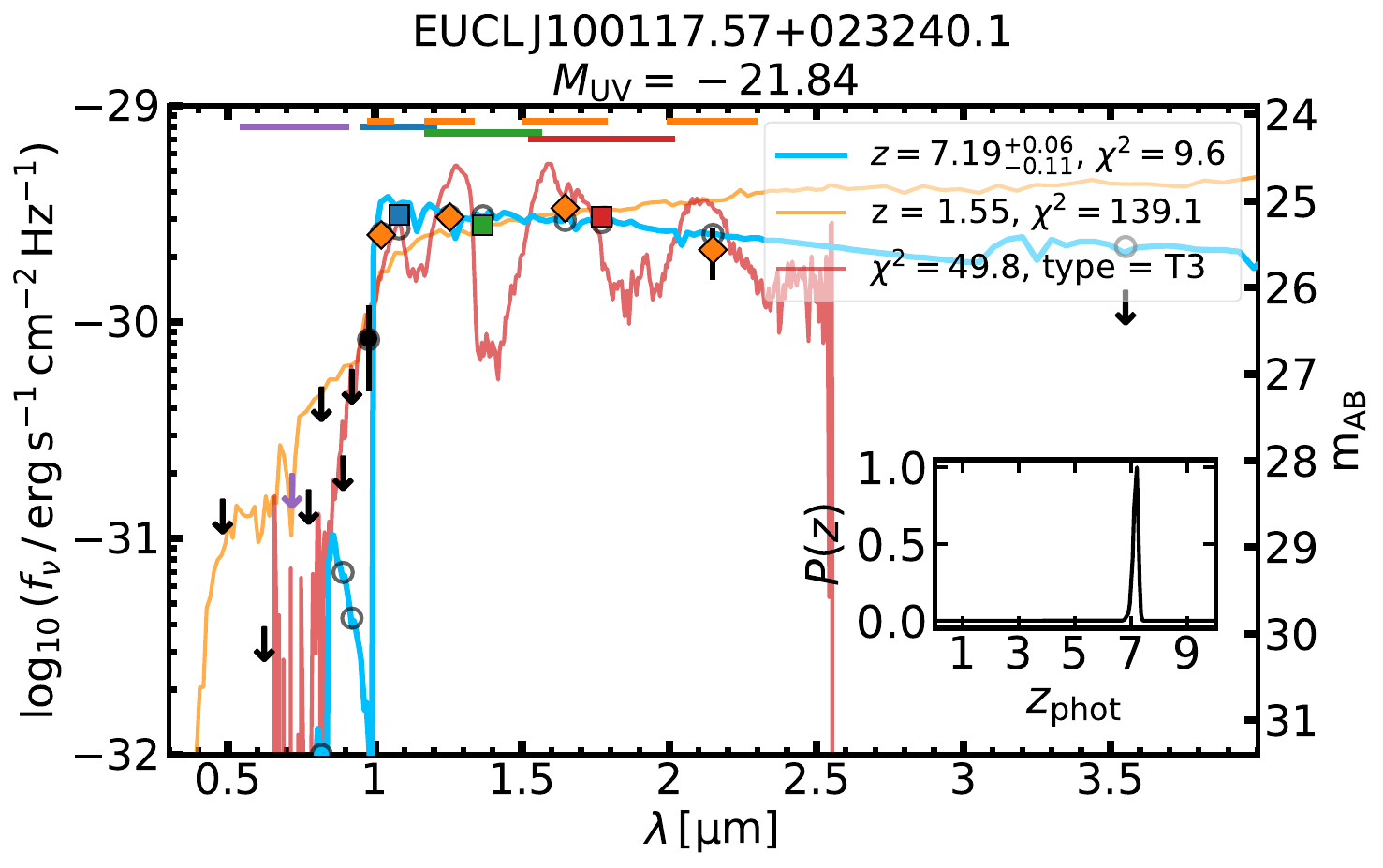}
    \includegraphics[width=0.47\linewidth]{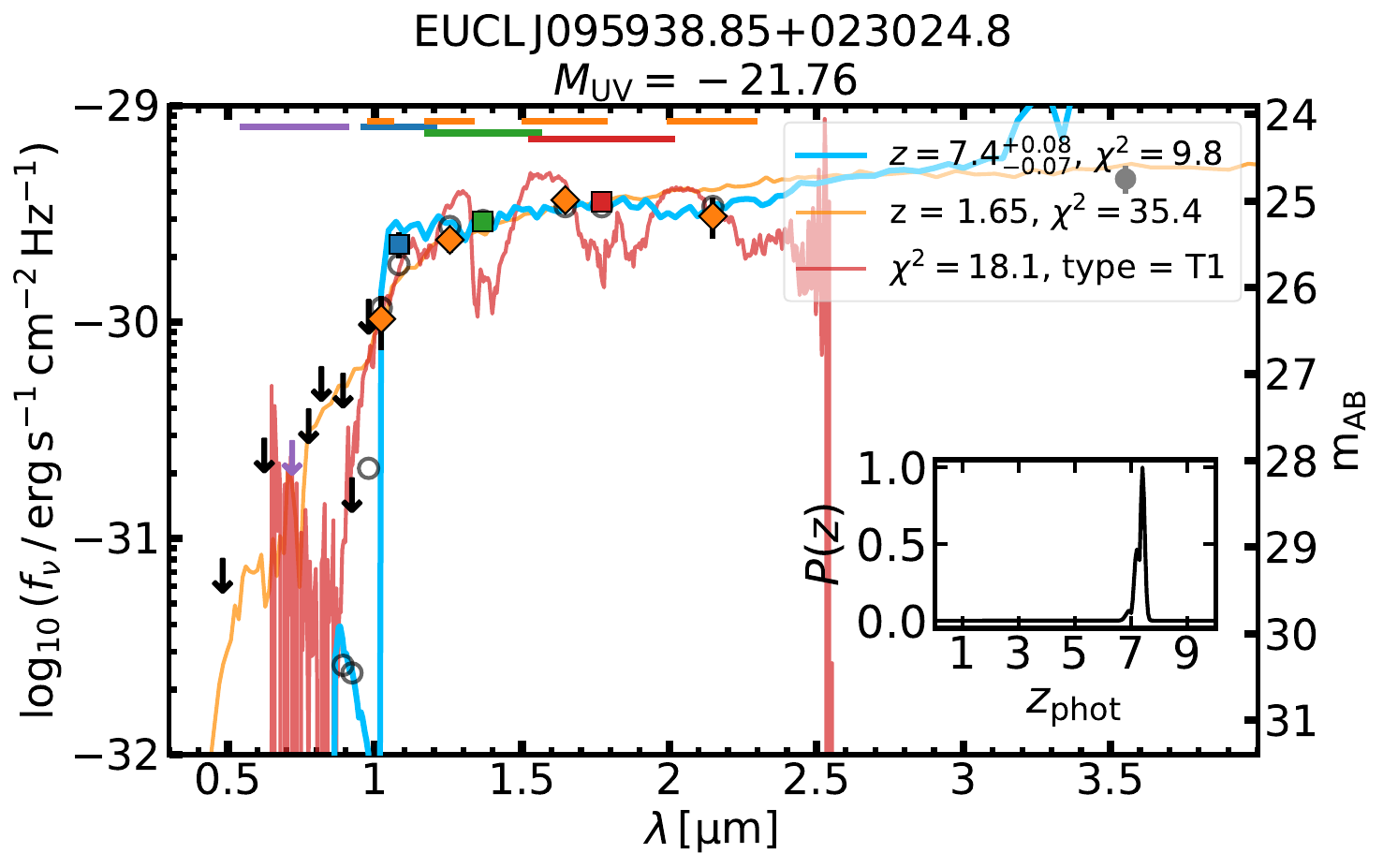}
    
    \includegraphics[width=0.44\linewidth]{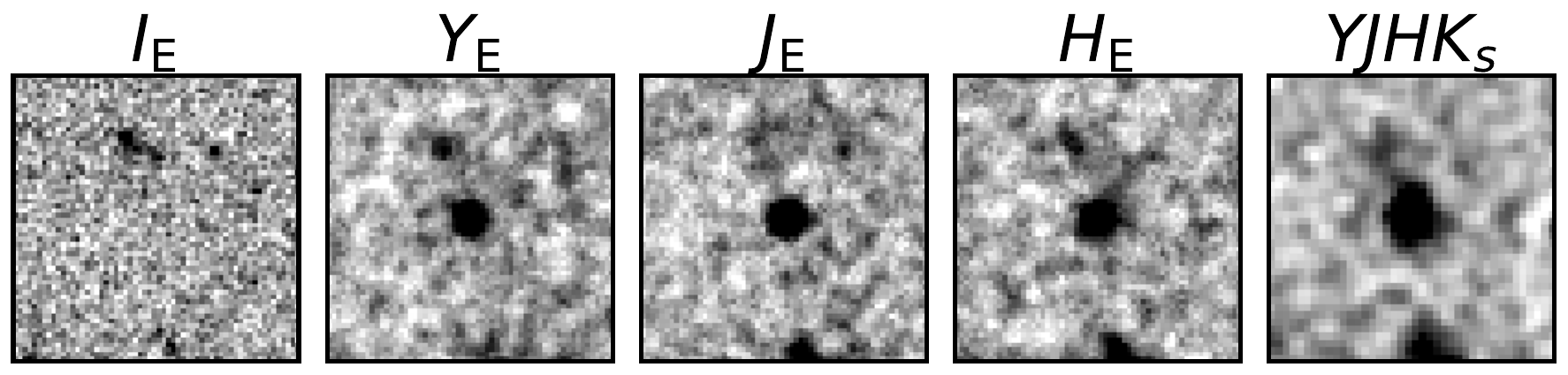}
    \hspace{12pt}
    \includegraphics[width=0.44\linewidth]{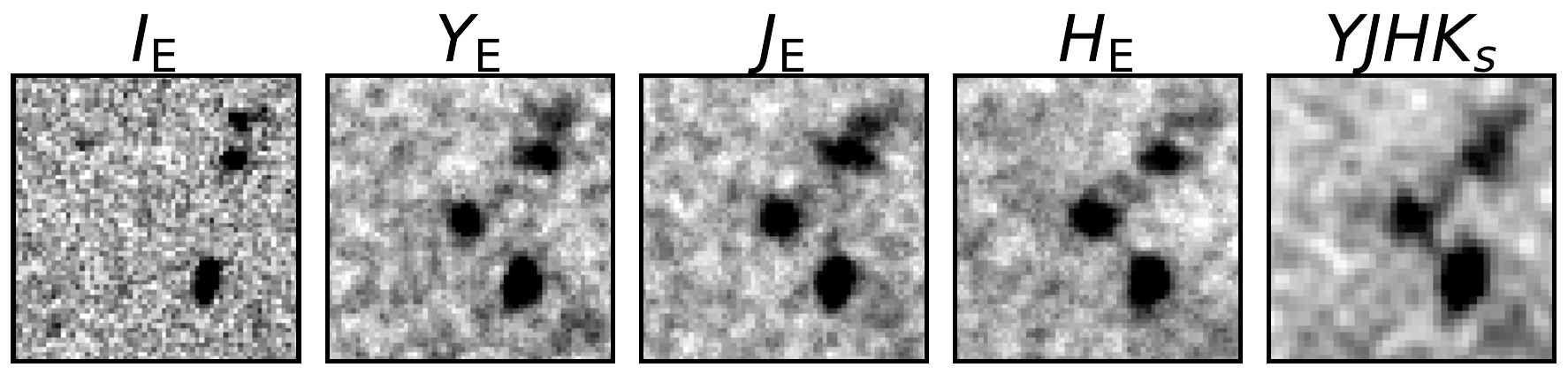}

    \vspace{12pt}

    \includegraphics[width=0.47\linewidth]{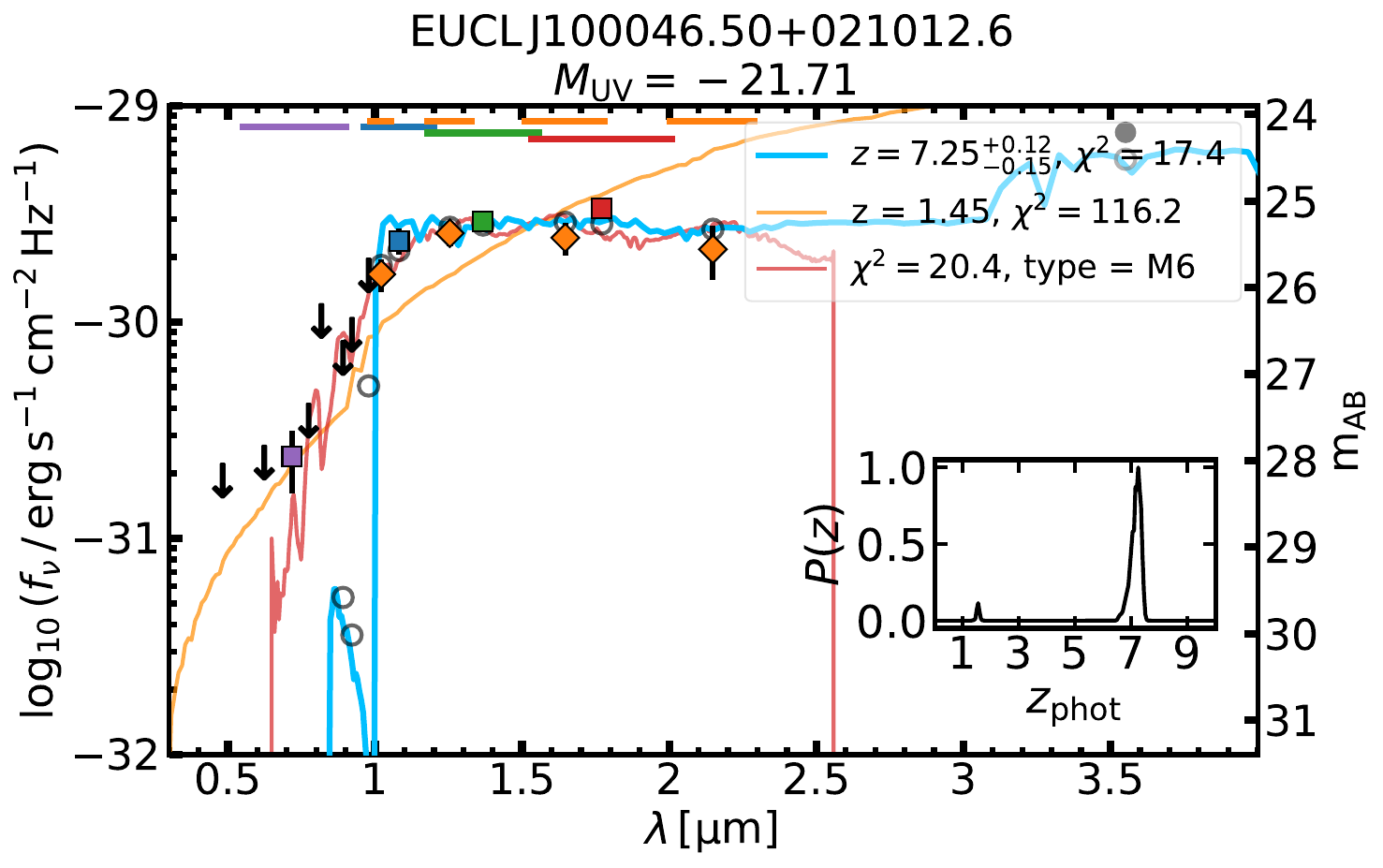}
    \includegraphics[width=0.47\linewidth]{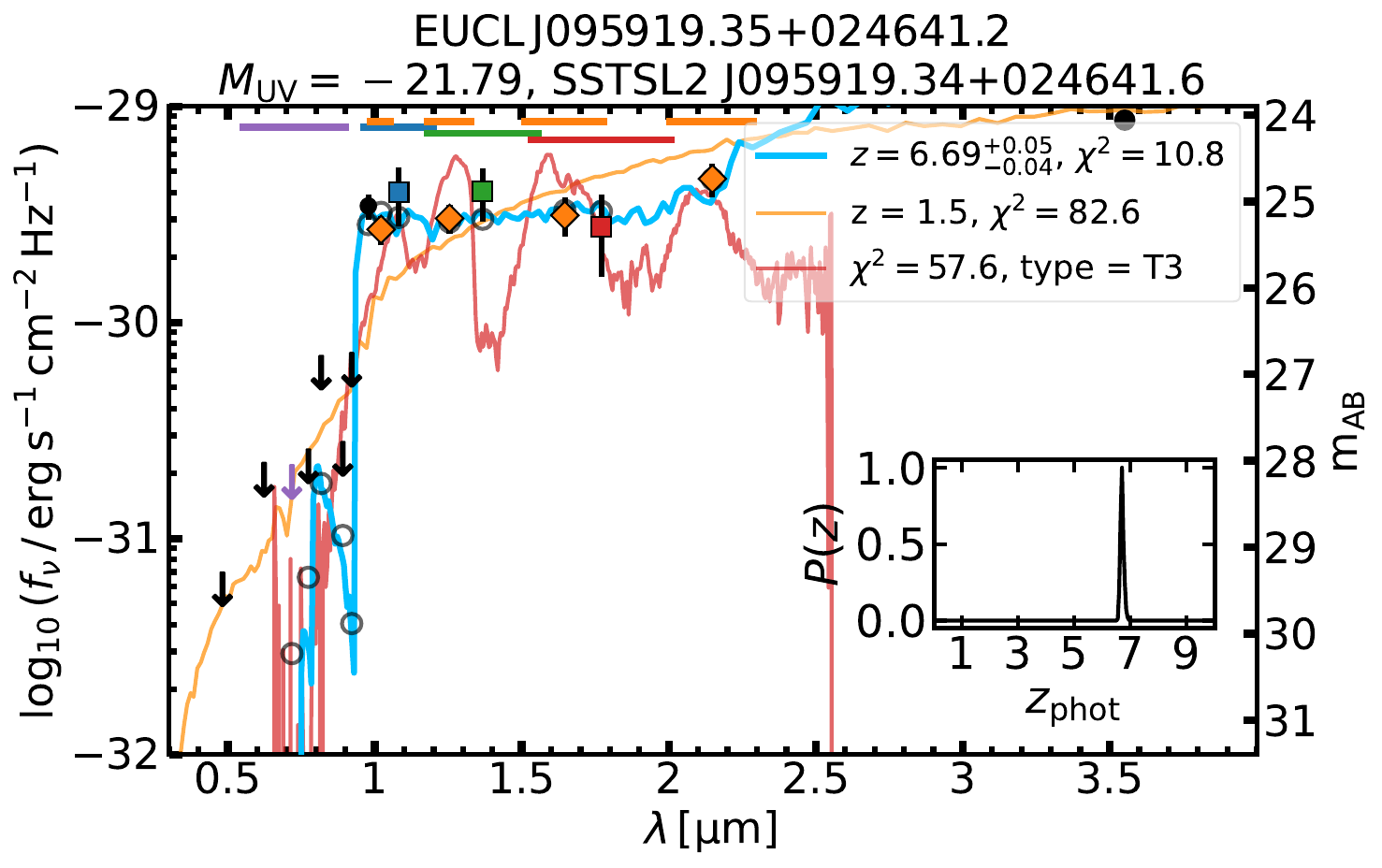}

    \includegraphics[width=0.44\linewidth]{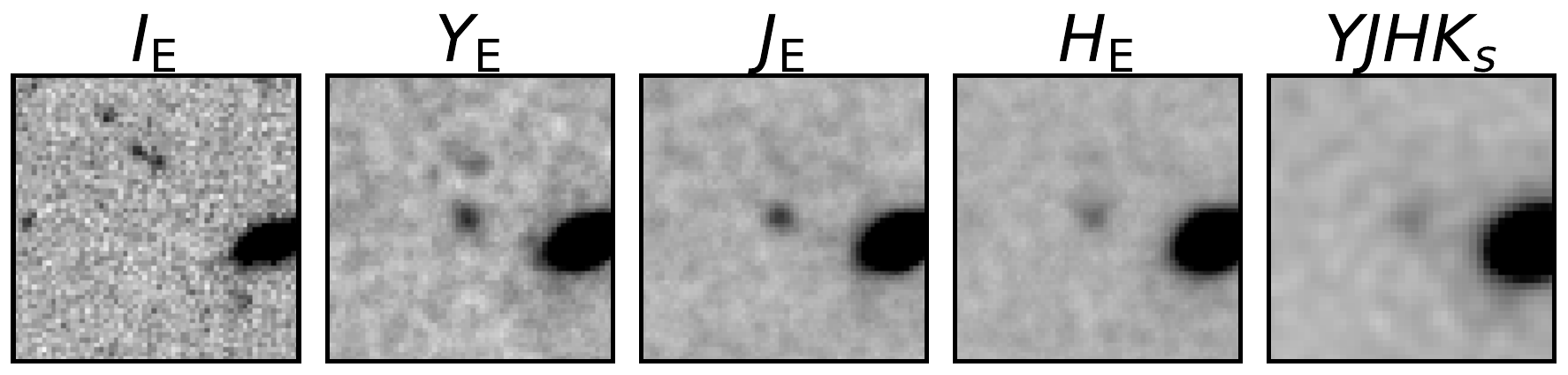}
    \hspace{12pt}
    \includegraphics[width=0.44\linewidth]{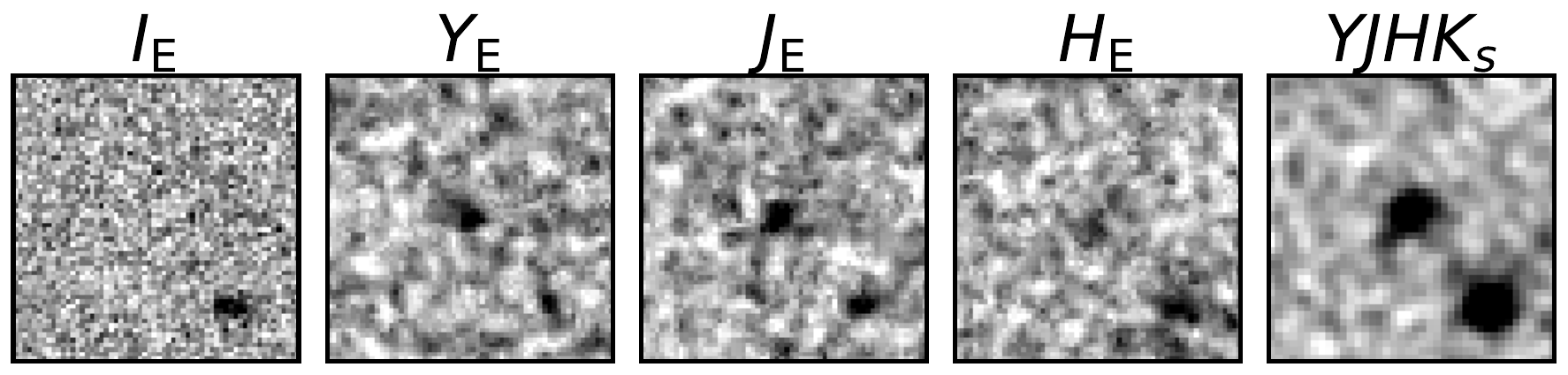}

    \vspace{12pt}

    \includegraphics[width=0.47\linewidth]{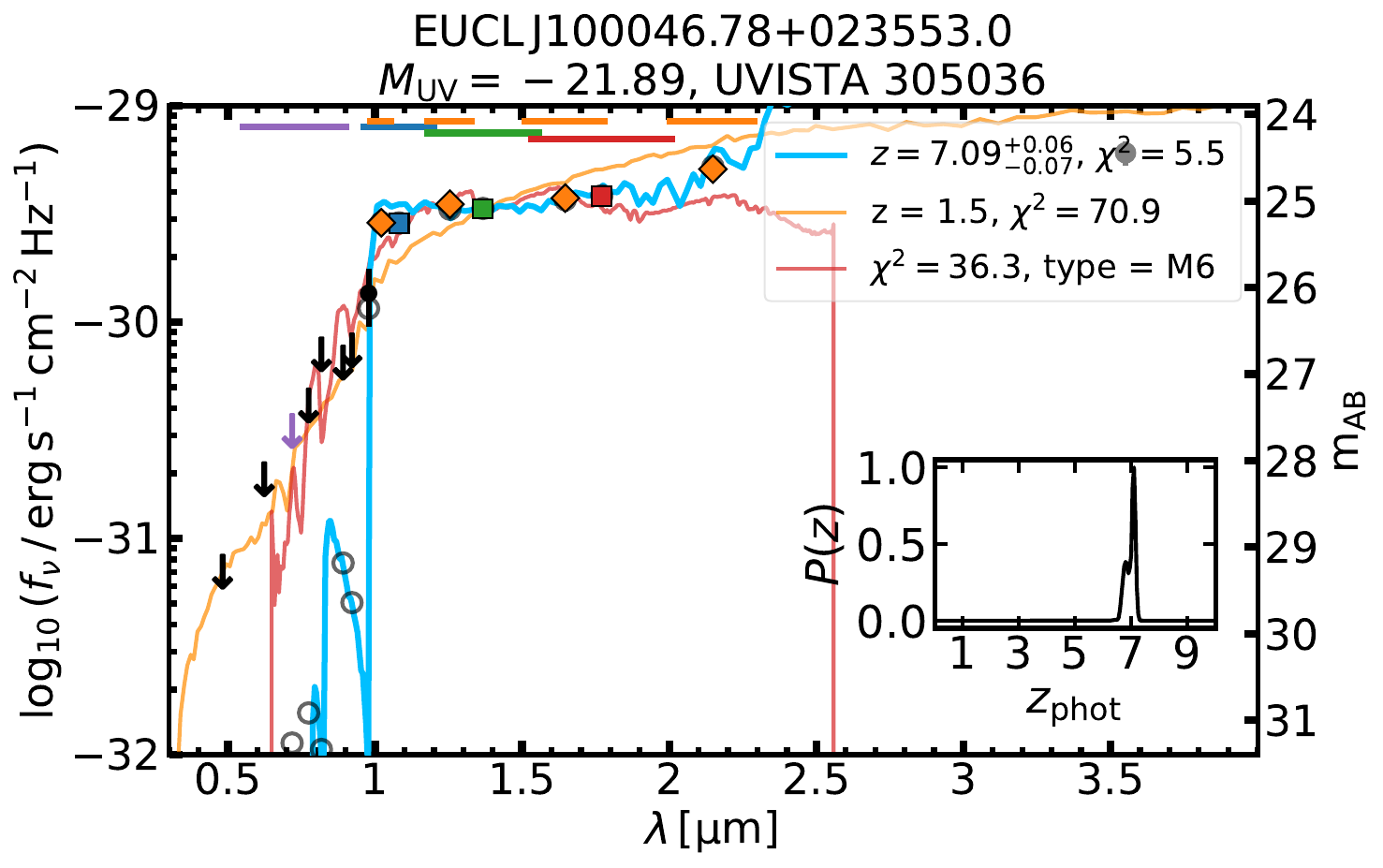}
    \includegraphics[width=0.47\linewidth]{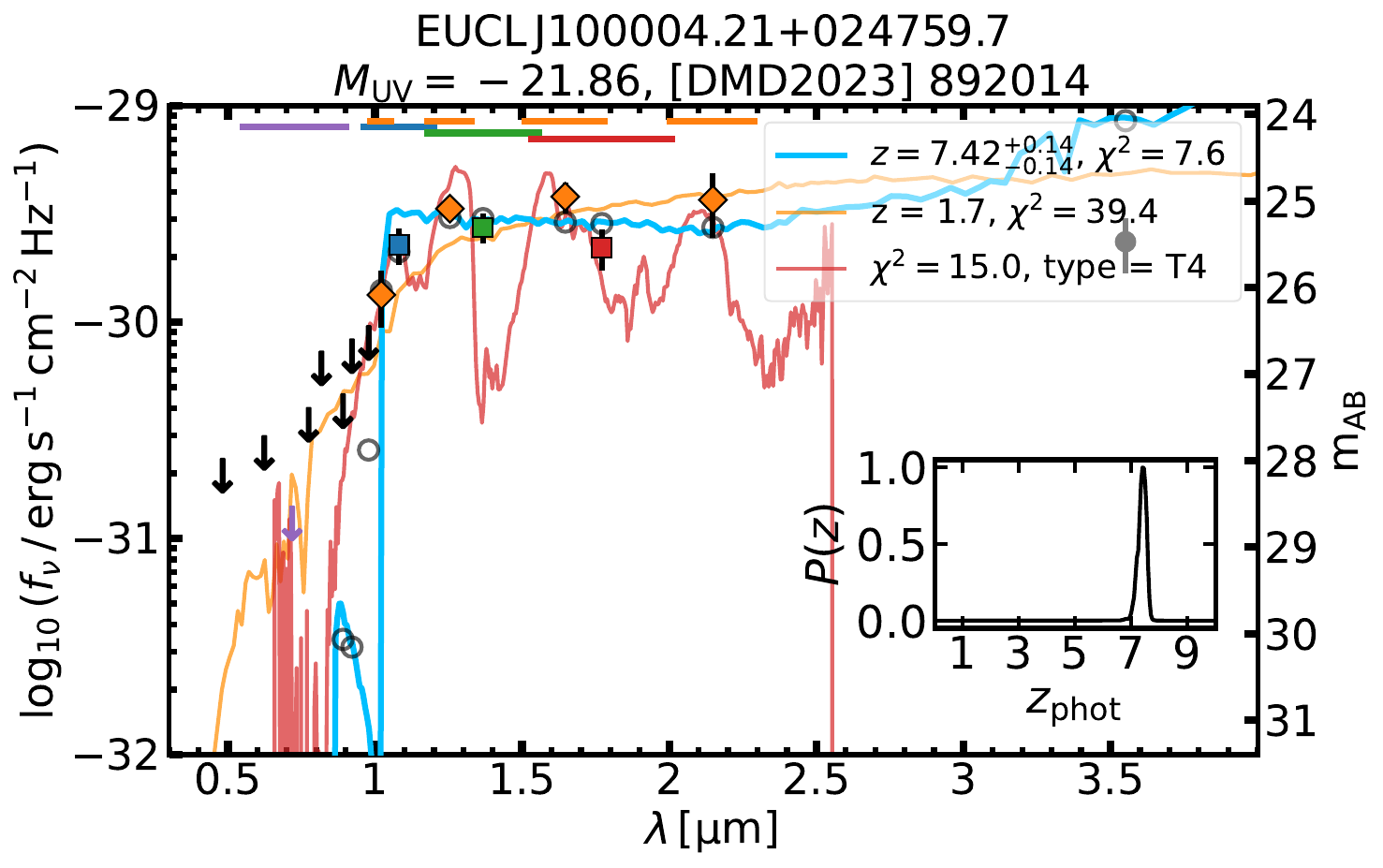}

    \includegraphics[width=0.45\linewidth]{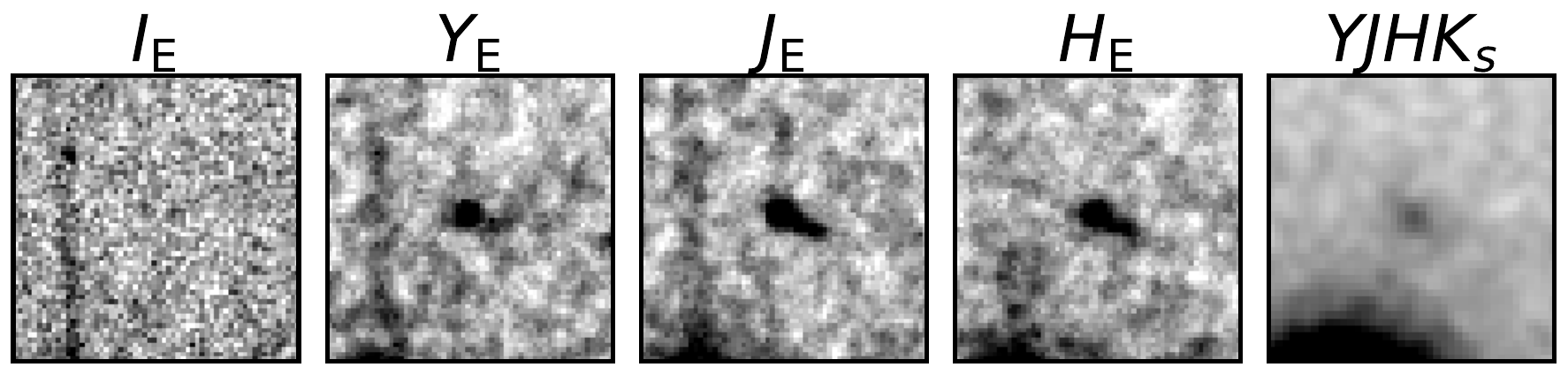}
    \hspace{12pt}
    \includegraphics[width=0.45\linewidth]{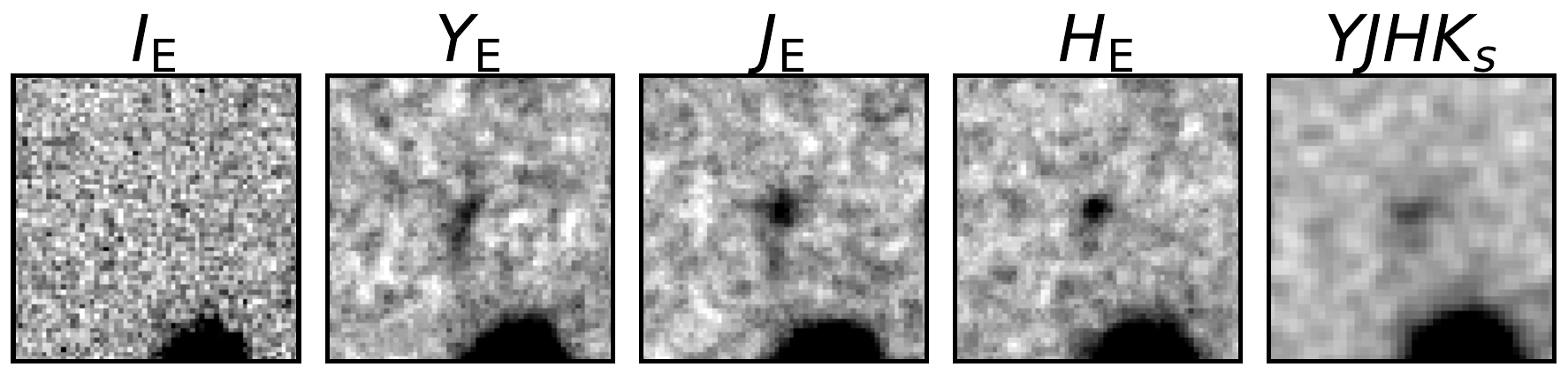}

\caption{Continued.}

\end{figure*}

\begin{figure*} 
    \ContinuedFloat

    \includegraphics[width=0.47\linewidth]{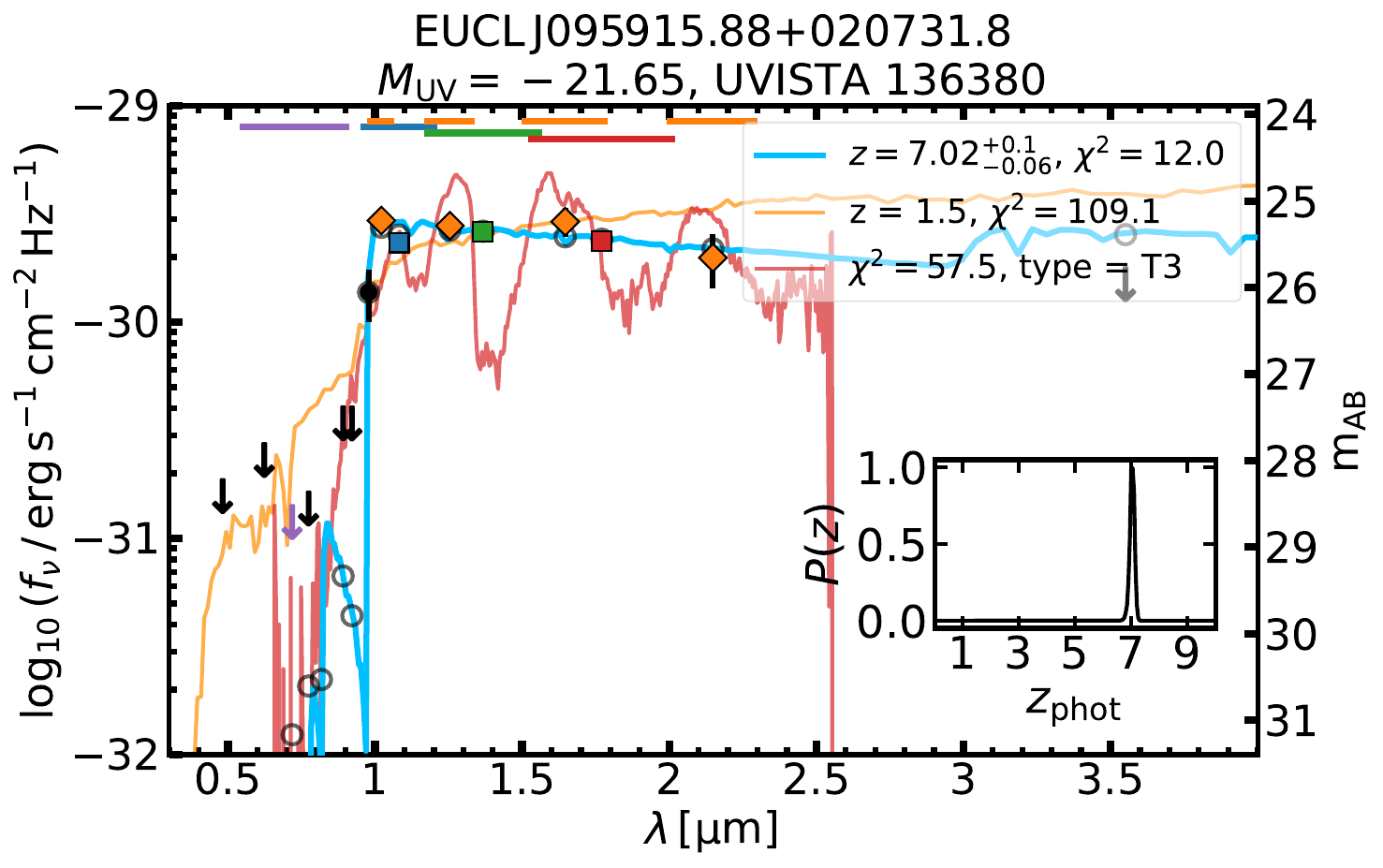}
    \includegraphics[width=0.47\linewidth]{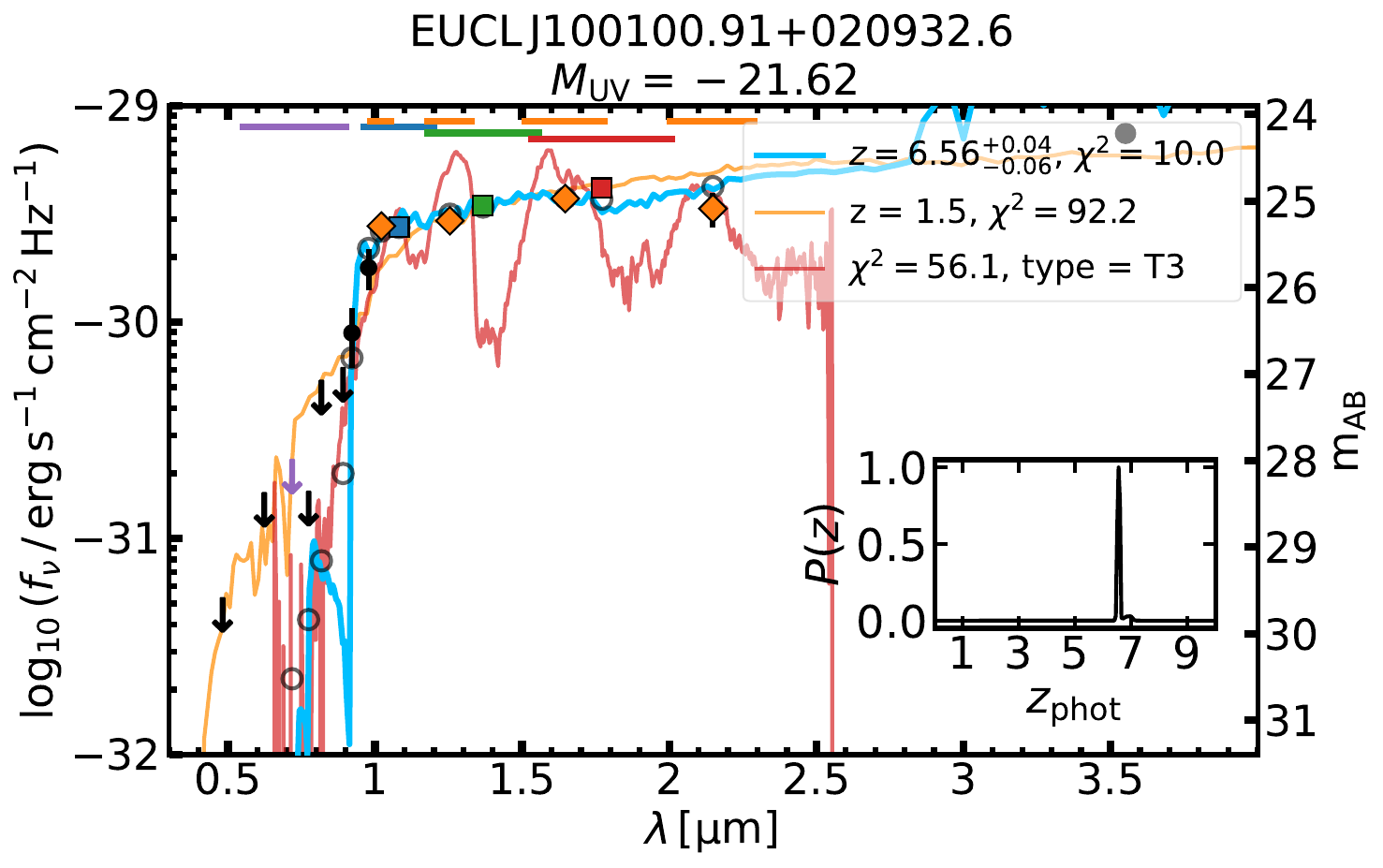}

    \includegraphics[width=0.44\linewidth]{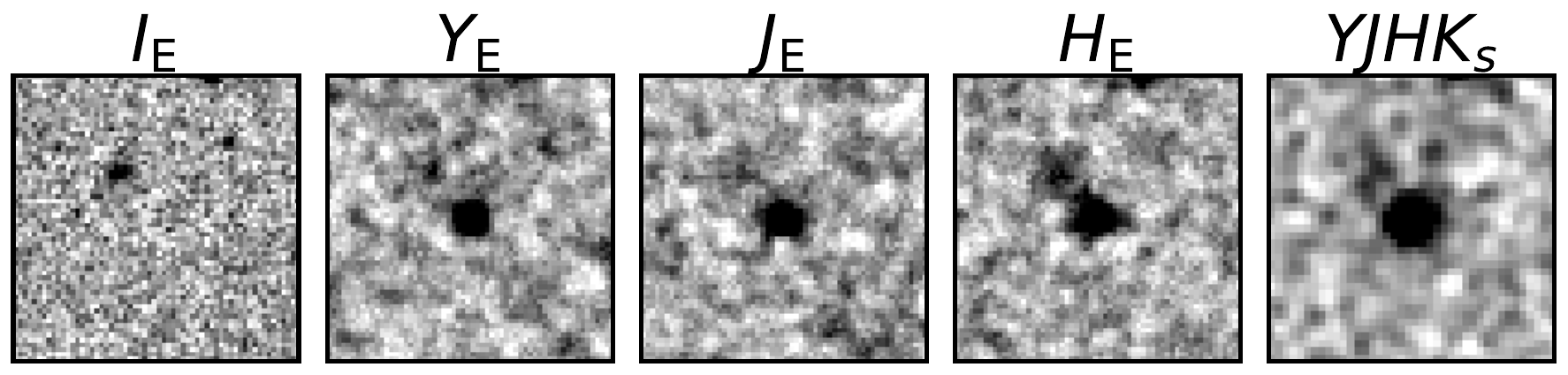}
    \hspace{12pt}
    \includegraphics[width=0.44\linewidth]{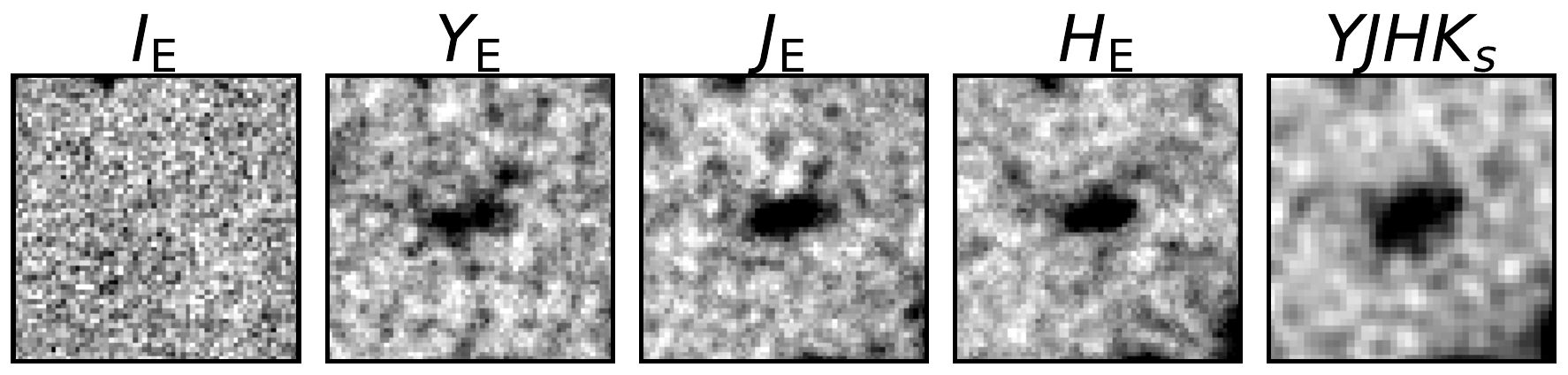}

    \vspace{12pt}

    \includegraphics[width=0.47\linewidth]{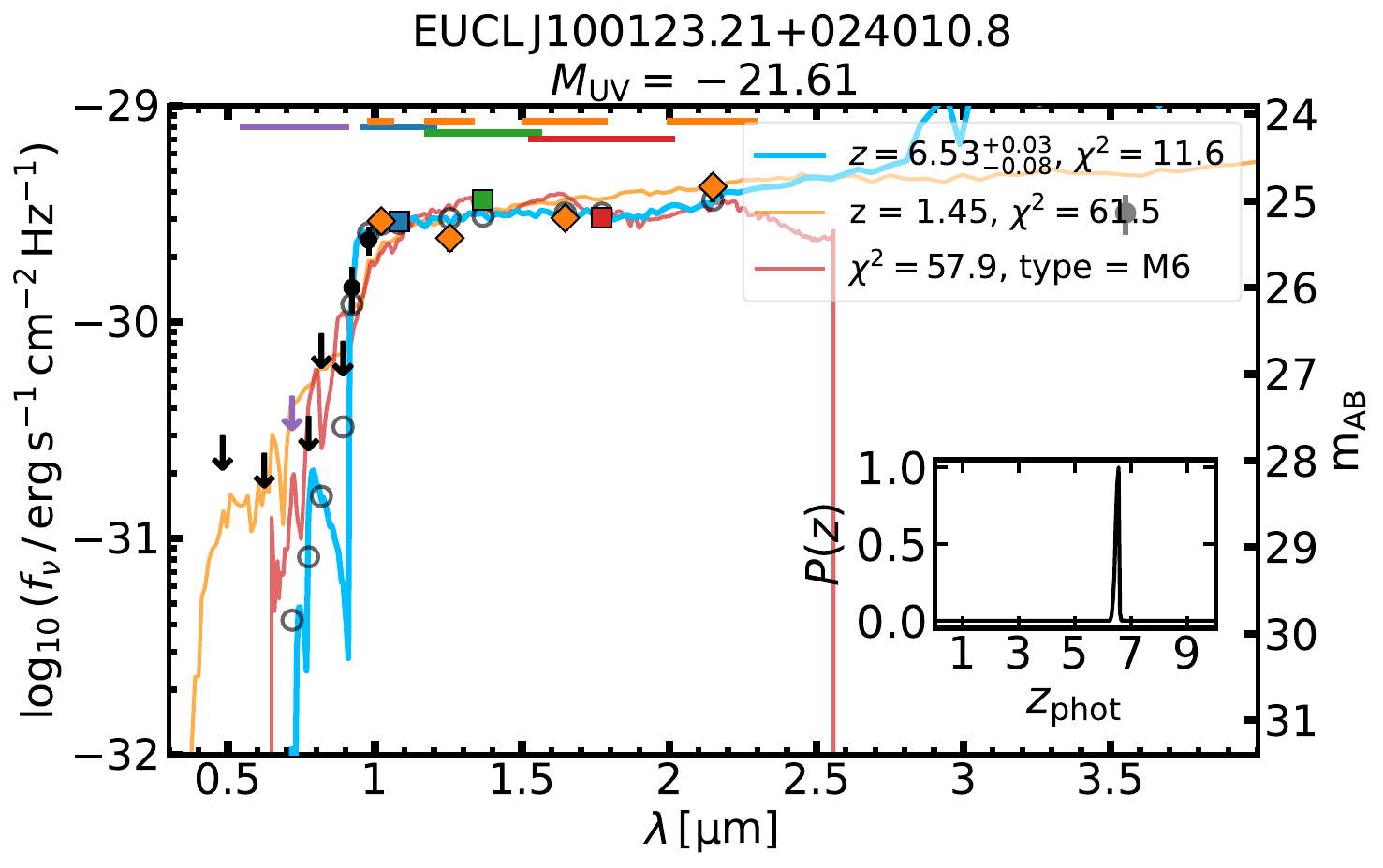}
    \includegraphics[width=0.47\linewidth]{stamps_SEDs/828001_SED_EUCLname.pdf}
    
    \includegraphics[width=0.44\linewidth]{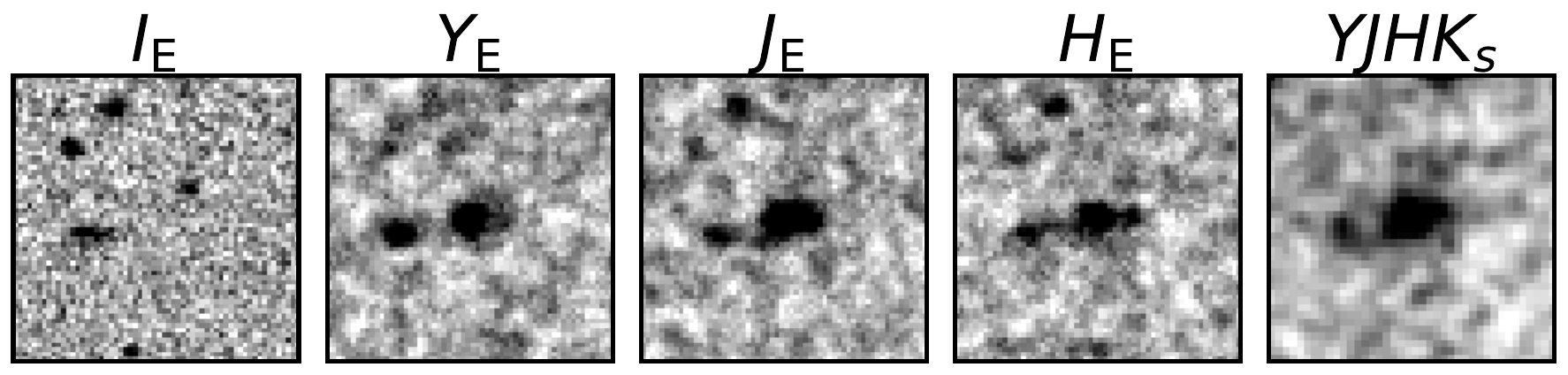}
    \hspace{12pt}
    \includegraphics[width=0.44\linewidth]{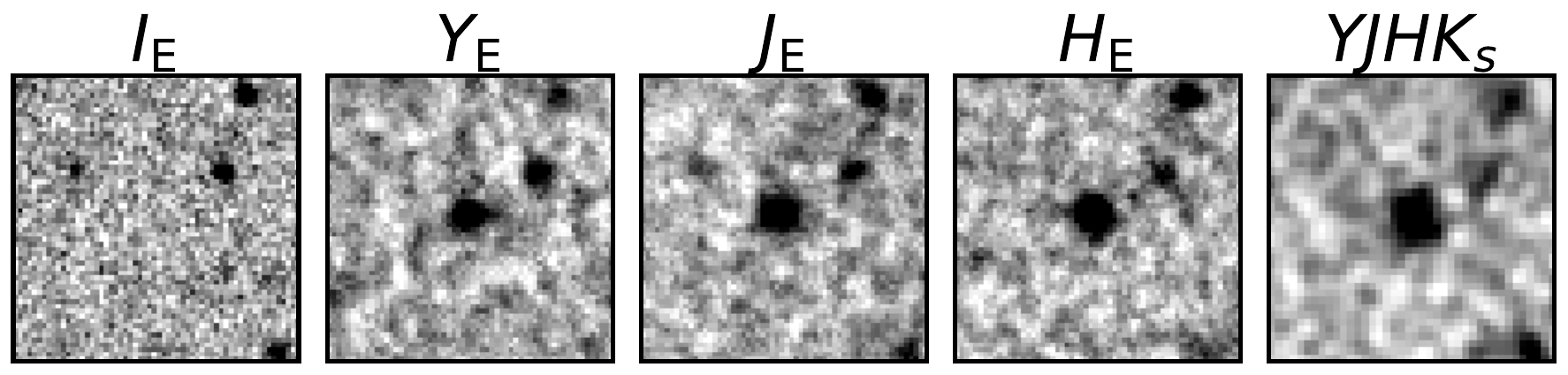}

    \vspace{12pt}

    \includegraphics[width=0.47\linewidth]{stamps_SEDs/510801_SED_EUCLname.pdf}
    \includegraphics[width=0.47\linewidth]{stamps_SEDs/1069823_SED_EUCLname.pdf}

    \includegraphics[width=0.44\linewidth]{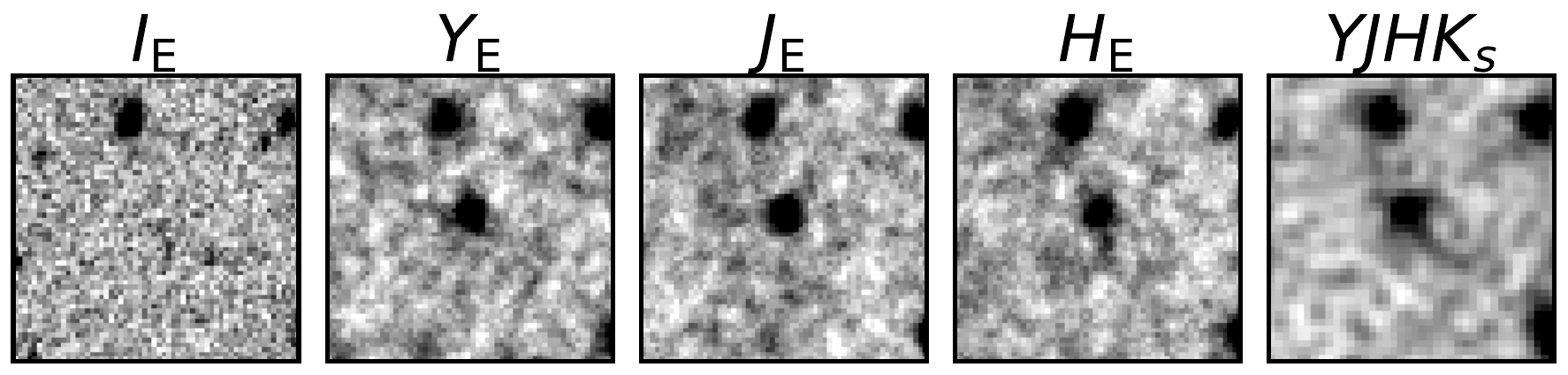}
    \hspace{12pt}
    \includegraphics[width=0.44\linewidth]{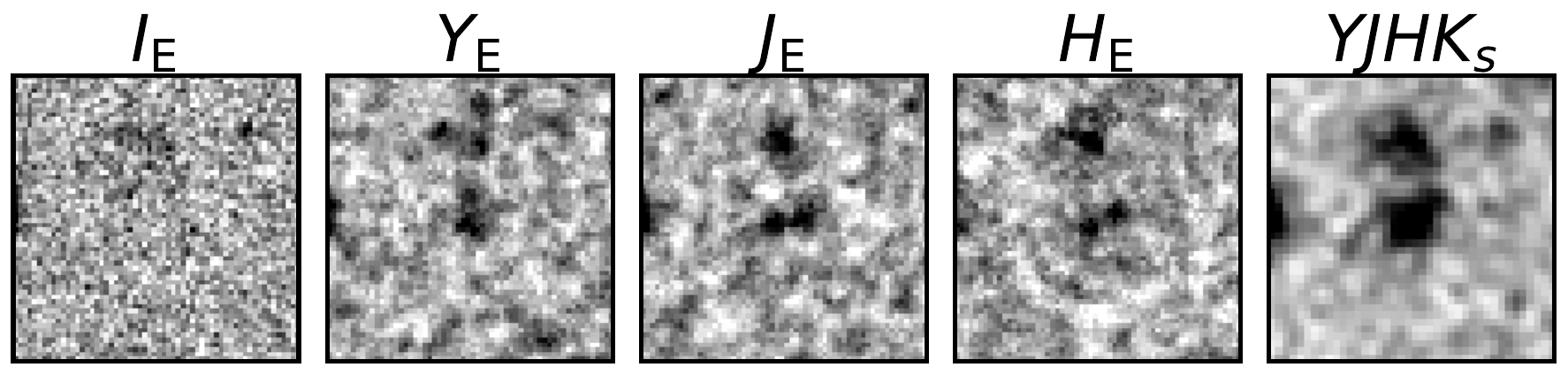}

\caption{Continued.}

\end{figure*}

\begin{figure*}
    \ContinuedFloat

    \includegraphics[width=0.47\linewidth]{stamps_SEDs/468417_SED_EUCLname.pdf}
    \includegraphics[width=0.47\linewidth]{stamps_SEDs/499162_SED_EUCLname.pdf}

    \includegraphics[width=0.44\linewidth]{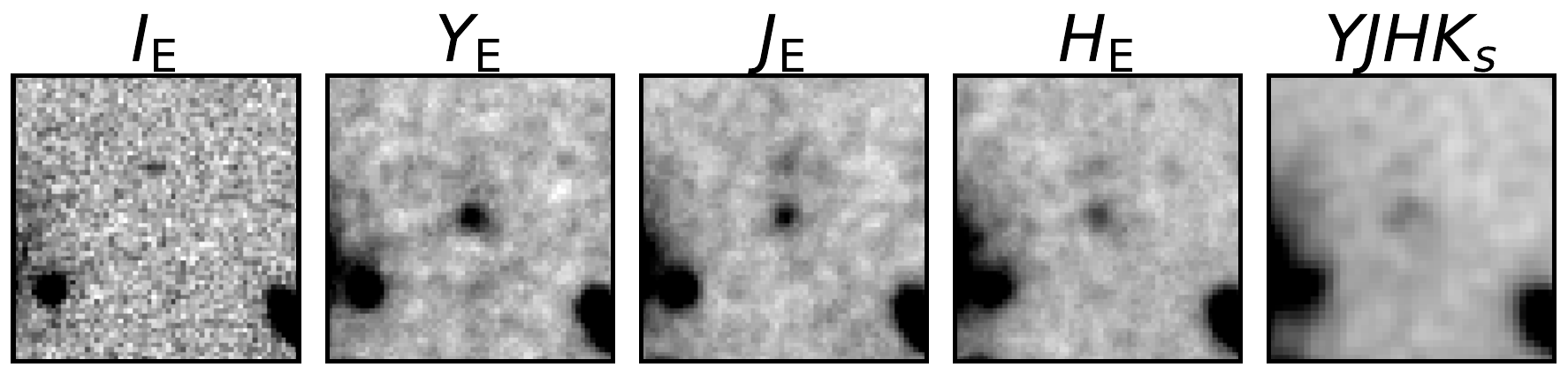}
    \hspace{12pt}
    \includegraphics[width=0.44\linewidth]{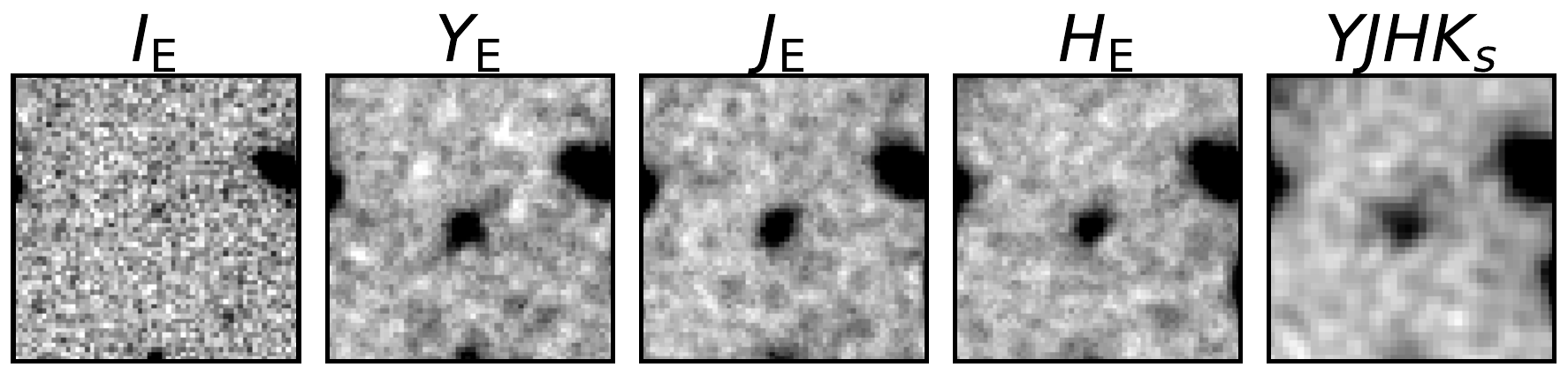}

    \vspace{12pt}

    \includegraphics[width=0.47\linewidth]{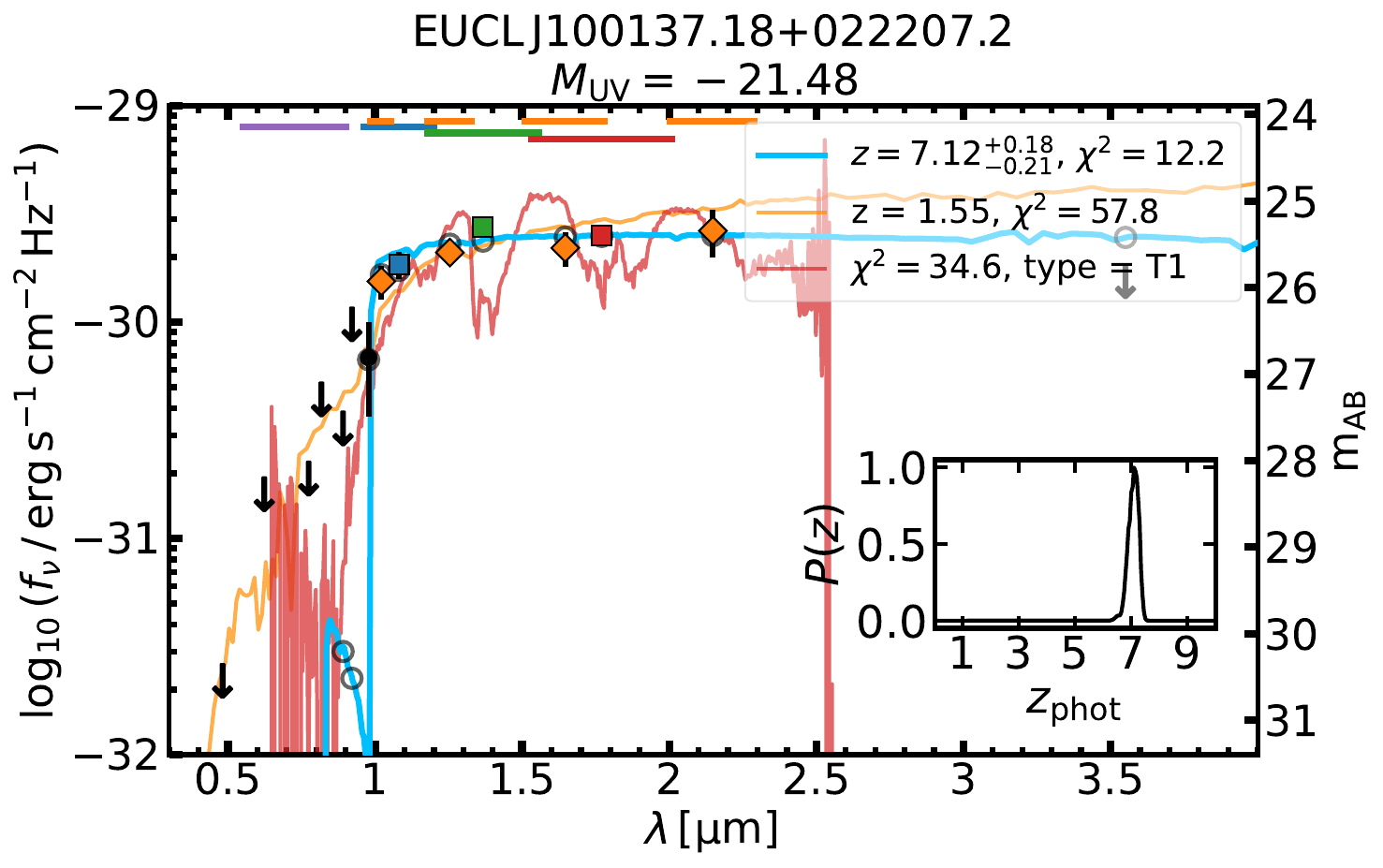}
    \includegraphics[width=0.47\linewidth]{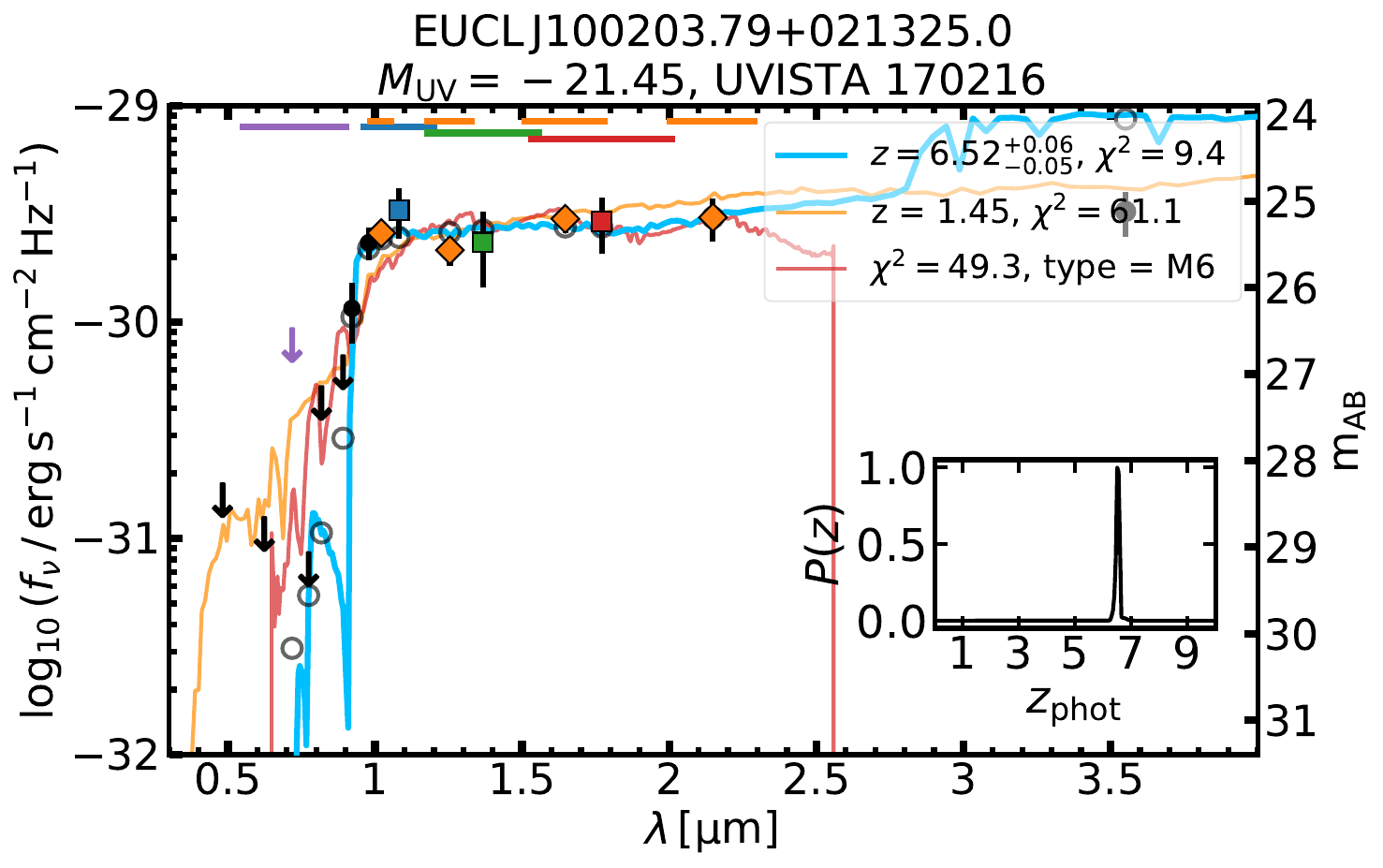}
    
    \includegraphics[width=0.44\linewidth]{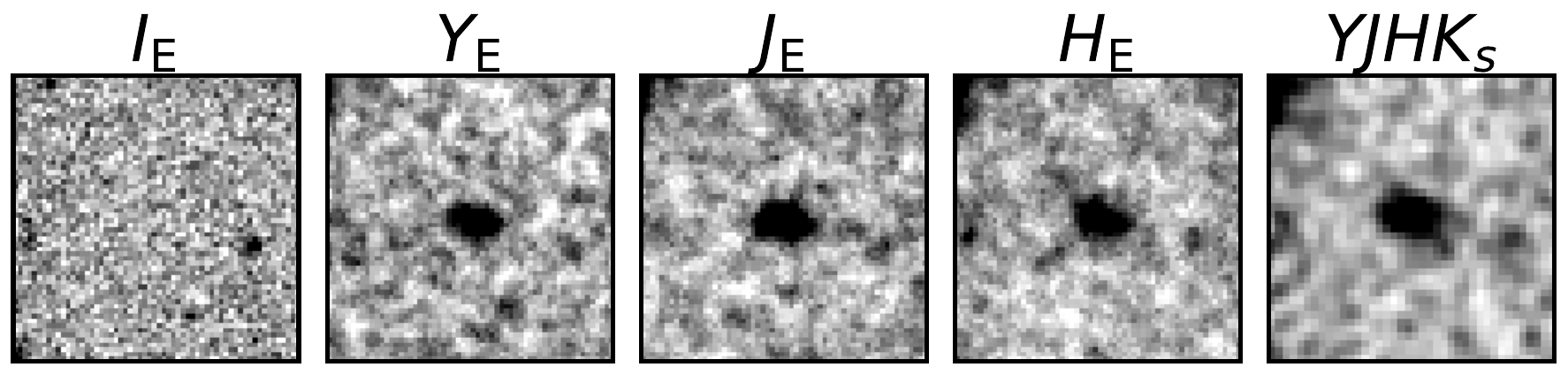}
    \hspace{12pt}
    \includegraphics[width=0.44\linewidth]{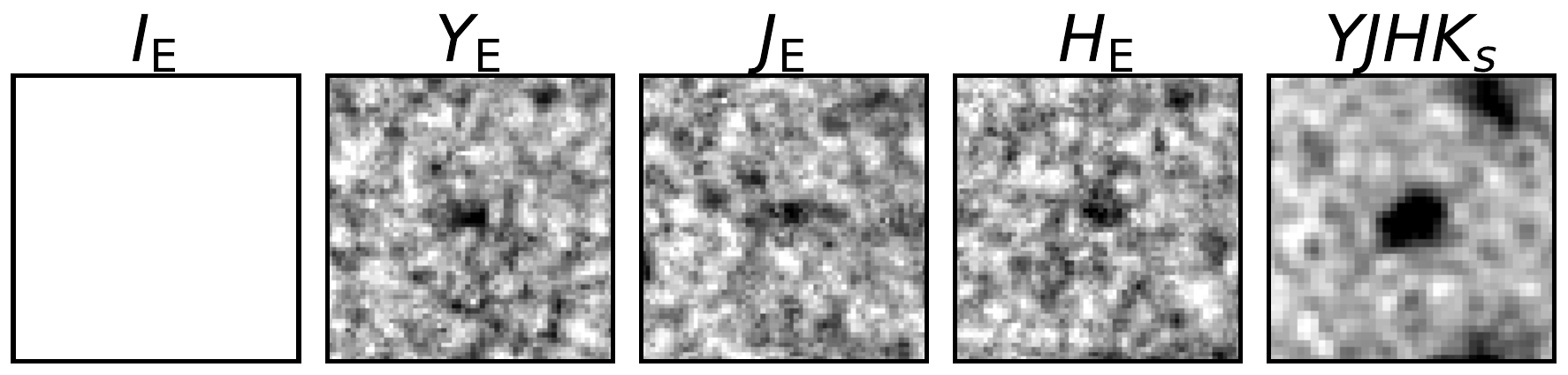}

    \vspace{12pt}

    \includegraphics[width=0.47\linewidth]{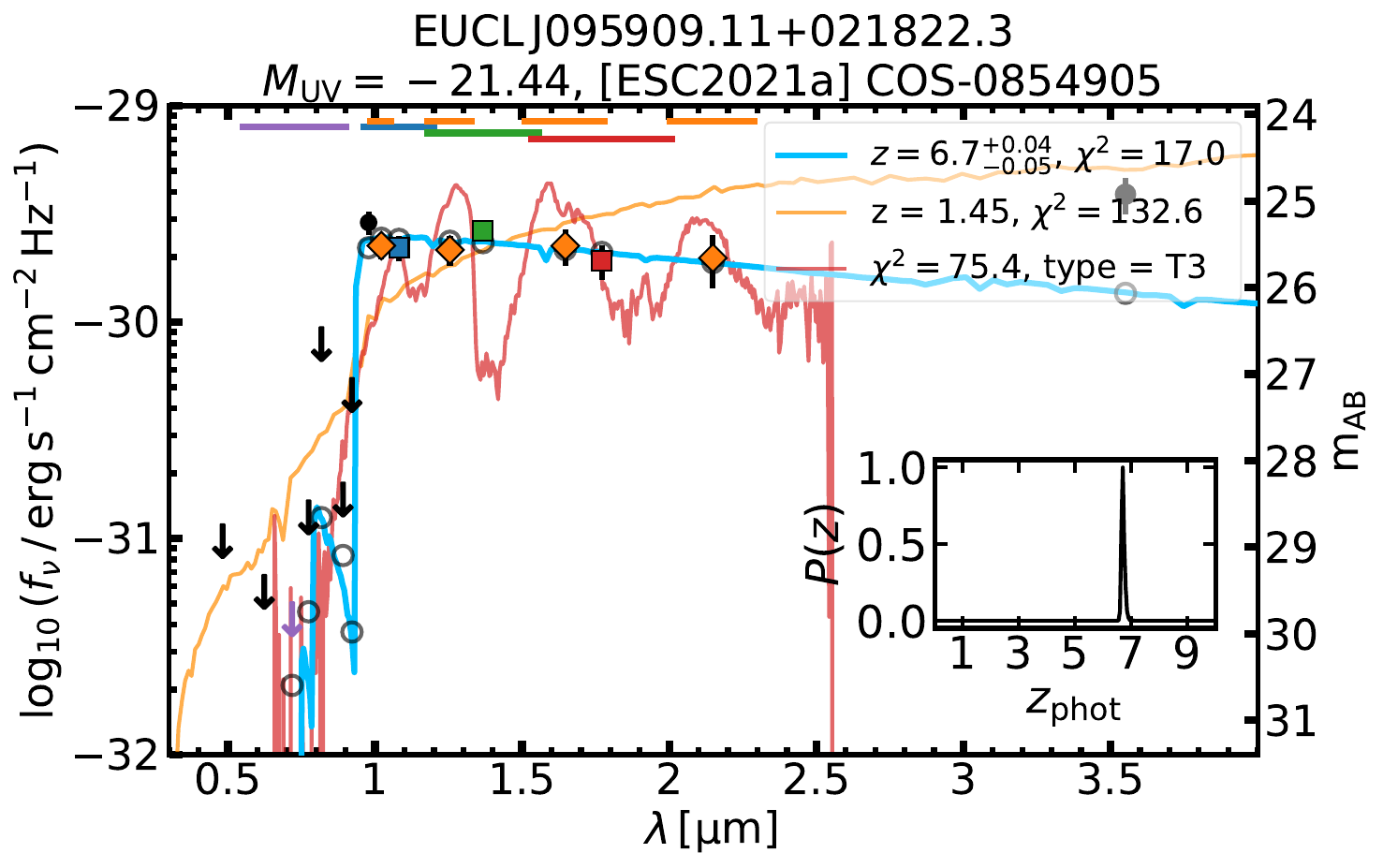}
    \includegraphics[width=0.47\linewidth]{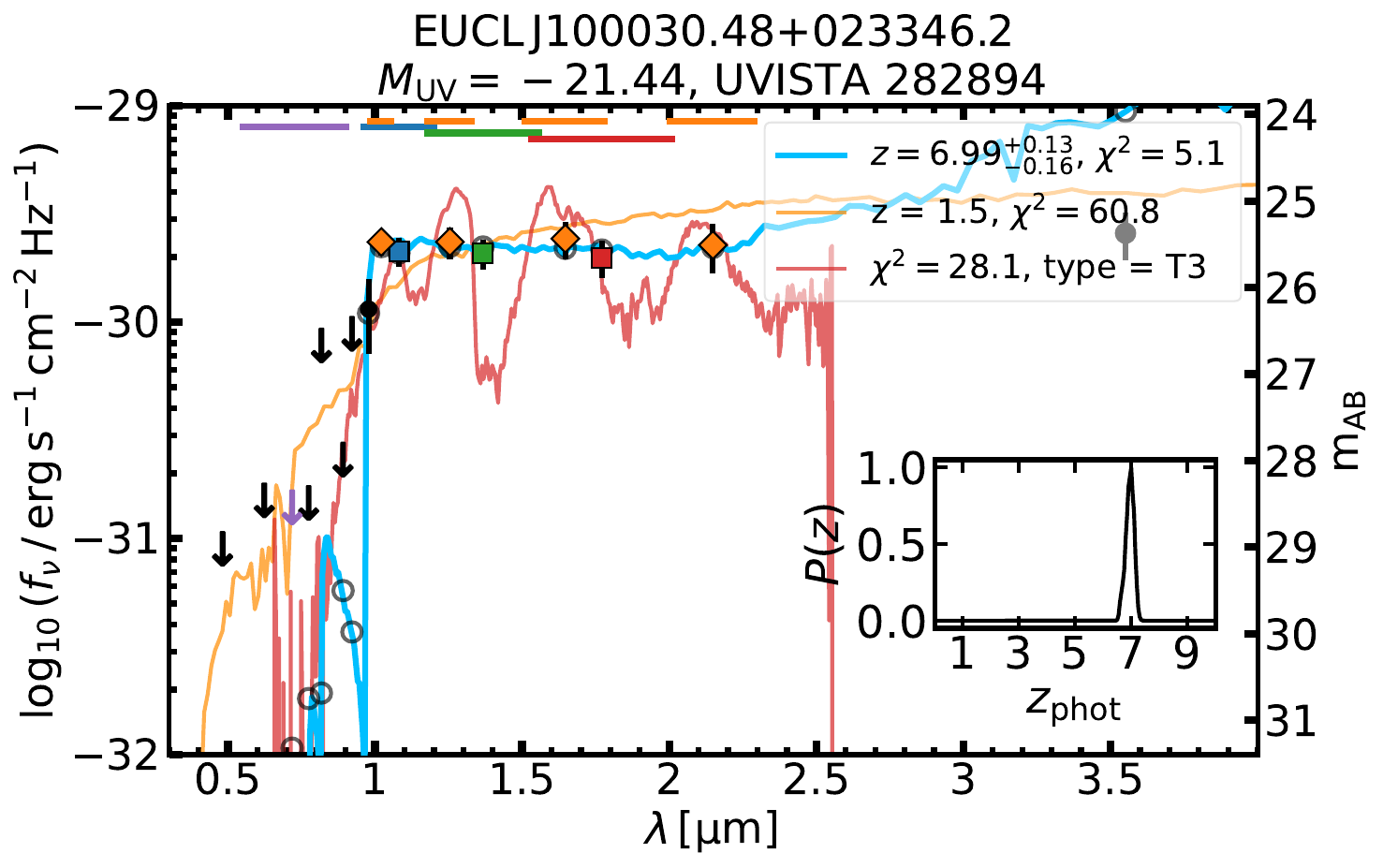}
    
    \includegraphics[width=0.44\linewidth]{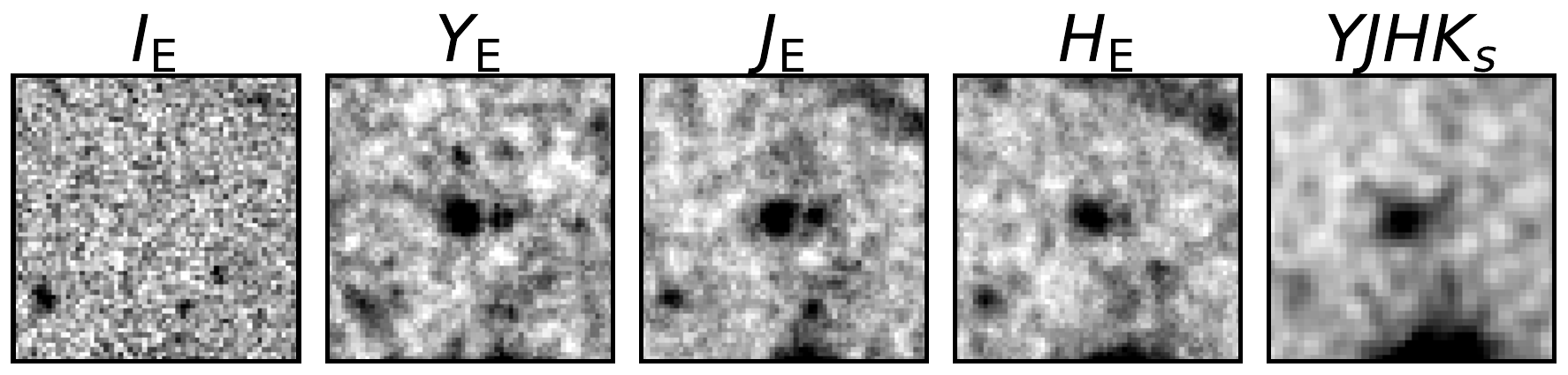}
    \hspace{12pt}
    \includegraphics[width=0.44\linewidth]{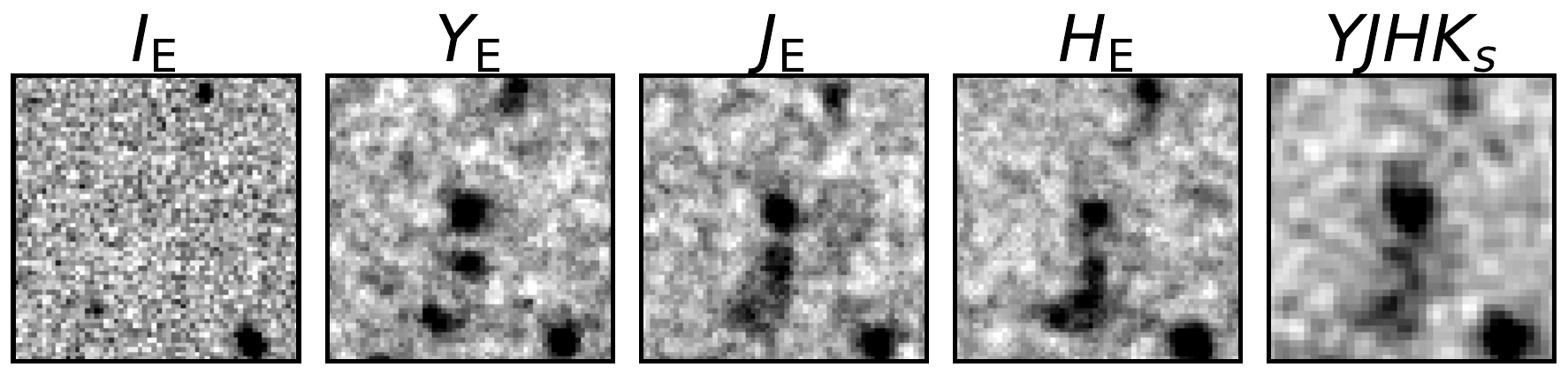}

\caption{Continued.}
\end{figure*}

\begin{figure*}
    \ContinuedFloat
    \includegraphics[width=0.47\linewidth]{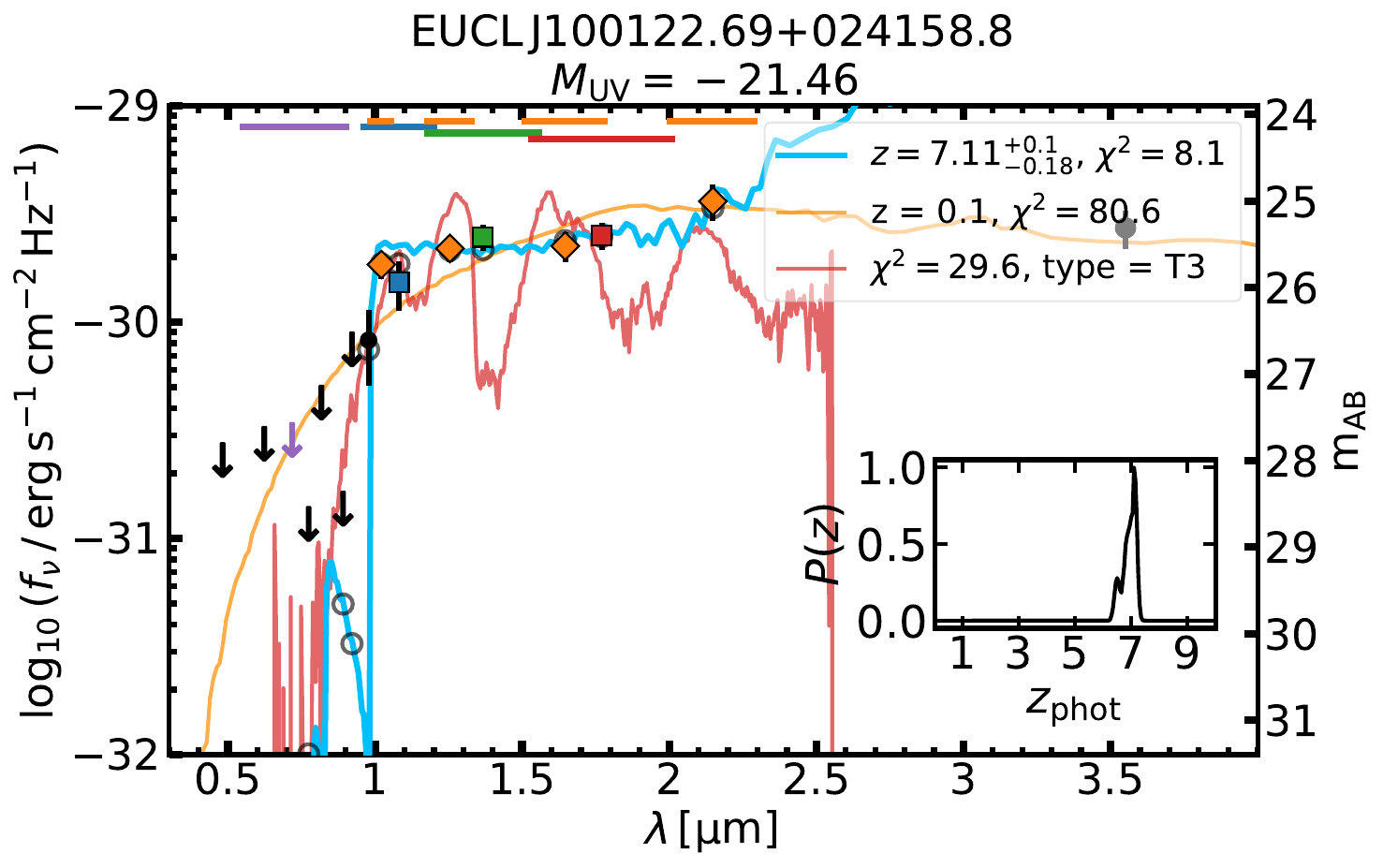}
    \includegraphics[width=0.47\linewidth]{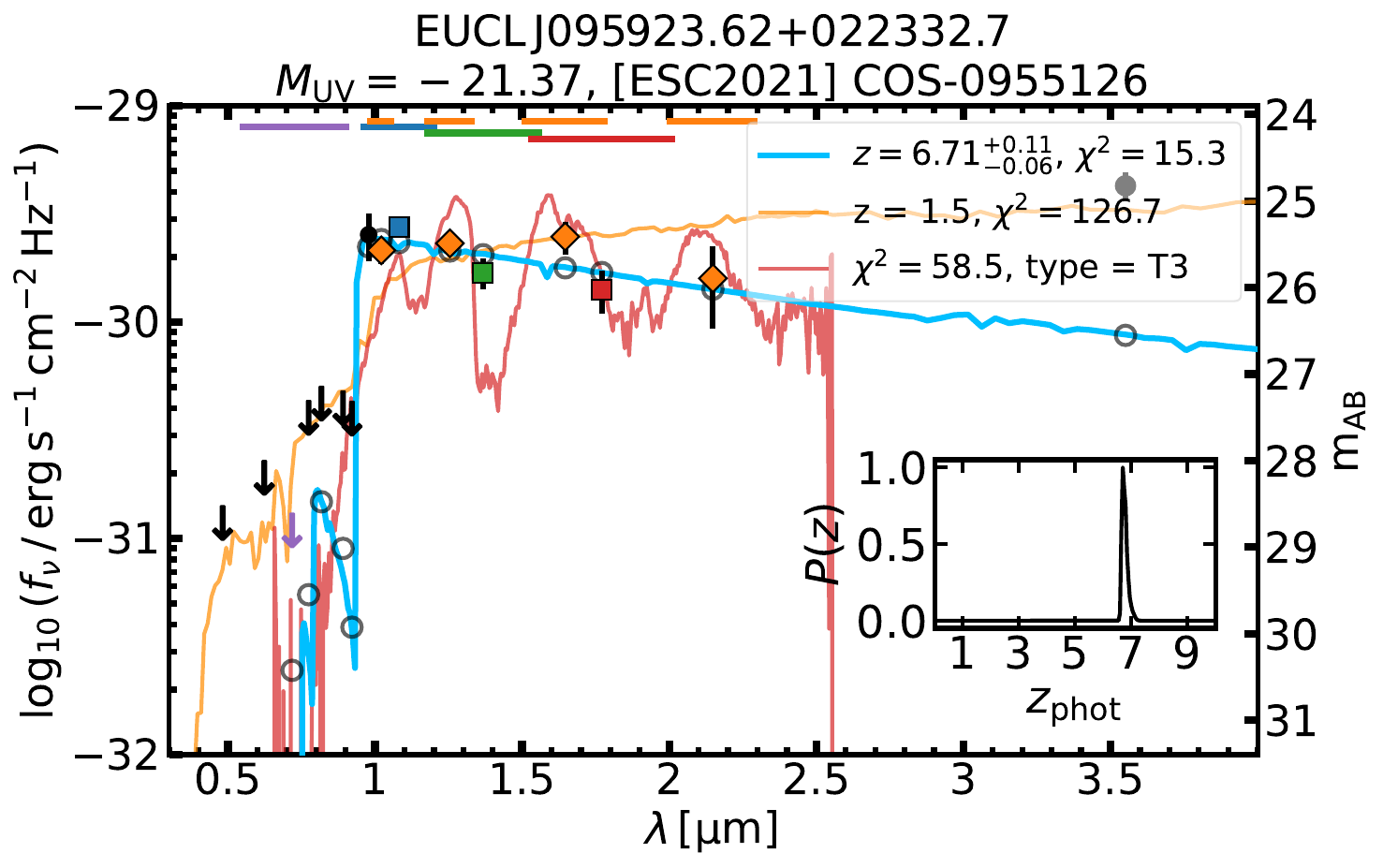}
    
    \includegraphics[width=0.44\linewidth]{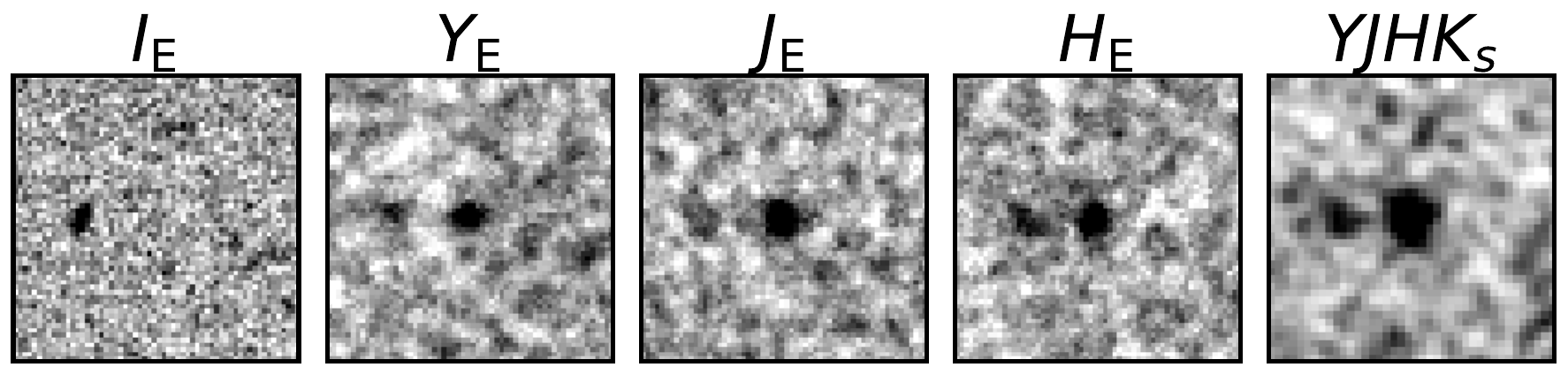}
    \hspace{12pt}
    \includegraphics[width=0.44\linewidth]{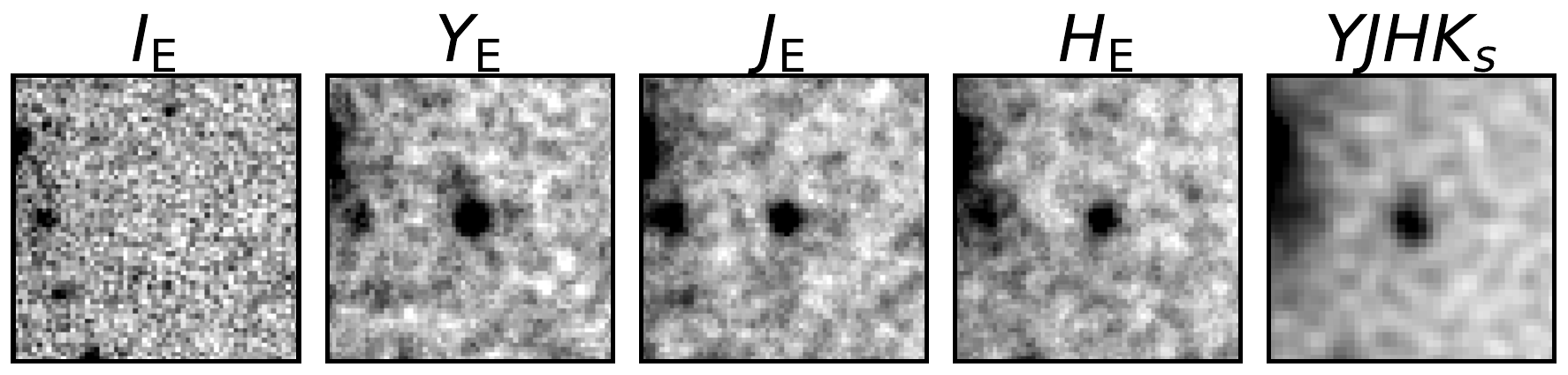}

    \vspace{12pt}

    \includegraphics[width=0.47\linewidth]{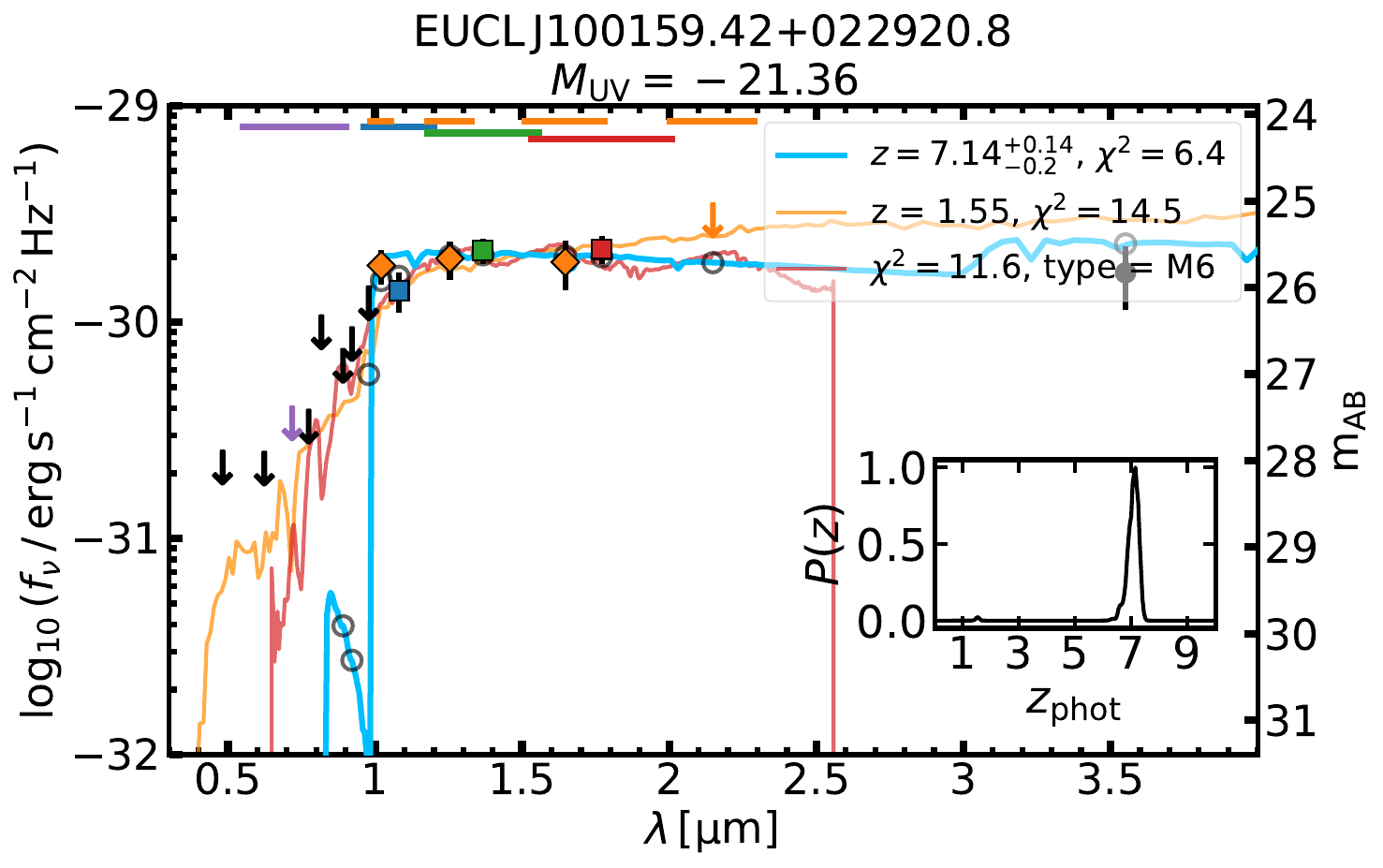}
    \includegraphics[width=0.47\linewidth]{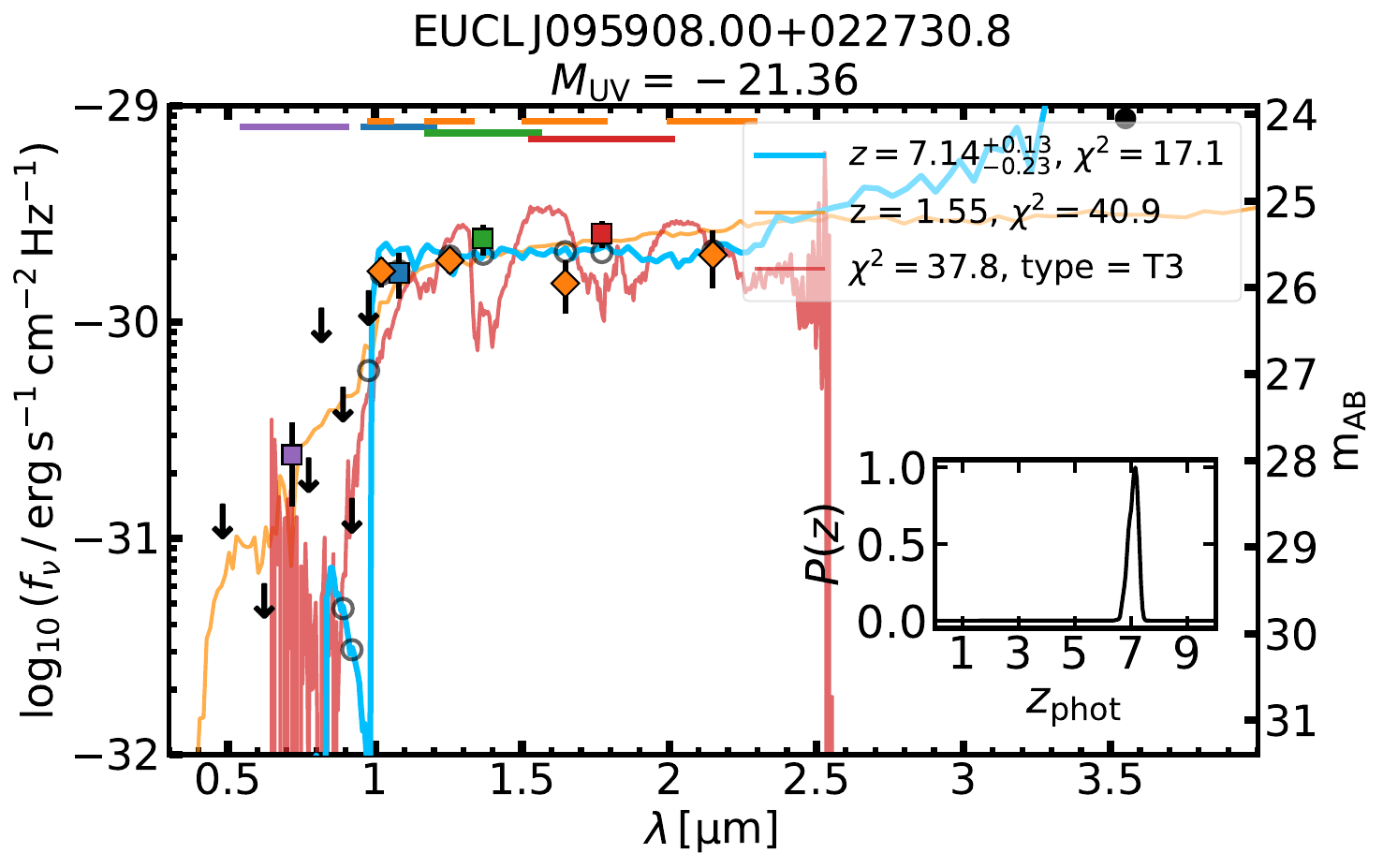}
    
    \includegraphics[width=0.44\linewidth]{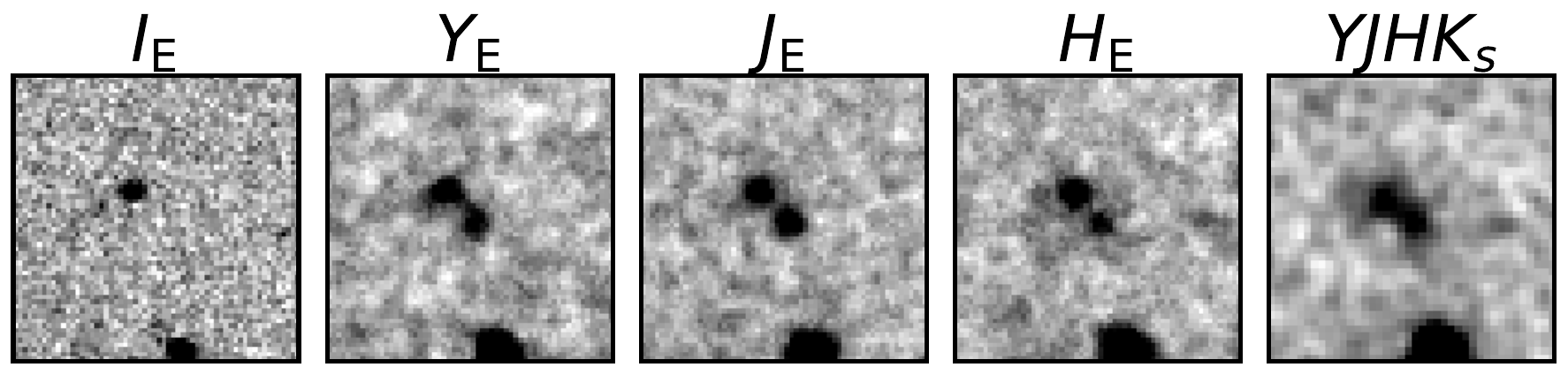}
    \hspace{12pt}
    \includegraphics[width=0.44\linewidth]{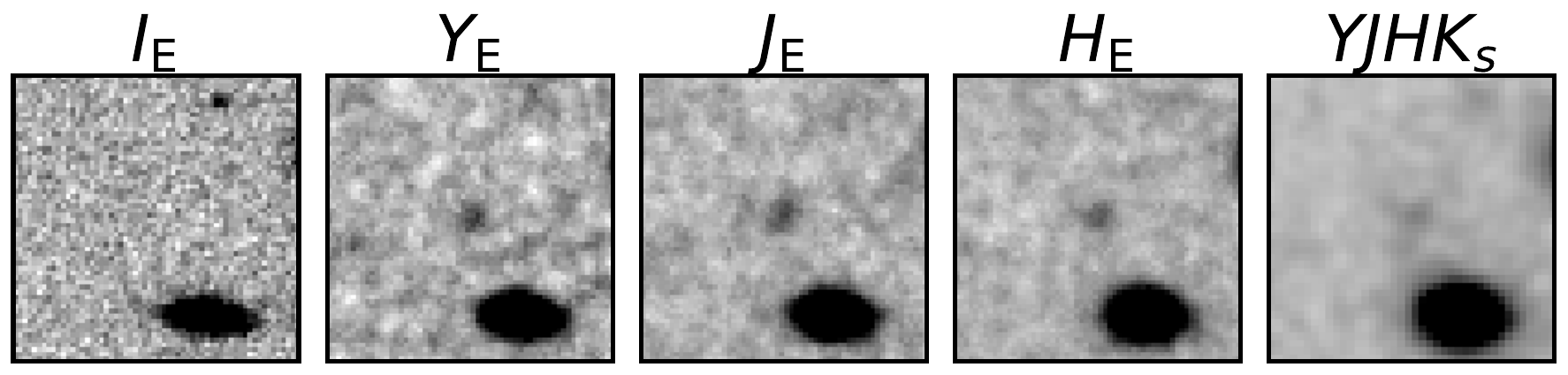}

    \vspace{12pt}

    \includegraphics[width=0.47\linewidth]{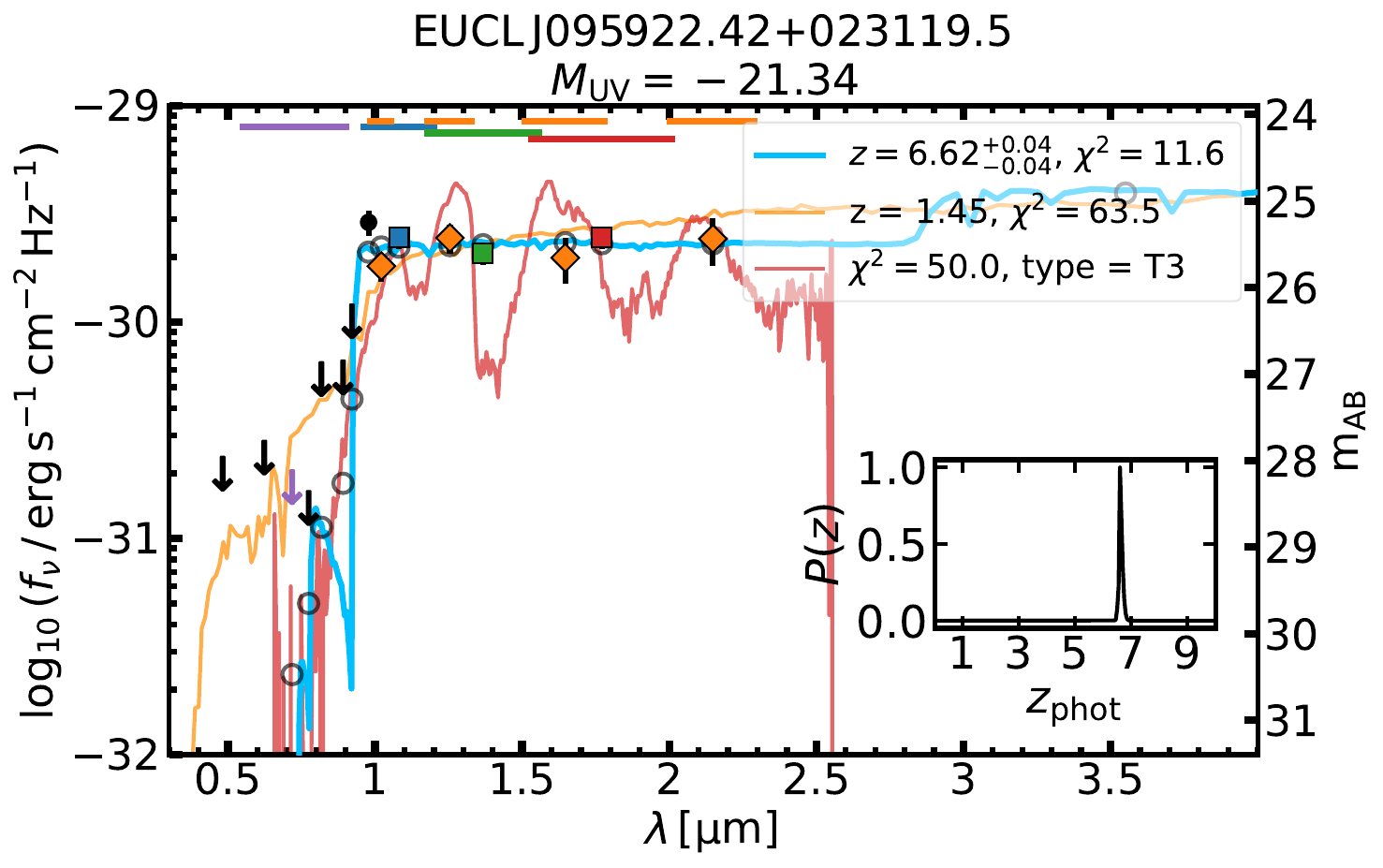}
    \includegraphics[width=0.47\linewidth]{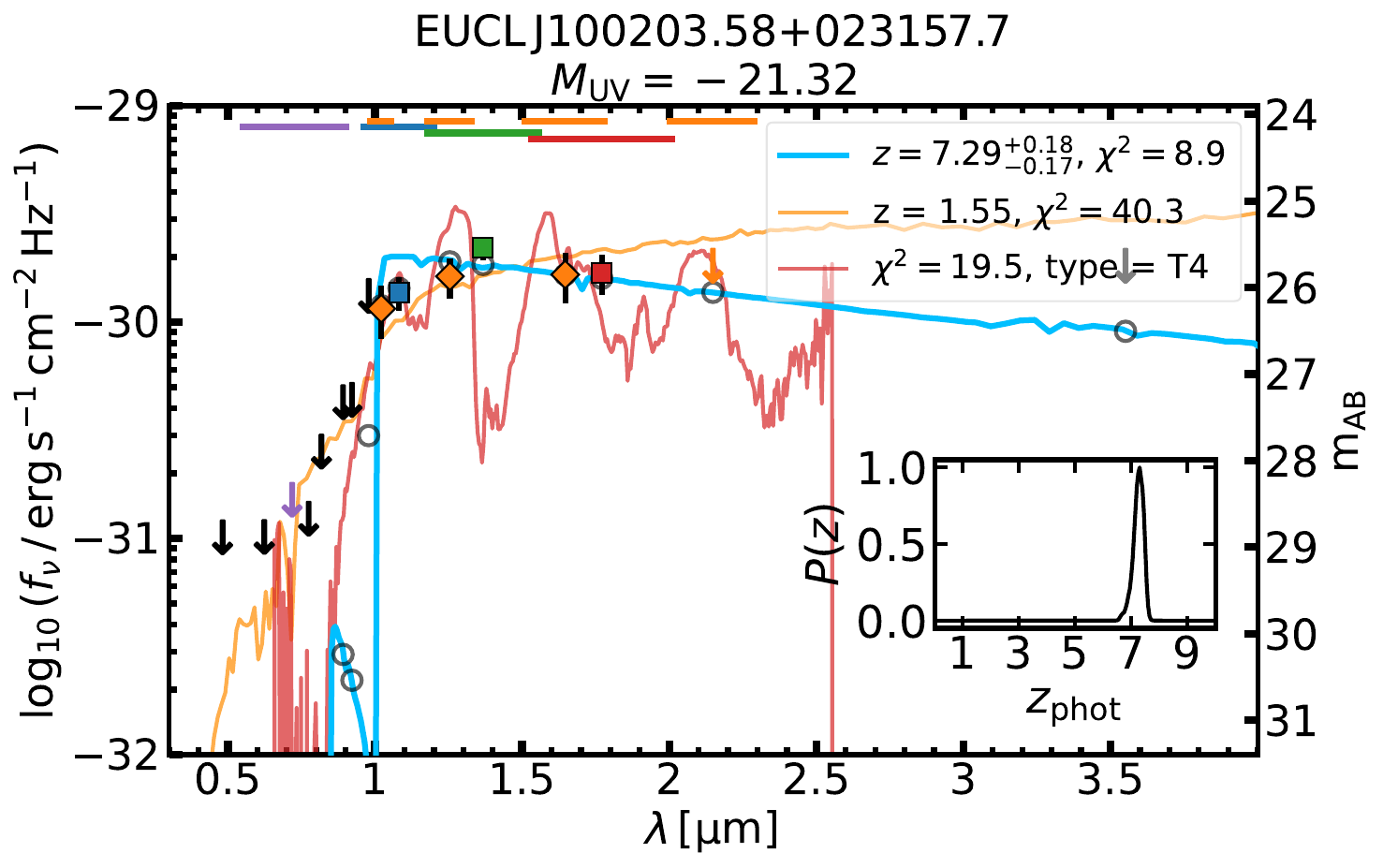}
    
    \includegraphics[width=0.44\linewidth]{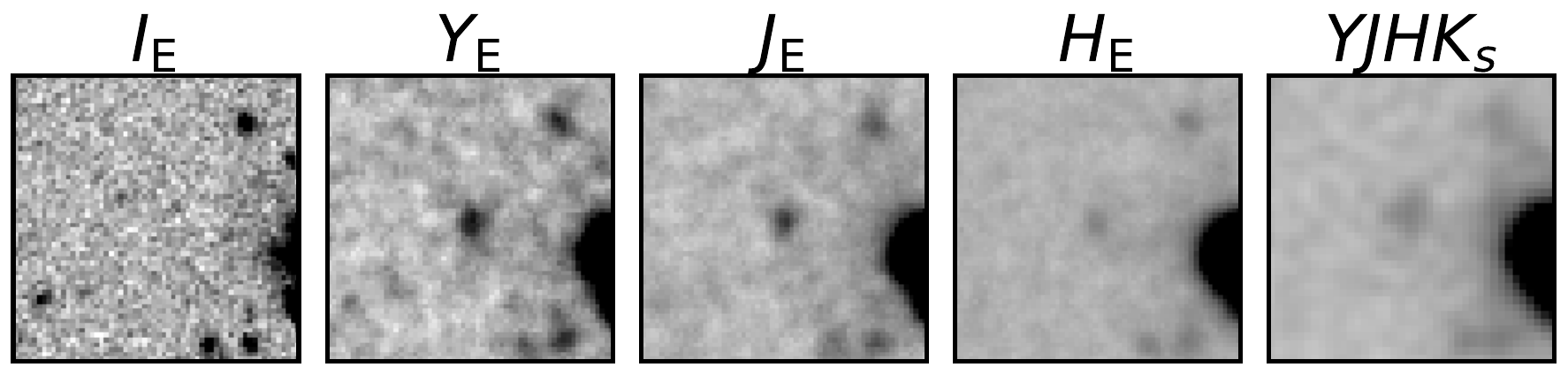}
    \hspace{12pt}
    \includegraphics[width=0.44\linewidth]{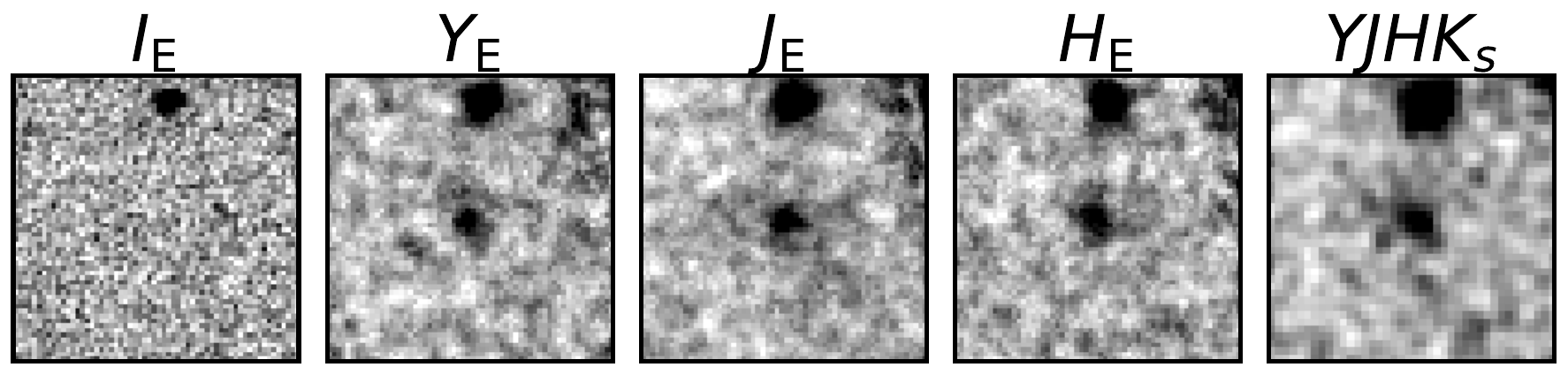}

\caption{Continued.}
\end{figure*}

\begin{figure*}
    \ContinuedFloat

    \includegraphics[width=0.47\linewidth]{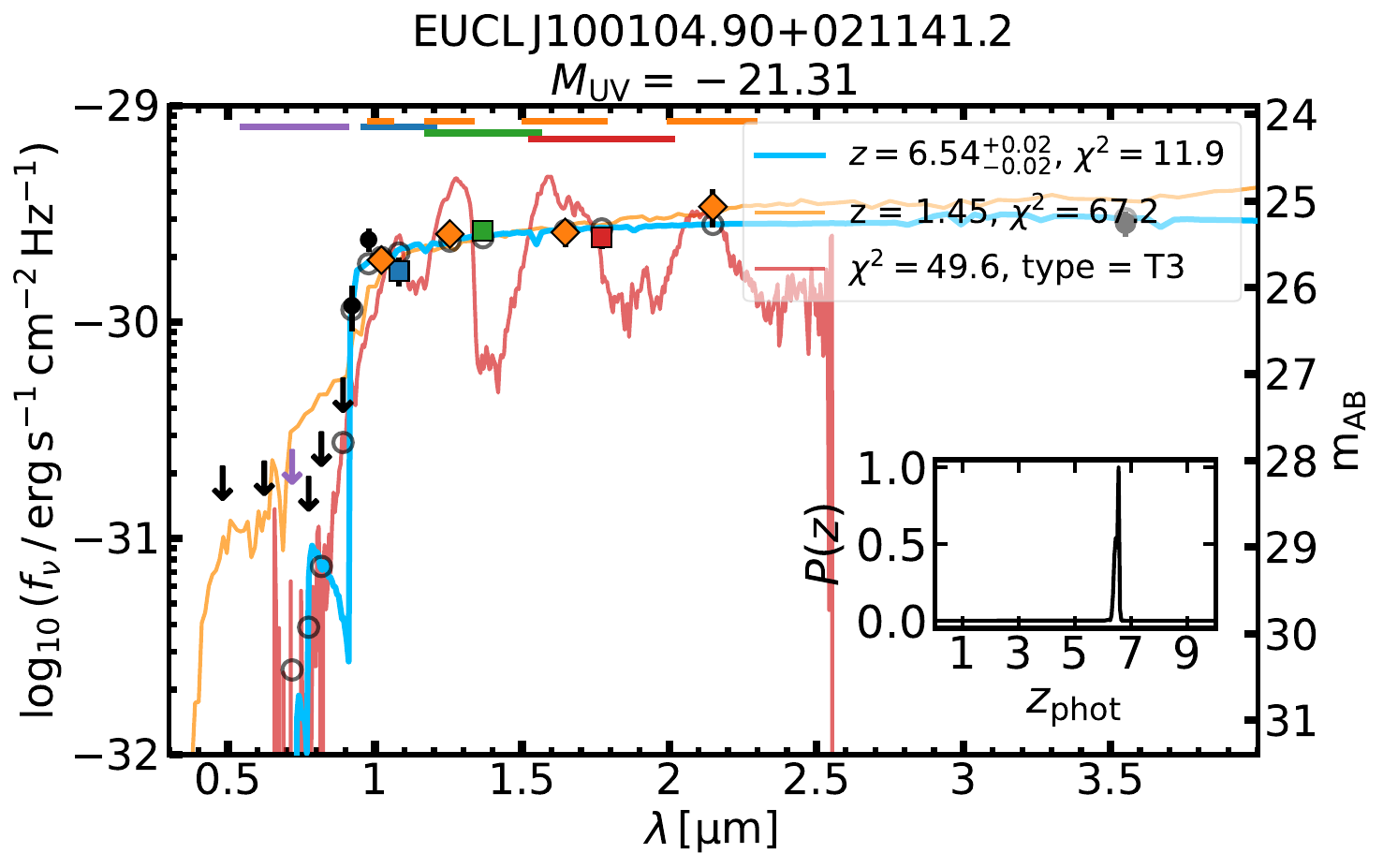}
    \includegraphics[width=0.47\linewidth]{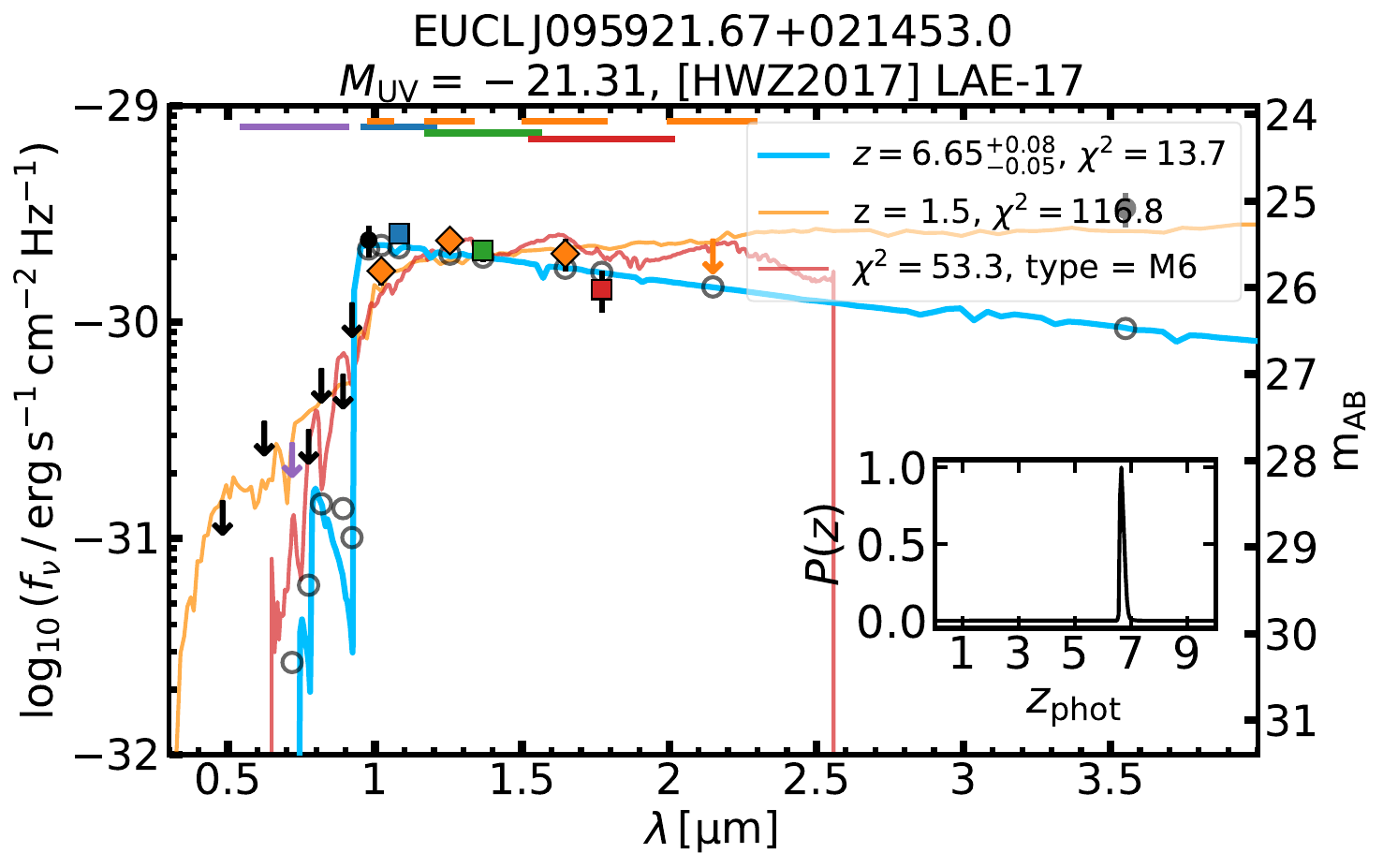}
    
    \includegraphics[width=0.44\linewidth]{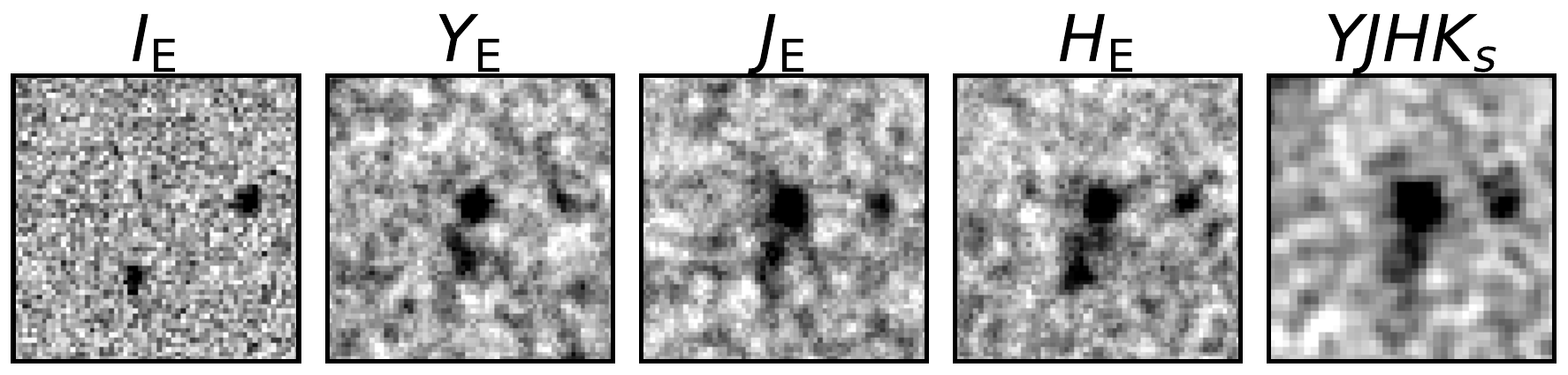}
    \hspace{12pt}
    \includegraphics[width=0.44\linewidth]{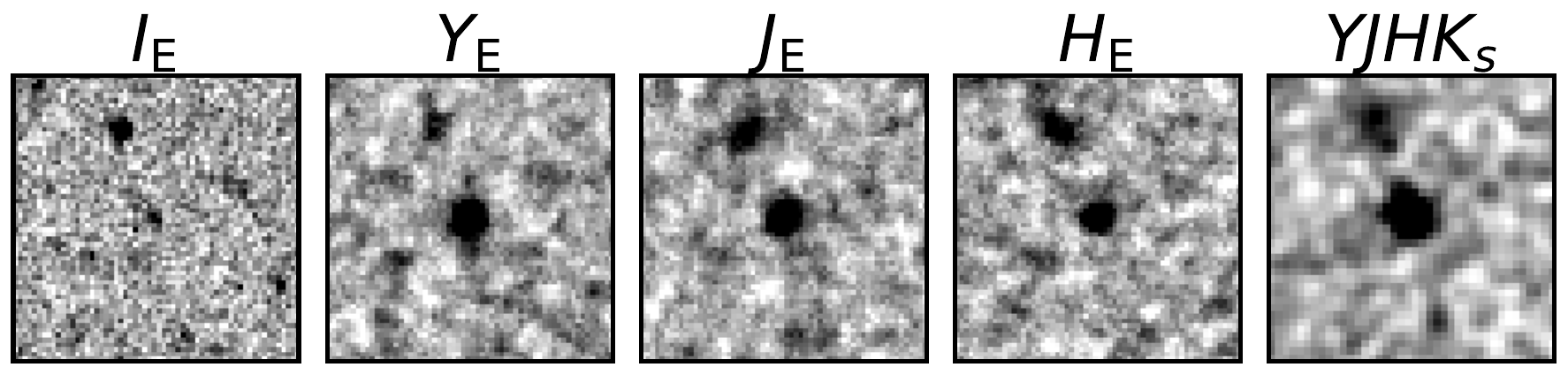}

\caption{Continued.}

\label{LastPage}
\end{figure*}

\end{appendix}

\end{document}